\documentclass[fleqn,usenatbib]{mnras}

\usepackage[T1]{fontenc}
\usepackage{ae,aecompl}

\usepackage{graphicx}	% Including figure files
\usepackage{amsmath}	% Advanced maths commands
\usepackage{amssymb}	% Extra maths symbols

\newcommand{\kms}{{km~s$^{-1}$}}

\newcommand{\oiii}{[O~{\sc iii}]}
\newcommand{\loiii}{$L_{\rm [O~III]}$}

\usepackage{newtxtext,newtxmath}

\title[Uncovering Radio-AGN in the Quasar Feedback Survey]{The Quasar Feedback Survey: Discovering hidden Radio-AGN and their connection to the host galaxy ionised gas}

\author[M.\,E.\,Jarvis et al.]{M.\,E.\,Jarvis,$^{1,2,3}$\thanks{E-mail: miranda.jarvis@gmail.com}
C.\,M.\,Harrison,$^{4}$\thanks{E-mail: christopher.harrison@newcastle.ac.uk} % 0000-0001-8618-4223
V.\,Mainieri,$^{2}$ 
D.\,M.\,Alexander,$^{5}$
F.\,Arrigoni\,Battaia,$^{1}$ \newauthor
G.\,Calistro\,Rivera,$^{2}$ 
C.\,Circosta,$^{6}$
T.\,Costa,$^{1}$ 
C.\,De\,Breuck,$^{2}$ 
A.\,C.\,Edge,$^{5}$
A.\,Girdhar,$^{2,3}$\newauthor
D.\,Kakkad,$^{7,8}$ 
P.\,Kharb,$^{9}$ 
G.\,B.\,Lansbury,$^{2}$ 
S.\,J.\,Molyneux,$^{10,2}$ 
D.\,Mukherjee,$^{11}$ \newauthor
J.\,R.\,Mullaney,$^{12}$ 
E.\,P.\,Farina,$^{1}$ 
Silpa S.,$^{9}$ 
A.\,P.\,Thomson,$^{13}$ 
S.\,R.\,Ward$^{2}$ 
\\
\\
$^{1}$Max-Planck Institut f\"ur Astrophysik, Karl-Schwarzschild-Str. 1, 85748 Garching, Germany \\
$^{2}$European Southern Observatory, Karl-Schwarzschild-Str. 2, 85748 Garching, Germany \\
$^{3}$Ludwig Maximilian Universit\"at, Professor-Huber-Platz 2, 80539 Munich, Germany\\
$^{4}$School of Mathematics, Statistics and Physics, Newcastle University, NE1 7RU, UK \\
$^{5}$Centre for Extragalactic Astronomy, Department of Physics, Durham University, South Road, Durham DH1 3LE, UK\\
$^{6}$Department of Physics \& Astronomy, University College London, Gower Street, London, WC1E 6BT, UK\\
$^{7}$European Southern Observatory, Alonso de Cordova, 3107, Vitacura Casilla 19001, Santiago, Chile\\
$^{8}$Department of Physics, University of Oxford, Denys Wilkinson Building, Keble Road, Oxford, OX1 3RH, UK\\
$^{9}$National Centre for Radio Astrophysics - Tata Institute of Fundamental Research, Pune University Campus, Post Bag 3, Ganeshkhind, Pune 411007, India\\ 
$^{10}$Astrophysics Research Institute, Liverpool John Moores University, 146 Brownlow Hill, Liverpool L3 5RF, UK\\ 
$^{11}$Inter-University Centre for Astronomy and Astrophysics, Post Bag 4, Pune - 411007, India\\
$^{12}$Department of Physics and Astronomy, The University of Sheffield, Hounsfield Road, Sheffield, S3 7RH, UK\\
$^{13}$Jodrell Bank Centre for Astrophysics, Department of Physics \& Astronomy, The Alan Turing Building, Upper Brook Street, Manchester M13 9PL, UK
}

\date{Accepted XXX. Received YYY; in original form ZZZ}

\pubyear{2020}

\begin{document}
\label{firstpage}
\pagerange{\pageref{firstpage}--\pageref{lastpage}}
\maketitle

\begin{abstract}
We present the first results from the Quasar Feedback Survey, a sample of 42 $z<0.2$, \oiii\ luminous AGN ($L_{\rm [O~III]}>10^{42.1}$~ergs~s$^{-1}$) with moderate radio luminosities (i.e.\ $L_{\rm 1.4GHz}>10^{23.4}$~W~Hz$^{-1}$; median $L_{\rm 1.4GHz}=5.9\times10^{23}$~W~Hz$^{-1}$). Using high spatial resolution ($\sim$0.3--1~arcsec), 1.5--6~GHz radio images from the Very Large Array, we find that 67~percent of the sample have spatially extended radio features, on $\sim$1--60~kpc scales. The radio sizes and morphologies suggest that these may be lower radio luminosity versions of compact, radio-loud AGN. By combining the radio-to-infrared excess parameter, spectral index, radio morphology and brightness temperature, we find radio emission in at least 57~percent of the sample that is associated with AGN-related processes (e.g.\ jets, quasar-driven winds or coronal emission). This is despite only 9.5--21~percent being classified as radio-loud using traditional criteria. The origin of the radio emission in the remainder of the sample is unclear. We find that both the established anti-correlation between radio size and the width of the \oiii\ line, and the known trend for the most \oiii\ luminous AGN to be associated with spatially-extended radio emission, also hold for our sample of moderate radio luminosity quasars. These observations add to the growing evidence of a connection between the radio emission and ionised gas in quasar host galaxies. This work lays the foundation for deeper investigations into the drivers and impact of feedback in this unique sample. 
%cut details of radi-agn selection Oct. 12 2020
\end{abstract}

%%%%%%%%%%%%%%%%%%%%%%%%%%%%%%%%%%%%%%%%%%%%%%%%%%

%%%%%%%%%%%%%%%%% BODY OF PAPER %%%%%%%%%%%%%%%%%%

\section{Introduction}

Active galactic nuclei (AGN) are amongst the most powerful phenomena in the observable universe and are generally accepted to inject significant energy into the gas in galaxies and the intergalactic medium and, consequently, play a vital role in galaxy evolution \citep[e.g.\ see reviews in][]{Alexander12,Fabian12,King15}. Over time our understanding of the physical processes in and around AGN has grown immensely. However, important questions remain unanswered. 

For AGN with powerful radio jets, particularly those with low accretion rates residing in massive galaxies in the local Universe, there is compelling evidence that AGN are able to regulate star formation through the jets injecting energy into the gaseous halos and regulating the cooling of gas onto the galaxy \citep[e.g.\ see review in][]{McNamara12,Hardcastle20}. On the other hand, there are many debated topics in the literature concerning radiatively efficient AGN such as quasars, which we define here as Type 1 or Type 2 AGN with bolometric luminosities of $L_{\rm AGN}\gtrsim10^{45}$\,erg\,s$^{-1}$ \citep[e.g.\ see review in][]{Harrison17}. For example, there is a significant debate about the dominant processes that produce the radio emission in typical `radio-quiet' AGN and through which mechanisms these AGN transfer energy to their host galaxies \citep[e.g.][]{Zakamska16,Kellermann16,Wylezalek18,Panessa19,Jarvis19}. Furthermore, the details of exactly how, or indeed if, these AGN impact upon the evolution of their host galaxies remains controversial and uncertain \citep[e.g.][]{Husemann16,VillarMartin16,Maiolino17,Harrison17,Cresci18,Perna18,Rosario18,Baron18,Scholtz18,Scholtz20,doNascimento19,Brownson20,Yesuf20,Greene20,Bluck20,Bischetti20}. Building a multi-wavelength quasar survey to address these questions is the focus of the Quasar Feedback Survey, which we introduce here.

The origin of the radio emission in the majority of AGN (i.e.\ those without powerful relativistic radio jets and that typically have radio luminosities $L_\textrm{1.4GHz}\lesssim10^{25}$~W Hz$^{-1}$) is widely debated and is the main focus of this paper. On one hand, the radio emission could be dominated by synchrotron emission caused by shocks from exploding supernovae and hence be directly related to the star formation in the host galaxy \citep[see e.g.][]{Condon92,Bonzini13,Condon13,Padovani15}. Alternatively, the majority of the radio emission could originate from the AGN \citep[see e.g.][]{Padovani17b,Jarvis19}. The three most likely ways in which AGN can generate radio emission are: i) synchrotron emission from jets \citep{Kukula98,Kharb15,Kharb17}, ii) coronal emission \citep{Laor08,Behar18} and iii) synchrotron from electrons accelerated at non-relativistic shocks that may result from wide angle sub-relativistic quasar winds \citep{Nims15,Zakamska16}, with magnetocentrifugal winds \citep{Blandford82,Everett05,Fukumura10}, thermally driven AGN winds \citep{Begelman83,Woods96,Mizumoto19} or any combination of the above also being possibilities. In this paper we identify sources in the Quasar Feedback Survey sample where AGN processes contribute significantly to the radio emission, and do not attempt to distinguish between the different AGN-related mechanisms.

Multiple studies of local and low-redshift AGN ($z<0.8$) have shown that the level of radio emission in AGN host galaxies is strongly connected to the presence of ionised outflows, which have primarily been identified through broad and asymmetric \oiii\ emission lines \citep[see e.g.][]{Heckman81,Wilson85,Veilleux91,Whittle92,Nelson96,ODea98,Holt08,Kim13,Mullaney13,Zakamska14,VillarMartin14,Santoro20}. Across these studies, this observation holds for a wide range of AGN luminosities (i.e. Seyferts and quasars) and orders of magnitude in radio luminosity (i.e. from typical `radio-quiet' AGN through to the most luminous `radio-loud' AGN). Nonetheless, some other studies, which used emission-line selected samples, have suggested that other factors (e.g.\ Eddington ratio and stellar mass) may be more important than the level of radio emission in determining the prevalence and properties of ionised outflows \citep[e.g.][]{Wang18,Rakshit18,Kauffmann19}. However, these studies are all based upon spatially unresolved data, and a more direct connection between radio emission can be observed when using spatially-resolved information. For example, \citet{Molyneux19} finds that extreme \oiii\ outflows (FWHM~$>$~1000~\kms) are more prevalent when the projected size of the radio emission is within, as opposed to extending beyond, the SDSS spectroscopic fibre (where the \oiii\ line width is measured). We further investigate the relationship between radio emission and ionised gas in this work.

We present multi-frequency radio observations of the 42 sources in the Quasar Feedback Survey, which we use in combination with archival data to identify sources where radio emission is associated with the AGN rather than star formation and to explore the radio -- outflow connection. In Section \ref{sec:survey}, we describe the survey and present the sample selection. In Section \ref{sec:observations} we describe our Karl G. Jansky Very Large Array (VLA) data, its reduction and imaging. Section \ref{sec:results} presents our results, which we then discuss in Section \ref{sec:discussion}. We summarise our conclusions in Section \ref{sec:conclusion}. In the supplementary material we provide figures showing all of the images used in this work, a summary of our observations for each target and a description of the relevant literature work on each target (Appendix A and B). 

We adopt $H_0=70$~km\,s$^{-1}$\,Mpc$^{-1}$, $\Omega_M=0.3$, $\Omega_\Lambda=0.7$ throughout, and define the radio spectral index, $\alpha$, using $S_\nu \propto \nu^{\alpha}$. We assume a \citet{Chabrier03} initial mass function (IMF).

\section{Survey Description}
\label{sec:survey}

In this paper we present the first results from our Quasar Feedback Survey, which is designed to study the spatially-resolved multi-wavelength properties of relatively low redshift ($z<0.2$) quasar host galaxies identified based on their optical emission lines. The main survey goals are to investigate: (1) the origin of radio emission; (2) the properties of multi-phase outflows and; (3) the impact that quasars have on their host galaxies. 

\subsection{Pilot Studies Summary}
\label{sec:pilot}
\begin{figure}
 \centering
 \includegraphics[width=\hsize]{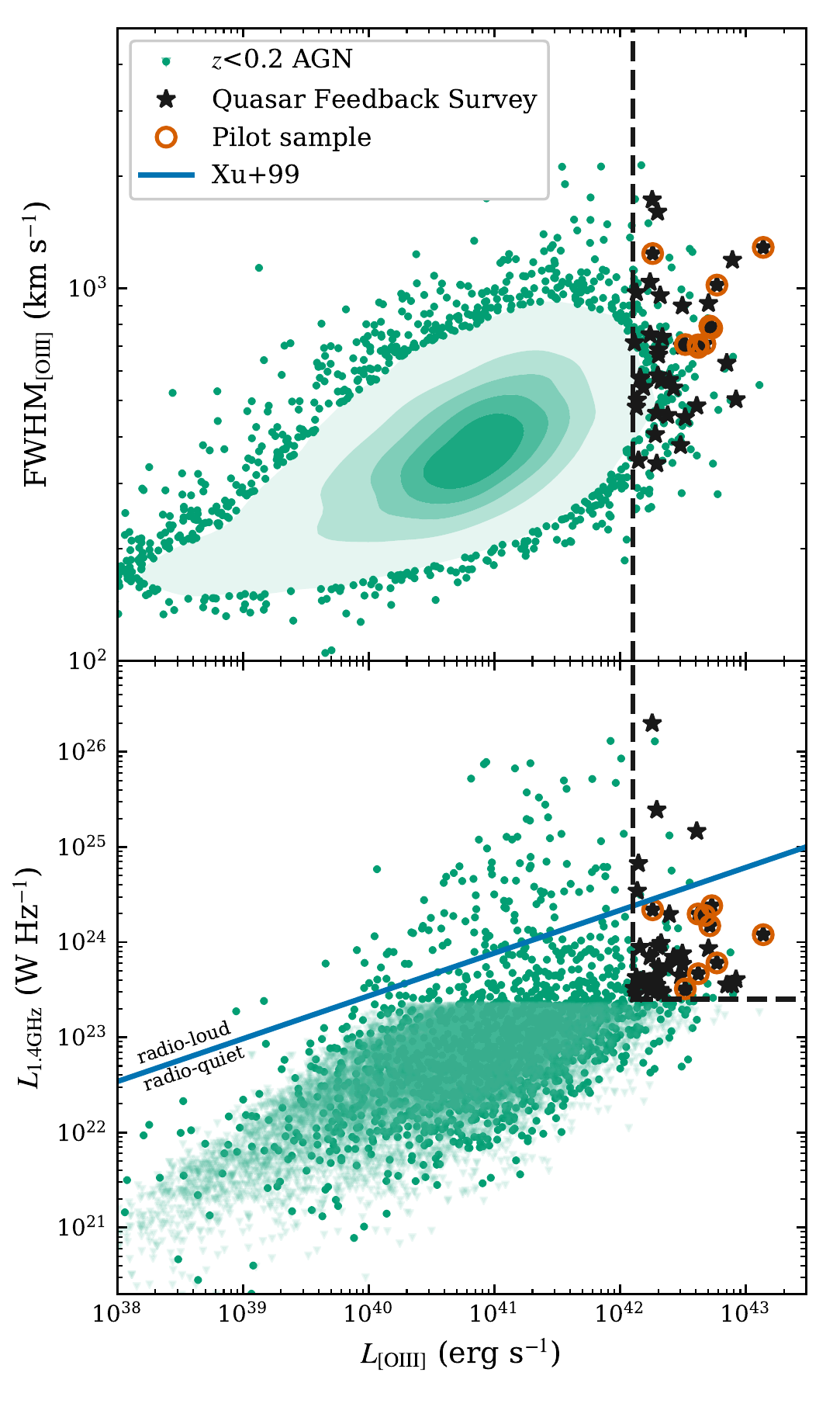}
 \caption{ The basic properties of the 42 sources in the Quasar Feedback Survey (black stars; see Table \ref{tab:sample_properties}) and the criteria used to select them (black dashed lines) compared to the overall \citet{Mullaney13} $z<0.2$ AGN population plotted as green circles and density contours. The sources in our pilot studies \citep[e.g.][]{Jarvis19} are marked with red circles in addition to the black stars denoting their membership in the Quasar Feedback Survey sample. \emph{Top:} The flux-weighted average FWHM from the \citet{Mullaney13} two Gaussian [O~{\sc iii}]$\lambda5007$ line fits versus the total observed \oiii\ luminosity. \emph{Bottom:} the 1.4~GHz radio luminosity from NVSS versus the total observed [O~{\sc iii}] luminosity (AGN with upper limits on their radio luminosity are shown as pale green triangles). The blue line marks the division between `radio-loud' and `radio-quiet' AGN from \citet{Xu99}, and the majority (37/42) of our quasars are classified as radio-quiet using this criterion. 
 }
  \label{fig:selection}
 \end{figure}

The Quasar Feedback Survey builds upon a series of papers where we presented pilot observations on a smaller sample, which were pre-selected to be quasars that were expected to host powerful ionised outflows based upon their SDSS spectra (i.e.\ \oiii\ emission-line components with full width half maximum -- FWHM~$>$~700\,km\,s$^{-1}$; see Fig.~\ref{fig:selection}; \citealt{Harrison14,Harrison15,Jarvis19}). This sample consisted of ten AGN, nine of which are part of this Quasar Feedback Survey and all of which are `radio-quiet' based on the \citet{Xu99} definition (see Section \ref{sec:radio_loud}). We used spatially resolved radio observations and IFS data to find kiloparsec-scale radio features, galaxy wide ionised gas outflows, and signatures of jet--gas interactions. In \citet{Jarvis19} we demonstrated that radio emission originating from the AGN can be effectively identified by combining spatially resolved radio observations and an observed radio excess beyond what is predicted from the radio--infrared (\emph{IR}) correlation for star-forming galaxies. In this work, we expand upon \citet{Jarvis19} by increasing our sample of sources with high resolution radio images by a factor of 4 (see Section~\ref{sec:target_selection}). 

In \citet{Jarvis20} we studied the CO emission and ultraviolet (\textit{UV}) -- far infrared (\textit{FIR}) spectral energy distributions (SEDs) for the nine Quasar Feedback Survey pilot targets. We found that, for at least seven of the targets, the host galaxies had high molecular gas fractions and short depletion times, which are consistent with those expected for the overall galaxy population with matched stellar masses and specific star formation rates. This is despite the presence of powerful quasars and ionised outflows in these sources. This work suggested that the AGN do not have an immediate appreciable impact on the global molecular gas content, but does not rule out a smaller scale impact over longer timescales. Finally, in \citet{Lansbury18} we studied the X-ray emission in one of our survey targets, J1430+1339 (nicknamed the ``Teacup AGN''; see Appendix~\ref*{sec:app:J1430+1339}), investigating both the X-ray emission from the nucleus and the spatially-resolved X-rays, which we found to be co-spatial with bubbles of radio and emission-line gas.

In the following sub-section we describe the target selection for the wider Quasar Feedback Survey sample presented in this work. Importantly, unlike the pilot studies described above, there is no selection criterion based on the width of the \oiii\ line (see Fig.~\ref{fig:selection}), and hence no explicit pre-selection on the presence or properties of AGN-driven ionised outflows. 

%-------------------------------------------------------------

 %-------------------------------------------------------------

\subsection{Target selection}
\label{sec:target_selection}
The targets presented in this work were selected from the parent sample of $\sim$24\,000 $z<0.4$ spectroscopically identified AGN from the Sloan Digital Sky Survey (SDSS) presented in \citet{Mullaney13}. We note that we re-calculated the relevant luminosities using the cosmological parameters adopted for this current paper. 

We restricted the table from \citet{Mullaney13}\footnote{\url{https://sites.google.com/site/sdssalpaka/}} to the 17\,431 sources  with a redshift of $z<0.2$. This redshift cut was designed to provide a sample of tens of powerful AGN with quasar-like luminosities (i.e.\ $L_{\rm AGN}\gtrsim10^{45}$\,erg\,s$^{-1}$; see below) whilst selecting galaxies that are still at low enough redshift to be studied with kiloparsec-scale resolution or better. Specifically, a typical range of resolutions that we expect from our high resolution multi-wavelength imaging will be approximately 0.2--0.7~arcsec \citep[see pilot studies in][]{Harrison14,Harrison15,Jarvis19}, which corresponds to 0.2--0.7\,kpc and 0.7--2.3\,kpc at the lowest redshift ($z=0.05$) and highest redshift ($z=0.2$) of the sample, respectively. 

We then restricted this $z<0.2$ AGN sample to the 220 targets with the highest [O~{\sc iii}]$\lambda5007$ luminosities, specifically those with $L_{\rm [O~III]}>10^{42.11}$~ergs~s$^{-1}$ calculated using the total flux values from the two Gaussian fits from \citet{Mullaney13}.\footnote{For the bulk of the discussion in this paper we round this value to log(\loiii/ergs~s$^{-1}$)~$>$~42.1.} This cut enabled us to select AGN with quasar level luminosities \citep[e.g.][see Section~\ref{sec:target_context}]{Reyes08}. For the last primary selection criterion for our initial sample, we selected the 67 quasars with a radio luminosity of $L_{\rm 1.4GHz}>10^{23.45}$~W~Hz$^{-1}$ (see Fig.~\ref{fig:selection} and Table \ref{tab:sample_properties}).\footnote{For the bulk of the discussion in this paper we round this value to log($L_{\rm 1.4GHz}$/W~Hz$^{-1}$)$>$23.4.} For this selection, we used the K-corrected radio luminosity values derived from the 1.4\,GHz flux densities from NRAO VLA Sky Survey \citep[NVSS;][]{Condon98} as matched in \citet{Mullaney13}. Our chosen radio luminosity cut was motivated to be above the NVSS detection limit (see Fig.~\ref{fig:selection}), where we, consequently, have a more robust understanding of the connection between the ionised gas properties and radio properties of the underlying population \citep[e.g.][]{Mullaney13,Molyneux19}.  

The final sample consists of all 42 remaining sources with a right ascension of 10~$<$~RA~$<$~300~degrees and a declination of 24~$<$~Dec or Dec~$>$~44~degrees. These sky position criteria were chosen to help facilitate observational scheduling at the VLA.\footnote{Specifically the RA cut removes two sources which are isolated in the sky making them difficult to observe and the Dec cut removes sources which would transit with elevations $>$80~degrees.} 

Based on the optical spectra of our targets, 25 are Type 2 AGN (60~percent) and the other 17 (40~percent) are Type 1 (broad line) AGN.\footnote{Determined using the H$\alpha$ line. We note that the Type~2 target J1347+1217 is misclassified as a Type~1 in the table of \citet{Mullaney13} and the Type~1 AGN J1355$+$2046 is misclassified as a non-AGN.} We show the SDSS spectra for our targets in Fig.~\ref{fig:app:J1553+4407} (for J1553$+$4407) and Figs \ref*{fig:app:J0749+4510}~--~\ref*{fig:app:J1715+6008} (for the other 41 targets). In Table~\ref{tab:sample_properties} we list all 42 targets with their sky positions, redshifts and AGN Type. 

Several of our targets have been the subject of multi-wavelength studies in the literature. We summarise the most relevant of these studies for each of the targets in Appendix~\ref*{app:objects} and, where relevant, we draw on this previous knowledge of these targets to aid with the analyses and discussion in this work. 

 %-------------------------------------------------------------
\begin{table*}
 \caption{ The Quasar Feedback Survey sources and their basic properties. (1) Source name; (2)--(3) Optical RA and Dec positions from SDSS (DR7) in the format hh:mm:ss.ss for RA and dd:mm:ss.s for Dec; (4) Redshift of the source from SDSS (DR7); (5) Rest-frame 1.4~GHz radio luminosities from NVSS using a spectral index of $\alpha=-0.7$ and assuming $S_\nu \propto \nu^{\alpha}$. The typical log errors are $\sim$0.03; (6) 1.4~GHz flux density of the target from NVSS; (7) Total observed [O~{\sc iii}]$\lambda5007$ luminosity calculated using the fluxes from \citet{Mullaney13}, the typical log errors are $\sim$0.01; (8) Flux-weighted average of the FWHM from the \citet{Mullaney13} two Gaussian fits to the [O~{\sc iii}] line profile; (9) AGN Type based on the width of the H$\alpha$ line from SDSS DR7 spectra from \citet{Mullaney13}, with 1 for broad line (Type 1) and 2 for narrow line (Type 2).
 \newline
 $^{\dagger}$ These nine sources have been previously studied by our group \citep[see e.g.][Fig,~\ref{fig:selection}]{Harrison14,Jarvis19,Jarvis20}.
 \newline
 $^*$ This source has also been published as 4C+54.27.
 \newline
 $^{\dagger\dagger}$ This source has also been published as Mrk783.
 \newline
 $^\ddagger$ This source has also been published as PKS1345+12 and 4C12.50.
 }
	\centering  
\begin{tabular}{llllcrcrc}
\hline
\multicolumn{1}{c}{Name} & \multicolumn{1}{c}{RA}& \multicolumn{1}{c}{Dec} & \multicolumn{1}{c}{$z$} & \multicolumn{1}{c}{log($L_\textrm{1.4GHz}$)} & \multicolumn{1}{c}{$S_\textrm{1.4GHz}$}& \multicolumn{1}{c}{log($L_\textrm{\oiii}$)} & \multicolumn{1}{c}{FWHM$_\textrm{\oiii}$} & AGN type \\
 &\multicolumn{1}{c}{ (J2000)} & \multicolumn{1}{c}{(J2000)}&  & \multicolumn{1}{c}{(/W Hz$^{-1}$)}  & \multicolumn{1}{c}{(mJy)} & \multicolumn{1}{c}{(/erg s$^{-1}$)} & \multicolumn{1}{c}{(\kms)}& \\
 \multicolumn{1}{c}{ (1) } &  \multicolumn{1}{c}{(2)}& \multicolumn{1}{c}{(3)} &\multicolumn{1}{c}{ (4) }& \multicolumn{1}{c}{(5)}&  \multicolumn{1}{c}{(6)} & \multicolumn{1}{c}{(7)}& \multicolumn{1}{c}{(8)} & \multicolumn{1}{c}{(9)}    \\
 \hline
J0749+4510 & 07:49:06.50 & +45:10:33.9 & 0.192 & 25.2 & 147.1$\pm$5.1 & 42.61 & 484$\pm$16 & 1 \\
J0752+1935 & 07:52:17.84 & +19:35:42.2 & 0.117 & 23.9 & 24.9$\pm$0.9 & 42.70 & 913$\pm$26 & 1 \\
J0759+5050 & 07:59:40.96 & +50:50:24.0 & 0.055 & 23.5 & 44.9$\pm$1.4 & 42.32 & 958$\pm$19 & 2 \\
J0802+4643 & 08:02:24.35 & +46:43:00.6 & 0.121 & 23.5 & 8.9$\pm$0.5 & 42.11 & 716$\pm$20 & 2 \\
J0842+0759 & 08:42:05.57 & +07:59:25.5 & 0.134 & 23.9 & 16.6$\pm$1.0 & 42.49 & 899$\pm$15 & 1 \\
J0842+2048 & 08:42:07.50 & +20:48:40.1 & 0.181 & 23.5 & 3.6$\pm$0.4 & 42.24 & 752$\pm$34 & 1 \\
J0907+4620 & 09:07:22.36 & +46:20:18.0 & 0.167 & 24.5 & 47.1$\pm$1.5 & 42.13 & 502$\pm$31 & 2 \\
J0909+1052 & 09:09:35.49 & +10:52:10.5 & 0.166 & 23.6 & 6.0$\pm$0.5 & 42.28 & 406$\pm$17 & 2 \\
J0945+1737$^{\dagger}$ & 09:45:21.33 & +17:37:53.2 & 0.128 & 24.3 & 45.6$\pm$1.4 & 42.67 & 711$\pm$12 & 2 \\
J0946+1319 & 09:46:52.57 & +13:19:53.8 & 0.133 & 23.6 & 7.9$\pm$0.5 & 42.89 & 1193$\pm$26 & 1 \\
J0958+1439$^{\dagger}$ & 09:58:16.88 & +14:39:23.7 & 0.109 & 23.5 & 10.9$\pm$0.5 & 42.52 & 707$\pm$33 & 2 \\
J1000+1242$^{\dagger}$ & 10:00:13.14 & +12:42:26.2 & 0.148 & 24.3 & 34.8$\pm$1.1 & 42.62 & 706$\pm$16 & 2 \\
J1010+0612$^{\dagger}$ & 10:10:43.36 & +06:12:01.4 & 0.098 & 24.3 & 92.4$\pm$3.3 & 42.26 & 1241$\pm$30 & 2 \\
J1010+1413$^{\dagger}$ & 10:10:22.95 & +14:13:00.9 & 0.199 & 24.1 & 11.1$\pm$0.5 & 43.14 & 1289$\pm$18 & 2 \\
J1016+0028 & 10:16:53.82 & +00:28:57.1 & 0.116 & 23.6 & 11.8$\pm$0.9 & 42.18 & 543$\pm$16 & 2 \\
J1016+5358 & 10:16:23.76 & +53:58:06.1 & 0.182 & 23.5 & 3.2$\pm$0.5 & 42.13 & 976$\pm$39 & 2 \\
J1045+0843 & 10:45:05.16 & +08:43:39.0 & 0.125 & 23.8 & 17.6$\pm$1.2 & 42.42 & 541$\pm$30 & 1 \\
J1055+1102 & 10:55:55.34 & +11:02:52.2 & 0.145 & 23.5 & 5.7$\pm$0.4 & 42.52 & 451$\pm$16 & 2 \\
J1100+0846$^{\dagger}$ & 11:00:12.38 & +08:46:16.3 & 0.100 & 24.2 & 59.8$\pm$1.8 & 42.71 & 793$\pm$16 & 2 \\
J1108+0659 & 11:08:51.03 & +06:59:01.4 & 0.181 & 24.0 & 11.1$\pm$0.5 & 42.32 & 566$\pm$11 & 2 \\
J1114+1939 & 11:14:23.81 & +19:39:15.8 & 0.199 & 24.0 & 8.4$\pm$0.5 & 42.30 & 584$\pm$19 & 2 \\
J1116+2200 & 11:16:25.34 & +22:00:49.3 & 0.143 & 23.7 & 10.5$\pm$0.5 & 42.38 & 457$\pm$18 & 2 \\
J1222-0007 & 12:22:17.85 & -00:07:43.7 & 0.173 & 23.6 & 4.5$\pm$0.4 & 42.85 & 631$\pm$37 & 2 \\
J1223+5409$^*$ & 12:23:13.21 & +54:09:06.5 & 0.156 & 25.4 & 387.6$\pm$11.6 & 42.29 & 339$\pm$11 & 1 \\
J1227+0419 & 12:27:39.83 & +04:19:32.4 & 0.180 & 23.8 & 6.8$\pm$0.5 & 42.49 & 701$\pm$23 & 1 \\
J1300+0355 & 13:00:07.99 & +03:55:56.5 & 0.184 & 24.3 & 21.6$\pm$0.8 & 42.39 & 568$\pm$10 & 1 \\
J1302+1624$^{\dagger\dagger}$ & 13:02:58.83 & +16:24:27.7 & 0.067 & 23.5 & 32.9$\pm$1.1 & 42.29 & 463$\pm$28 & 1 \\
J1316+1753$^{\dagger}$ & 13:16:42.90 & +17:53:32.5 & 0.150 & 23.8 & 10.3$\pm$0.5 & 42.77 & 1022$\pm$28 & 2 \\
J1324+5849 & 13:24:18.25 & +58:49:11.6 & 0.192 & 23.8 & 7.1$\pm$0.5 & 42.24 & 1040$\pm$72 & 1 \\
J1347+1217$^\ddagger$ & 13:47:33.36 & +12:17:24.3 & 0.121 & 26.3 & 5397.2$\pm$161.9 & 42.25 & 1730$\pm$35 & 2 \\
J1355+2046 & 13:55:50.20 & +20:46:14.5 & 0.196 & 23.6 & 4.2$\pm$0.5 & 42.30 & 1605$\pm$59 & 1 \\
J1356+1026$^{\dagger}$ & 13:56:46.10 & +10:26:09.0 & 0.123 & 24.4 & 62.9$\pm$1.9 & 42.73 & 783$\pm$6 & 2 \\
J1430+1339$^{\dagger}$ & 14:30:29.88 & +13:39:12.0 & 0.085 & 23.7 & 26.5$\pm$0.9 & 42.62 & 695$\pm$25 & 2 \\
J1436+4928 & 14:36:07.21 & +49:28:58.5 & 0.128 & 23.6 & 9.4$\pm$0.9 & 42.16 & 572$\pm$30 & 2 \\
J1454+0803 & 14:54:34.35 & +08:03:36.7 & 0.130 & 23.7 & 12.7$\pm$0.6 & 42.30 & 664$\pm$20 & 1 \\
J1509+1757 & 15:09:13.79 & +17:57:10.0 & 0.171 & 24.0 & 11.8$\pm$0.5 & 42.30 & 685$\pm$26 & 1 \\
J1518+1403 & 15:18:56.27 & +14:03:19.0 & 0.139 & 23.6 & 8.6$\pm$0.9 & 42.13 & 481$\pm$25 & 2 \\
J1553+4407 & 15:53:15.94 & +44:07:49.3 & 0.197 & 23.6 & 4.2$\pm$0.5 & 42.48 & 379$\pm$14 & 2 \\
J1555+5403 & 15:55:01.44 & +54:03:26.9 & 0.180 & 23.5 & 3.4$\pm$0.5 & 42.33 & 741$\pm$52 & 1 \\
J1655+2146 & 16:55:51.37 & +21:46:01.8 & 0.154 & 23.6 & 6.6$\pm$0.5 & 42.92 & 504$\pm$27 & 1 \\
J1701+2226 & 17:01:58.24 & +22:26:41.9 & 0.197 & 24.8 & 64.0$\pm$2.6 & 42.14 & 346$\pm$5 & 1 \\
J1715+6008 & 17:15:44.05 & +60:08:35.6 & 0.157 & 23.9 & 13.8$\pm$0.6 & 42.16 & 576$\pm$8 & 2 \\

	\hline   
	\end{tabular}
    
\label{tab:sample_properties} 

	\end{table*}
 %-------------------------------------------------------------

\subsection{Our sample in context}
\label{sec:target_context}

In Table~\ref{tab:sample_properties} we present the radio luminosities and \oiii\ luminosities of our sample which are visually presented in the context of the parent population in Fig.~\ref{fig:selection}. Our \oiii\ luminosity cut of $L_{\rm [O~III]}>10^{42.11}$~ergs~s$^{-1}$ represents the 1.3~percent most luminous targets from the parent sample within our redshift range of interest (i.e.\ $z<0.2$). The \oiii\ luminosity range of our targets (i.e.\ $L_{\rm [O~III]}=10^{42.1}$--$10^{43.1}$~ergs~s$^{-1}$)\footnote{We note that if we use the reddening corrected values from \citet{Mullaney13} the range in \oiii\ luminosities is $L_{\rm [O~III]}=10^{42.3}$--$10^{45.1}$~ergs~s$^{-1}$.} corresponds to an AGN luminosity of $L_{\rm AGN}=10^{45.6}$--$10^{46.6}$\,erg\,s$^{-1}$, assuming a bolometric correction of 3500 \citep{Heckman04}. Since it is known that \oiii\ often over-predicts the AGN luminosity by about an order of magnitude \citep{Schirmer13,Hainline13} this roughly corresponds to a cut of $L_{\rm AGN}\gtrsim10^{45}$\,erg\,s$^{-1}$, which can be verified using optical--far infrared SED analyses \citep[e.g.][]{Jarvis19,Jarvis20}. The range of luminosities covered by our sample is designed to be representative of typical AGN luminosities found during the peak epoch of cosmic black hole growth. Specifically, the knee of the bolometric AGN luminosity function is $L_{\rm AGN}=10^{45.8}$\,erg\,s$^{-1}$ at $z=0.8$ and $L_{\rm AGN}=10^{46.6}$\,erg\,s$^{-1}$ at $z=2$, following \citet{Shen20} (see also \citealt{Hopkins07}), which overlap with the AGN luminosities in this sample.

\begin{figure*}
\centering
\includegraphics[width=16cm]{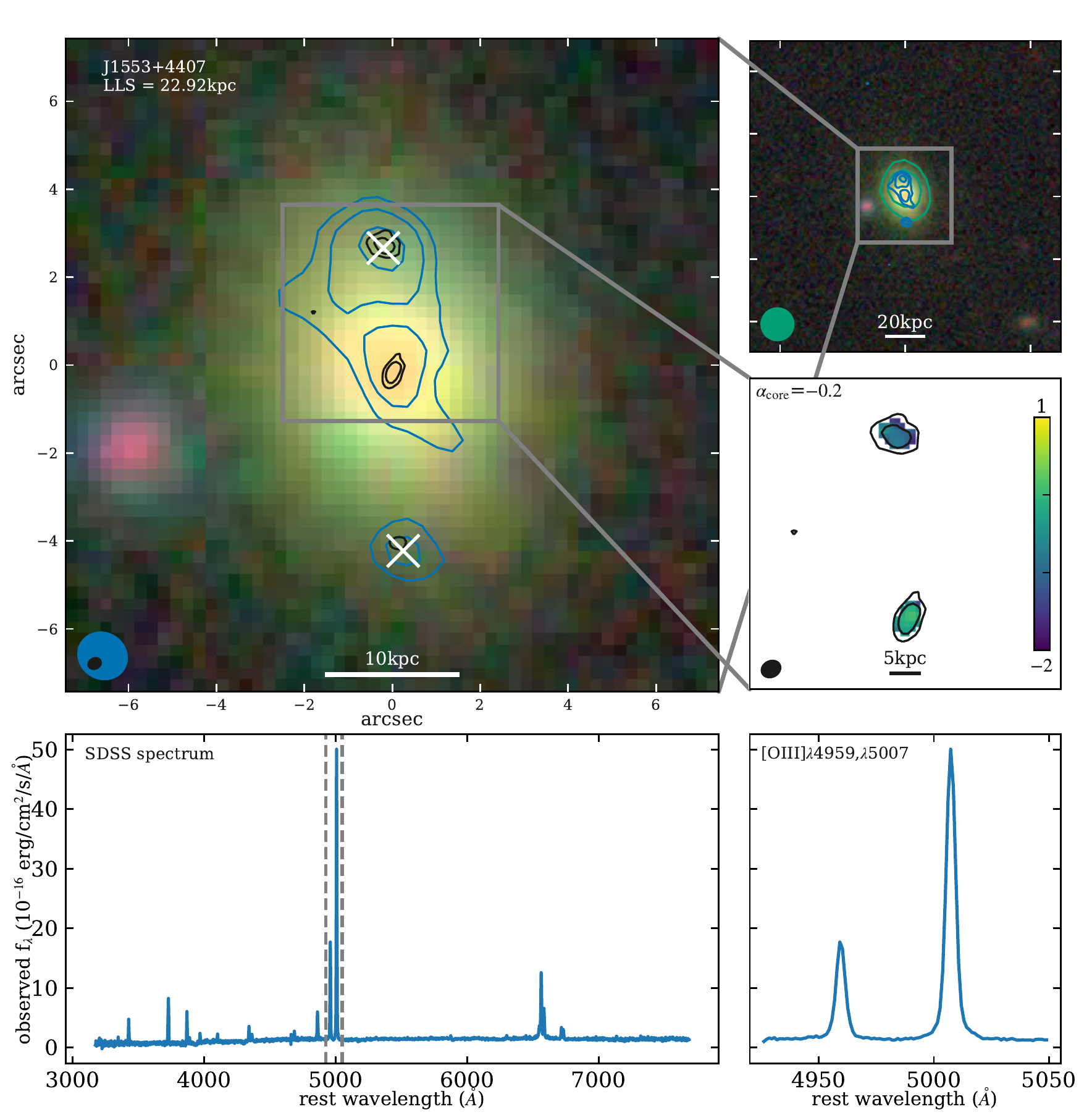}
\caption{ This figure highlights the observed properties of J1553$+$4407 and the data used in this paper; similar figures for the entire sample are shown in Appendix \ref*{app:objects}. In the upper left is an rgb image from the DESI Legacy Imaging Survey in the (\textit{z,r,g}) bands, with contours from our VLA L-band data (1.5~GHz; $\sim$1~arcsec resolution; MR image) in blue and our C-band data (6~GHz; $\sim$0.3~arcsec resolution; HR image) in black. North is up and east is to the left. In the top right is a zoom-out with FIRST survey radio contours overlaid (green; 1.4~GHz; $\sim$5~arcsec resolution; LR image) in addition to the L-band (MR) contours in blue. The white X's mark the radio peaks used to calculate the size for this source in Table \ref{tab:radio_properties} (in this case using the L-band [MR] data; see Section \ref{sec:size} for details). The middle right panel shows a zoom-in with the in-band spectral index map (colour bar inset to the right) from our C-band data (i.e.\ 4--8~GHz), with flux density contours (as to the left) overlaid. The value measured for the core spectral index (i.e.\ $\alpha_\textrm{core}$; see Section \ref{sec:core}) is given in the top left of this panel. Radio contours for all data are plotted at levels of $\pm$[4, 8, 16, 32, 64, 128]$\sigma$ with dashed lines for the negative contours (in this case there are no negative features with at least $4\sigma$ significance). The beams for the relevant radio images are shown in the bottom left of each sub-plot in the matching colours. The bottom row shows the SDSS spectrum of the source with vertical grey dashed lines marking the region of the \oiii\ doublet for which a zoom-in is shown to the right (the y-axis for both spectra is the same). 
}
 \label{fig:app:J1553+4407}
\end{figure*}

In Table~\ref{tab:sample_properties} we also give the flux-weighted average of the FWHM of the two \oiii\ emission-line Gaussian components as fit by \citet{Mullaney13}.\footnote{We note that we take the absolute value of the tabulated FWHM parameters from \citet{Mullaney13}.} These values for our full sample are represented in Fig.~\ref{fig:selection}, compared to the parent sample, where it can be seen that these emission-line widths are representative of the parent sample with FWHM values spanning 340--1730\,km\,s$^{-1}$. Sampling the full range in \oiii\ emission-line profiles is a crucial step forward for our wider Quasar Feedback Survey presented here, compared to our previous work that focused on sources with broad \oiii\ line widths \citep[e.g.][see Section~\ref{sec:pilot}]{Harrison14,Jarvis19}. In Fig.~\ref{fig:app:J1553+4407} (for J1553$+$4407) and Figs~\ref*{fig:app:J0749+4510}--\ref*{fig:app:J1715+6008} we show zoom-ins of the \oiii\ emission-line profile for each source revealing the broad diversity in profiles across the sample. This includes narrow and symmetric profiles (e.g.\ Fig.~\ref*{fig:app:J1055+1102}), double peaked profiles (e.g.\ Fig.~\ref*{fig:app:J1316+1753}) and broad asymmetric profiles (e.g.\ Fig.~\ref*{fig:app:J1509+1757}). 

%-------------------------------------------------------------
 %-------------------------------------------------------------

Our survey compliments several other ongoing and previous observational campaigns that also aim to study the drivers of ionised outflows and the broader impact of bolometrically luminous AGN (including quasars) on their host galaxies. We focus our comparison on surveys that aim to study the \oiii\ emission because this was the basis of our sample selection (see Fig.~\ref{fig:selection}). 

At higher redshifts (i.e.\ larger than about 0.4--4.0) there are large IFS studies of powerful radio galaxies \citep{Nesvadba17a} and X-ray selected AGN \citep{Harrison16,Circosta18} as well as a number of smaller studies of the spatially-resolved gas kinematics in luminous AGN \citep[e.g.][]{Harrison12a,Liu13b,Liu14,Carniani15,Perna15,Brusa15,Kakkad16,Zakamska16b,Vietri18}. The advantage of studying sources at high redshift is that they cover the peak epoch of cosmic black hole growth, and, consequently there are a larger number of powerful sources where the impact of AGN/quasars might be most important \citep[e.g.][]{Hopkins07,McCarthy11}. However, although advantages can be gained by the use of adaptive optics \cite[e.g.][]{Fischer19,Davies20,Kakkad20}, it is typically not possible to reach kiloparsec-scale resolution observations at high redshift. The exception is by studying rare gravitationally lensed AGN host galaxies \citep[e.g.][]{Fischer19,Chartas20}. The strong effect of surface brightness dimming also makes it very observationally expensive to study high redshift targets in comparable detail to their lower redshift counterparts. This is a very important consideration; for example, the faint radio emission associated with `radio-quiet' AGN and quasars can be extended on sub-kiloparsec scales and the kinematic signatures of outflows can be hard to disentangle from host galaxy dynamics without high spatial resolution and high signal-to-noise ratio data \cite[e.g.][]{Jarvis19,Venturi18}. 

By studying AGN in the very local Universe (i.e.\ $z\lesssim0.05$), excellent spatial resolution and sensitivity can be achieved. There are many spatially-resolved studies investigating ionised, and multi-phase, outflows from samples of local galaxies hosting AGN \citep[e.g.][]{HusemannCARS,Ramakrishnan19,Mingozzi19,Schonell19,Wylezalek20,Davies20RIC,Venturi21}. However, these surveys are naturally dominated by lower power AGN than our survey where we aim to focus on powerful quasars. 

Overall, by focusing on sources at $z\approx0.1$, our survey allows for both a reasonable spatial resolution ($\sim$1\,kpc) and sensitivity, whilst also allowing us to build up a sample of bolometrically luminous AGN \citep[also see e.g.][]{Husemann13,Karouzos16,Rupke17,Rose18,Rackshit18,Balmaverde20,Santoro20}. Importantly, we have selected our quasars from a well studied parent sample (Section~\ref{sec:target_selection}) and, as is usually not the case for bolometrically luminous `radio-quiet' quasars, we will start with a detailed study of the spatially-resolved radio properties.  

\subsection{Defining the targets as radio-loud or radio-quiet}
\label{sec:radio_loud}

The majority (37/42) of the Quasar Feedback Survey sample are `radio-quiet' as defined by the \oiii\ and radio luminosity division of \citet{Xu99} (see Fig.~\ref{fig:selection}). These classifications are tabulated in Table \ref{tab:radio_properties}. This fraction of 88~percent `radio-quiet' in our sample, is consistent with the `radio-quiet' fraction of the overall AGN population \citep[i.e.\ $\sim$90~percent;][]{Zakamska04}. 

We use the criterion of \citet{Xu99} as our primary definition of `radio-quiet' because it can be applied to both the Type~1 and Type~2 AGN in our sample. However, for the Type~1 sources we additionally consider the commonly used radio-loudness parameter $R$: the ratio of radio to optical brightness \citep{Kellermann89}. Specifically, we follow the prescription of \citet{Ivezic02} which calculates $R$ using the radio flux density at 1.4~GHz and the SDSS \emph{i} band magnitude. We calculate $R$ using: 
\begin{equation}
R=0.4 (m_i-t)    ,
\end{equation}
where $m_i$ is the \emph{i} band magnitude from SDSS DR16 \citep{Ahumada20}\footnote{\url{http://skyserver.sdss.org/dr16/en/tools/crossid/crossid.aspx}} using either the de Vaucouleurs or exponential profile fits to the galaxy luminosity profile (whichever provides a better fit) and $t$ is the `AB radio magnitude' calculated using: 
\begin{equation}
    t=-2.5\log\left(\frac{S_\textrm{1.4GHz}}{3631 \textrm{Jy}}\right),
\end{equation}
where $S_\textrm{1.4GHz}$ is the NVSS flux density as tabulated in Table \ref{tab:sample_properties}. Of the 17 Type 1 AGN in our sample, we find that seven would be classified as radio-loud, with $R>1.0$. This includes the three Type 1 AGN that were also radio-loud following the \citet{Xu99} criterion. We note that the four Type 1 targets, which did not meet the \citet{Xu99} criterion, but with $R>1.0$ have extremely modest values of $1.1\leq R \leq 1.4$ and three additional sources lie right on the boundary with $R=1$. This is compared to the typical $R$ value for radio-loud sources from \citet{Ivezic02} of 2.8. The sources in our sample which meet both `radio-loud' criteria have values of $R=1.9$--2.5. The values of $R$ calculated for all of the Type 1 sources in our sample are tabulated in Table \ref{tab:radio_properties}.

This result is consistent with recent works finding an overlap in the $R$ parameters measured for Seyfert galaxies and FRI radio galaxies, suggesting that these form a continuous population \citep{Kharb14}. Our work demonstrates that this also extends to the most bolometrically luminous AGN.

The majority of this paper is dedicated to employing more physically motivated tests to establish how many of the `radio-quiet' sources also have radio emission associated with the AGN, which we refer to as `Radio-AGN' to distinguish from the more traditional `radio-loud' and `radio-quiet' divisions.\footnote{We note that all classical radio-loud AGN would be classified as Radio-AGN as we define the term.}

\section{Observations and data reduction} 
\label{sec:observations}

We observed our entire sample with the VLA to produce spatially resolved radio maps with around 1 and 0.3~arcsec resolution at 1.5 and 6~GHz, respectively. The VLA data for nine of the sample have previously been presented in \citet{Jarvis19} \citep[highlighted in Table \ref{tab:sample_properties}; see also][]{Harrison15}. These observations were taken under proposal ID.\ 13B-127 [PI.\ Harrison], with observations carried out between 2013 December 1 and 2014 May 13 in four configuration -- frequency combinations: (1) A-array in L-band (1--2\,GHz; $\sim$1.0~arcsec\,resolution); (2) A-array in C-band (4--8\,GHz; $\sim$0.3~arcsec\,resolution); (3) B-array in L-band (1--2\,GHz; $\sim$4.0~arcsec\,resolution) and (4) B-array in C-band (4--8\,GHz; $\sim$1.0~arcsec\,resolution). For more details about these observations and the reduction of these data see \citet{Jarvis19}. 

The remainder of the sample were observed under proposal ID.\ 18A-300 [PI.\ Jarvis] using the A-array in the L- and C-bands (around 1.0 and 0.3~arcsec resolution at 1--2 and 4--8~GHz, respectively), and were observed between 2018 March 2 -- 22. We observed for 7.1~hours in the L-band (5--10~minutes per source) and 9.5~hours in C-band (5--25~minutes per source). To improve the \emph{uv} coverage, each source was observed in at least two scans spread across the 1--2~hour observing blocks. We began each observing block with a $\sim$10\,minute scan of a standard calibration source (3C147, 3C286 or 3C138). We performed 3\,minute scans (including slew-time) of nearby (typically within 10$^\circ$) phase calibrators every 10--15~minutes. The details of the VLA observations for each source (phase calibrator and observation date) are given in Appendix \ref*{app:obs}.

\subsection{Data reduction and imaging}
\label{sec:reduction_and_imaging}

We reduced and imaged the VLA 18A-300 data using {\sc casa} version 5.1.2-4. Specifically, we used the VLA {\sc casa} Calibration Pipeline default VLA recipe for Stokes I continuum. We then split each of the science targets into their own measurement set using 3~s time binning. To remove radio frequency interference from the L-band data (where it is stronger than in the C-band), we then used the {\sc casa} task `flagdata' in `tfcrop' mode to apply additional flags to each L-band measurement set. After inspecting the pipeline calibrated data, we performed additional
manual flagging of bad data using the {\sc casa} tool `plotms' for the C-band observations of J1045$+$0843 and J1108$+$0659, specifically flagging all of the data from antenna 6 in the first scan of each source.

To provide consistent analysis with the remainder of the sample, we re-imaged the VLA 13B-127 data for this work, rather than using the images which were created for \citet{Jarvis19}. In order to take into account the broad and varying bandwidths of our observations, all of the VLA images we present were made using the Multi-Frequency Synthesis (MFS) mode of the `clean' function. We weighted the baselines using the Briggs weighting scheme with a robustness parameter of 0.5 \citep{Briggs95}. For the C-band data we performed this imaging using two Taylor terms to model the frequency dependence of the sky emission in order to simultaneously generate in-band (i.e.\ 4--8~GHz) spectral index maps as well as maps depicting the reliability of the spectral index. The details of the new VLA images presented in this work (noise and beam size) are given in Appendix \ref*{app:obs}. All of these images, together with cutouts from the VLA `Faint Images of the Radio Sky at Twenty-cm' survey \cite[FIRST;][]{Becker95}\footnote{\url{https://third.ucllnl.org/cgi-bin/firstcutout}} and archival optical images from the Dark Energy Spectroscopic Instrument (DESI) Legacy Imaging Survey in the (\textit{z,r,g}) bands \citep{Dey19}\footnote{\url{https://www.legacysurvey.org/}} are shown in Appendix \ref*{app:objects}.

Throughout this work we use three sets of images for each source which we refer to as low resolution (LR), medium resolution (MR) and high resolution (HR) throughout, defined as: 
\begin{itemize}
    \item LR: We used 2~arcmin cutouts from FIRST to characterise the low resolution ($\sim$5~arcsec; 1.4\,GHz) radio properties of our sample. 
    
    \item MR: Images using L-band (1.5\,GHz) A-array using Briggs weighting and a robustness parameter of 0.5 ($\sim$1~arcsec resolution).

    \item HR: The highest resolution images used in this work are the C-band (6\,GHz) A-array VLA data imaged with Briggs weighting and a robustness parameter of 0.5 ($\sim$0.3~arcsec resolution).   
    
\end{itemize}

We note additionally, that J1347$+$1217 is a VLA calibrator and so we did not observe it with the sample presented here, relying on the extensive archival information available (see Appendix~\ref*{sec:app:J1347+1217}). In particular we use L- and C-band A-array images with similar beam sizes to those from the remainder of the sample (i.e.\ around 1 and 0.3~arcsec respectively) from the NRAO VLA Archive Survey Images Pilot Page.\footnote{\href{http://www.aoc.nrao.edu/~vlbacald/read.shtml}{http://www.aoc.nrao.edu/$\sim$vlbacald/read.shtml}}

\section{Analyses and Results}
\label{sec:results}

The primary goals of this paper are to establish what fraction of the Quasar Feedback Survey targets have radio emission with an AGN origin (as opposed to star formation) and to explore the relationship between the radio and ionised outflow properties of `radio-quiet' quasars. To that end, in the following sub-sections, we describe our radio size measurements followed by the four separate measurements that we made to search for radio emission associated with an AGN. In the final sub-section we summarise the final identification of `Radio-AGN' across the sample.

\subsection{Radio sizes}
\label{sec:size}

We measure the largest linear size (LLS; specifically the observed / projected physical size in kpc) following two common approaches in the literature (e.g.\ \citealt{Kunert-Bajraszewska10,Doi13}; also see \citealt{Jarvis19}).\footnote{We note that the sizes presented in this work (see Table \ref{tab:radio_properties}) for the sample that was also studied in \citet{Jarvis19} differ in some cases from those presented in that work, due to slightly different images and methods used (see Appendix\,\ref*{app:objects} for details of the specific sources).} 

Firstly, we define LLS as the distance between the farthest peaks in the lowest resolution image where the source is not featureless (see Section~\ref{sec:morph}). We identified the location of these peaks using the Photutils `find\_peaks' function \citep{Bradley19} and these locations are shown with cross symbols in the images in Appendix \ref*{app:objects}. These radio features typically were detected with $\geq$8$\sigma$ significance. In some cases lower signal-to-noise features were considered as significant if their presence is supported by similar features in other resolutions (e.g.\ 1553$+$4407; see Fig.~\ref{fig:app:J1553+4407}). We note that secondary radio features in J1300$+$0355, J1509$+$1757 and J1518$+$1403 have been ignored in the determination of their size and morphology (see Section \ref{sec:morph}) since to the best of our knowledge, they are associated with companion or background galaxies (see Appendix~\ref*{app:objects}). Furthermore, for many sources which were classified as having jet-like morphology or have closely blended central components (see Section \ref{sec:morph})\footnote{Specifically, J0752$+$1935, J0802$+$4643, J1055$+$1102, J1316$+$1753, J1355$+$2046, J1518$+$1403 and J1555$+$5403.} the secondary peak(s) were identified using the residual image, after roughly fitting the core emission with a Gaussian, whose shape was fixed to the beam major and minor axes sizes.\footnote{We note that for J1108$+$0659 and J1222$-$0007 (see Figs \ref*{fig:app:J1108+0659} and \ref*{fig:app:J1222-0007}) we used our HR data to measure the size, since our MR image, although extended, did not show distinct peaks.}

Secondly, in the case where the source is featureless in all three spatial resolutions considered, we used the major axis size, deconvolved from the beam, that we measured using {\sc casa} `imfit'. Where `imfit' determined the target was a point source we use the upper limits on the sizes provided by `imfit', if one was determined, otherwise, we use half the beam size \citep[following e.g.][]{Doi13}. 

\begin{figure*}
 \centering
 \includegraphics[width=18cm]{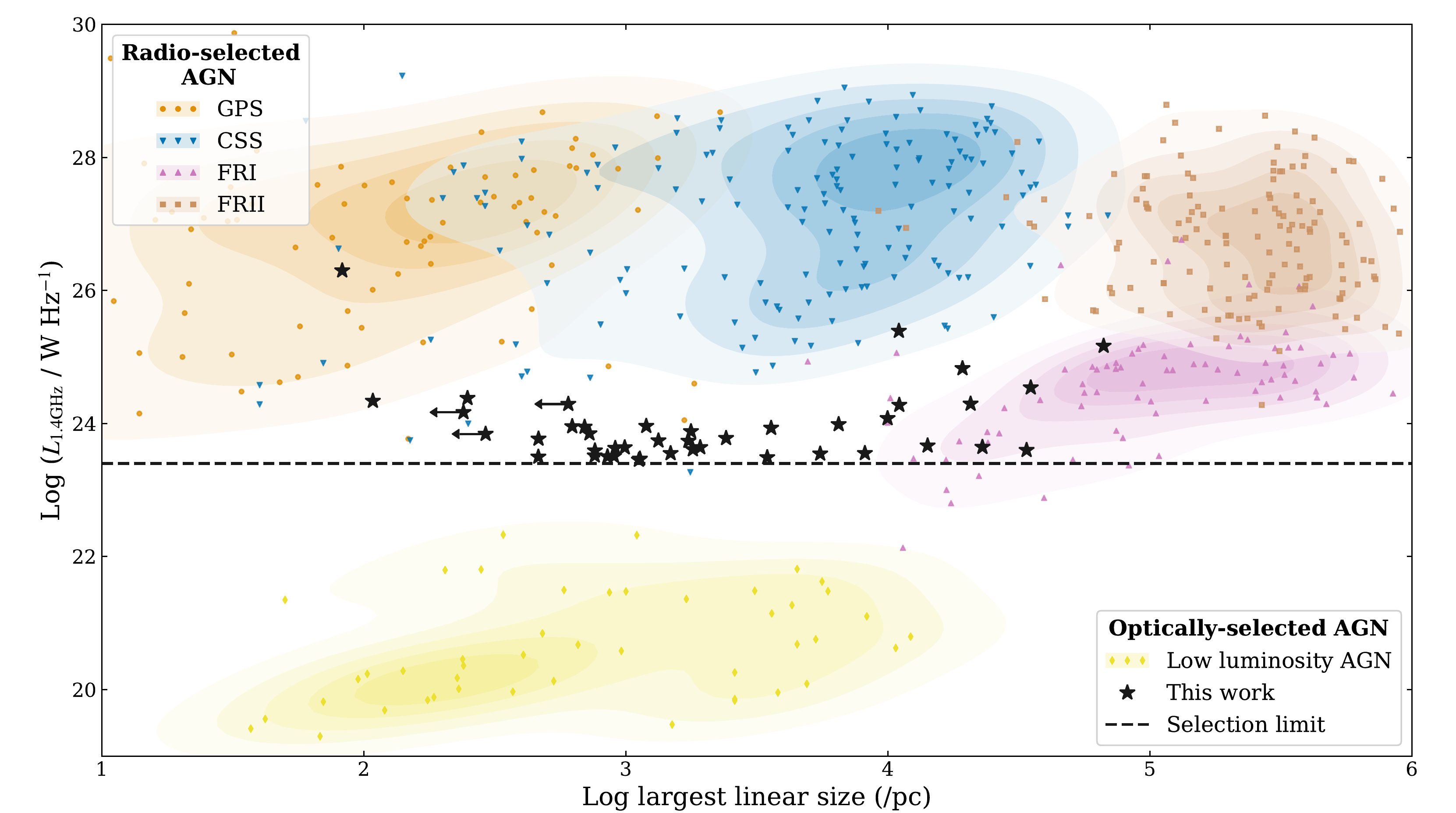}
 \caption{ Radio luminosity versus largest linear size for the complete sample presented here (black stars) compared to the radio-selected AGN compiled by \citet{An12} (points and density contours of the appropriate shape / colour; see also \citealt{Jarvis19}). Seyfert and LINER galaxies \citep[`low luminosity AGN';][]{Gallimore06,Baldi18} are shown for additional comparison. The black dashed line marks the radio luminosity selection criterion for our sample (log[$L_{\rm 1.4GHz}$~/~W~Hz$^{-1}$]~$>23.4$). Our quasars share properties with the lowest luminosity compact radio galaxies (CSS / GPS) and the most compact, low luminosity, FRI radio galaxies. We suggest that at least some of the separation of the different populations are driven by selection effects (see Section \ref{sec:radioSamples_comp}).
  }
  \label{fig:size_vs_lum}
 \end{figure*}

The largest linear size calculated for each source is listed in Table \ref{tab:radio_properties} and plotted along with comparison samples in Fig.~\ref{fig:size_vs_lum} (discussed further in Section \ref{sec:radioSamples_comp}). The measured projected sizes of our sources range from 0.08--66.5~kpc, with only three of the sources being unresolved in all images. We note that we see no significant difference in the sizes between the Type~1 and Type~2 sources, with mean / median values of 7.9/1.8\,kpc or the Type~1s and 7.4/1.2\,kpc for the Type~2s.

%-------------------------------------------------------------
\begin{table*}
 \caption{ Summary of the radio properties used throughout this work. (1) Source name; (2) Radio-loudness classification following \citet{Xu99} (see Fig.~\ref{fig:selection}), with `RL' for the sources which are radio-loud and `RQ' for the rest; (3) the `$R$' parameter quantifying radio-loudness following \citet{Ivezic02} for the Type 1 AGN (see Section \ref{sec:radio_loud}); (4) Largest linear size (LLS) in kpc (see Section \ref{sec:size}); (5) In-band (4--8~GHz) spectral index ($\alpha_\textrm{core}$), of the component best identified as the core, defined as $S_\nu \propto \nu^{\alpha}$ (see Section \ref{sec:core}); (6) Radio excess parameter $q_\textrm{IR}$ (see Section \ref{sec:qir}); (7) Log brightness temperature ($T_\textrm{B}$ / K) of the component best identified as the core (see Section \ref{sec:Tb}); (8--10) Morphological classifications from: the FIRST image (LR; column 8), our VLA MR image (column 9) and our VLA HR image (column 10), where the possible classifications are: C = compact, D = double, T = triple, J = one sided jet, I = irregular / complex, and U = undetected (see Section \ref{sec:morph}); (11) Our final verdict if this source hosts a Radio-AGN, with `yes' for Radio-AGN based on our measurements, `yes*' where the Radio-AGN classification was aided by extra information from the literature, or `maybe' when the data are inconclusive (see Section \ref{sec:final_classifications}). In columns 5--10, the values that are indicative of radio emission from the AGN that are used in our final classification in column 11 are marked in bold.  }
	\centering  

\begin{tabular}{lcccrrccccc}
	\hline 
\multicolumn{1}{c}{Name } & Xu+99 & $R$ & LLS (kpc) & \multicolumn{1}{c}{$\alpha_\textrm{core}$} &  \multicolumn{1}{c}{$q_\textrm{IR}$} & log($T_\textrm{B}$/K)&  FIRST (LR) & MR & HR &   Radio-AGN\\

\multicolumn{1}{c}{(1)} & (2) & \multicolumn{1}{c}{(3)} & \multicolumn{1}{c}{(4)} & \multicolumn{1}{c}{(5)}& \multicolumn{1}{c}{(6)} &\multicolumn{1}{c}{(7)}& (8) & (9) & (10) & (11)  \\
	\hline 
J0749+4510 & RL & 2.1 & 66.5 & \textbf{0.31$\pm$0.07} & -- & \textbf{$>$5.4} & D & I & C & yes \\
J0752+1935 & RQ & 0.8 & 3.58 & -1.2$\pm$0.1 & -- & 3.4$^{+0.1}_{-0.2}$ & C & C & I & maybe \\
J0759+5050 & RQ & -- & 0.85 & -1.22$\pm$0.06 & \textbf{1.28$\pm$0.04} & 3.82$^{+0.1}_{-0.13}$ & C & C & J & yes \\
J0802+4643 & RQ & -- & 0.76 & -1.135$\pm$0.002 & -- & 3.49$\pm$0.05 & C & C & J & maybe \\
J0842+0759 & RQ & 1.2 & 1.77 & \textbf{-0.4$\pm$0.3} & -- & 3.56$^{+0.03}_{-0.04}$ & C & C & C & yes \\
J0842+2048 & RQ & 1.1 & 0.46 & -0.7$\pm$0.02 & -- & 3.8$^{+0.04}_{-0.05}$ & C & C & C & maybe \\
J0907+4620 & RL & -- & 35.05 & -0.6$\pm$0.1 & -- & 4.24$^{+0.07}_{-0.09}$ & D & I & I & yes* \\
J0909+1052 & RQ & -- & 0.99 & -1.12$\pm$0.06 & -- & 3.29$^{+0.04}_{-0.05}$ & C & C & C & maybe \\
J0945+1737 & RQ & -- & 11.04 & -0.89$\pm$0.07 & \textbf{1.25$^{+0.03}_{-0.04}$} & 4.5$^{+0.2}_{-0.4}$ & C & D & I & yes \\
J0946+1319 & RQ & 0.4 & 1.66 & -1.0$\pm$0.3 & \textbf{1.19$\pm$0.04} & 3.2$^{+0.09}_{-0.12}$ & C & C & J & yes \\
J0958+1439 & RQ & -- & 0.9 & -1.23$\pm$0.01 & -- & 3.52$^{+0.07}_{-0.08}$ & C & C & \textbf{D} & yes \\
J1000+1242 & RQ & -- & 20.66 & -0.73$\pm$0.05 & \textbf{$<$1.05} & \textbf{5.3$^{+0.2}_{-0.3}$} & I & I & \textbf{T} & yes \\
J1010+0612 & RQ & -- & 0.11 & -1.11$\pm$0.02 & \textbf{0.83$^{+0.04}_{-0.05}$} & \textbf{5.9$^{+0.1}_{-0.2}$} & C & C & C & yes \\
J1010+1413 & RQ & -- & 9.96 & -0.9$\pm$0.04 & -- & 4.6$^{+0.2}_{-0.4}$ & C & D & I & yes* \\
J1016+0028 & RQ & -- & 33.79 & -- & -- & -- & \textbf{D} & D & U & yes \\
J1016+5358 & RQ & -- & 1.12 & -1.3$\pm$0.1 & -- & 2.67$^{+0.07}_{-0.08}$ & C & C & J & maybe \\
J1045+0843 & RQ & 1.0 & $<$0.29 & -1.05$\pm$0.05 & -- & $>$3.8 & C & C & C & maybe \\
J1055+1102 & RQ & -- & 3.46 & -1.11$\pm$0.09 & -- & 2.62$\pm$0.05 & C & J & D & maybe \\
J1100+0846 & RQ & -- & $<$0.24 & -1.04$\pm$0.04 & \textbf{1.04$^{+0.01}_{-0.02}$} & $>$4.6 & C & C & C & yes \\
J1108+0659 & RQ & -- & 6.48 & -1.4$\pm$0.1 & 1.83$^{+0.07}_{-0.08}$ & 3.1 & C & I & I & yes* \\
J1114+1939 & RQ & -- & 0.62 & -0.83$\pm$0.1 & -- & 3.98$\pm$0.04 & C & C & C & maybe \\
J1116+2200 & RQ & -- & 1.33 & -1.22$\pm$0.05 & -- & 3.67$\pm$0.04 & C & C & J & maybe \\
J1222-0007 & RQ & -- & 8.16 & -0.87$\pm$0.1 & -- & 2.51$^{+0.07}_{-0.08}$ & C & I & \textbf{T} & yes \\
J1223+5409 & RL & 2.5 & 11.0 & \textbf{-0.5$\pm$0.3} & -- & 4.2$^{+0.2}_{-0.6}$ & C & D & \textbf{T} & yes \\
J1227+0419 & RQ & 0.9 & 0.46 & -1.02$\pm$0.04 & -- & 3.81$^{+0.03}_{-0.04}$ & C & C & C & maybe \\
J1300+0355 & RQ & 1.4 & $<$0.6 & \textbf{0.2$\pm$0.1} & -- & \textbf{$>$4.7} & C & C & C & yes \\
J1302+1624 & RQ & 1.1 & 5.51 & -0.87$\pm$0.07 & \textbf{$<$1.16} & 3.74$^{+0.04}_{-0.05}$ & C & I & C & yes \\
J1316+1753 & RQ & -- & 2.41 & -1.18$\pm$0.04 & -- & 3.4$^{+0.2}_{-0.3}$ & C & C & \textbf{T} & yes \\
J1324+5849 & RQ & 1.0 & 0.73 & -0.94$\pm$0.02 & -- & 3.61$\pm$0.04 & C & C & C & maybe \\
J1347+1217 & RL & -- & 0.08 & \textbf{-0.4$\pm$0.2} & \textbf{-0.39$\pm$0.02} & \textbf{8.53$\pm$0.02} & C & C & C & yes \\
J1355+2046 & RQ & 0.2 & 1.92 & -0.92$\pm$0.03 & $<$2.04 & 3.09$^{+0.06}_{-0.08}$ & C & C & J & maybe \\
J1356+1026 & RQ & -- & 0.25 & -1.09$\pm$0.06 & \textbf{1.09$\pm$0.04} & \textbf{$>$5.0} & C & C & C & yes \\
J1430+1339 & RQ & -- & 14.16 & -1.2$\pm$0.1 & \textbf{1.21$^{+0.02}_{-0.03}$} & 3.7$^{+0.2}_{-0.4}$ & J & I & D & yes \\
J1436+4928 & RQ & -- & 0.76 & -1.1$\pm$0.1 & -- & 3.29$\pm$0.04 & C & C & C & maybe \\
J1454+0803 & RQ & 0.9 & 1.73 & -1.1$\pm$0.1 & -- & 3.12$\pm$0.02 & C & C & C & maybe \\
J1509+1757 & RQ & 1.0 & 1.2 & -1.4$\pm$0.1 & 1.91$^{+0.05}_{-0.06}$ & 3.1$\pm$0.02 & C & C & C & maybe \\
J1518+1403 & RQ & -- & 0.91 & -1.2$\pm$0.1 & -- & 2.7$\pm$0.04 & C & C & J & maybe \\
J1553+4407 & RQ & -- & 22.92 & \textbf{-0.2$\pm$0.1} & -- & $>$2.2 & C & \textbf{T} & T & yes \\
J1555+5403 & RQ & 0.8 & 1.13 & -1.0$\pm$0.1 & -- & 2.58$^{+0.08}_{-0.1}$ & C & C & J & maybe \\
J1655+2146 & RQ & 0.4 & 1.8 & -1.15$\pm$0.04 & -- & 2.94$^{+0.05}_{-0.06}$ & C & C & J & maybe \\
J1701+2226 & RL & 1.9 & 19.24 & \textbf{-0.12$\pm$0.02} & -- & \textbf{4.8$^{+0.2}_{-0.5}$} & I & D & I & yes \\
J1715+6008 & RQ & -- & 0.69 & -1.08$\pm$0.08 & -- & 3.81$\pm$0.02 & C & C & C & yes* \\
	\hline   
	\end{tabular}
    
\label{tab:radio_properties} 

	\end{table*}
 %-------------------------------------------------------------

%-------------------------------------------------------------

 %-------------------------------------------------------------

\subsection{Radio morphology}
\label{sec:morph}

\begin{figure}
 \centering
 \includegraphics[width=\hsize]{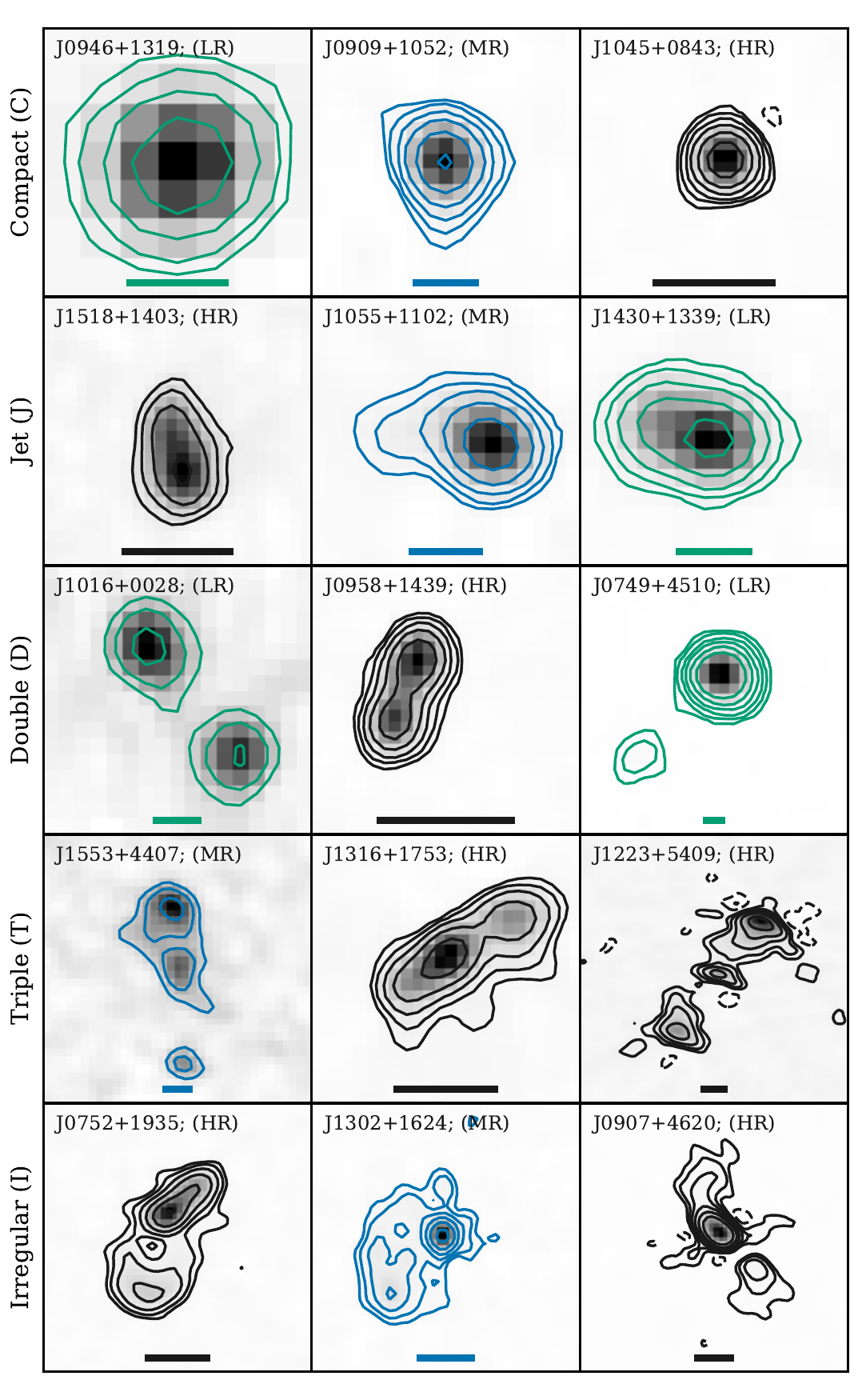}
 \caption{ Examples of each of the morphological classifications used in this work as defined in Section \ref{sec:morph}. Each radio map is plotted both in grey-scale and with contours whose colour denotes the data displayed: green for FIRST (LR), blue for our MR data and black for our HR data. Contours are plotted at $\pm$[4, 8, 16, 32, 64, 128]$\sigma$, with negative contours marked by dashed lines. The size bar in each frame represents lengths of 11, 4 or 2~kpc for the LR, MR and HR data, respectively. These data are displayed in the context of the other radio data considered and the optical images for each source in Appendix \ref*{app:objects}.
  }
  \label{fig:morph}
 \end{figure}

We classified the morphology observed in all three sets of radio images (LR, MR and HR; see Section~\ref{sec:reduction_and_imaging}) roughly following \citet{Baldi18} and \citet{Kimball11a}. The classifications we have used are illustrated with examples from this work in Fig.~\ref{fig:morph} and are described below: 
\begin{itemize}
    \item Compact (C): if the source shows no visibly spatially-resolved features (i.e.\ has the appearance of a single two-dimensional Gaussian). We note that these sources may still be extended or elongated compared to the beam. 
    
    \item Jet (J): if the source is composed of one contiguous feature, visibly spatially extended in one direction. We note that this definition is purely to describe the morphology, and may not be physically associated with an AGN-driven jet.

    \item Double (D): if the source shows two distinct peaks in the radio emission (i.e.\ the radio emission has the appearance of two two-dimensional Gaussian components).
    
    \item Triple (T): if the sources has three distinct radio peaks.
    
    \item Irregular (I): sources with spatially extended (irregular) radio morphologies that do not fit within the above categories.
    
     \item Undetected (U): if the source is not detected in the image. 
    
\end{itemize}

Classifications were given by eye using the figures presented in Appendix \ref*{app:objects} by five of the authors.\footnote{M.E.J., C.M.H., V.M., A.G., and S.J.M.} The final classification given to each source in each of the three images considered was taken to be the majority classification (in each case a minimum of three people agreed). These classifications are given in Table \ref{tab:radio_properties}.

As expected, Table~\ref{tab:radio_properties} shows that many more extended structures are seen in our highest resolution (HR) images compared to either our MR images or the LR images, with 36/42 classified as Compact in the LR images (i.e.\ the FIRST data) compared to just 18/42 in our HR images. On the other hand, it is important to note that some sources (e.g.\ J0907$+$4620 and J1016$+$0028; see Figs \ref*{fig:app:J0907+4620} and \ref*{fig:app:J1016+0028}) exhibit spatially diffuse extended features, visible at low resolutions, which are resolved out in our higher resolution images. Overall, we find that 28/42 (67~percent) of our sources show visibly extended features (i.e.\ any morphological classifications except for Compact) in at least one of the images. We discuss this fraction in the context of other samples in Section~\ref{sec:Loiii}.

%-------------------------------------------------------------

 %-------------------------------------------------------------

The radio emission from star formation and AGN processes (jets, winds and coronal emission) can often be hard to distinguish, particularly for marginally resolved radio features, in galaxies undergoing mergers and in the absence of secondary spatially-resolved star-formation tracers \citep[see e.g.][]{Condon92,Bondi16,Alexandroff+16,Panessa19,Chen20,Smith20}. Therefore, we are conservative with our use of morphology to classify Radio-AGN. We only consider  \emph{symmetric} triple and double structures as they are very likely to be associated with a jet or loosely collimated wind \citep[see e.g.][]{Alexandroff+16,Baldi18,Kimball11a}. Specifically, all five sources which were classified as triple (T; see Table \ref{tab:radio_properties}) are classified as Radio-AGN and J0958$+$1439 and J1016$+$0028 (see Figs \ref*{fig:app:J0958+1439} and \ref*{fig:app:J1016+0028}) are classified as Radio-AGN because of their symmetric double structures \citep[see also][]{Jarvis19}.

\subsection{Core spectral index}
\label{sec:core}

The identification of a flat spectrum radio core ($\alpha \gtrsim-$0.5; using $S_\nu \propto \nu^{\alpha}$) is considered a particularly strong criterion to positively identify an AGN with an active radio jet \citep[see e.g.][]{Orienti14,Panessa19}. Therefore, we measured the spectral index of the radio component best identified as the candidate `core' from our HR images (4--8~GHz). For sources where there are multiple peaks in the image, we defined the candidate core to be either the brightest peak or the component with $\alpha>-$0.6, if one exists. 

To measure the spectral indices we used the spectral index maps, and their respective error maps (see Section \ref{sec:reduction_and_imaging}). We took the weighted average value of the $\alpha$ image within the 16$\sigma$ contour from the total intensity images, weighting by the squared inverse of the spectral index error map. We use the statistical error on the weighted average as the error on the final spectral index value. In a few cases, we were not able to use the 16$\sigma$ contour to reliably isolate the candidate core and so another $\sigma$ level was chosen (discussed on a case by case basis in Appendix \ref*{app:objects}). We note that using these different contour levels does not cause any of the measured values of $\alpha$ to change from a steep ($\alpha<-0.5$) to a flat ($\alpha$$\geq$$-0.5$) classification. For J1347+1217, since the available C-band VLA images did not contain spectral information, we instead took advantage of the fact that this source is unresolved at $\sim$0.3~arcsec scales and use the total spectral index from archival low resolution data (see Appendix \ref{sec:app:J1347+1217}). 

The core spectral index for each source is listed in Table \ref{tab:radio_properties}.\footnote{We have no measurement of the spectral index for J1016$+$0028 since this source was undetected in our HR image (see Fig.~\ref*{fig:app:J1016+0028}).} Overall, we can state that at the flux and resolution limits of our data, we detect flat spectrum ($\alpha$$\geq$$-0.5$) cores in seven of our 42 sources. Unsurprisingly, all but one of the sources classified as `radio-loud' following \citet{Xu99} (Section~\ref{sec:radio_loud}) exhibit a flat spectral index and the exception is close to the flat/steep boundary (i.e.\ $\alpha=-0.6$ for J0907+4629). However, a further three of the sources with flat spectrum cores are classified as `radio-quiet', demonstrating the limitations of traditional `radio-quiet' versus `radio-loud' definitions for identifying emission from jets in AGN \citep[see also Section \ref{sec:RL_diagnostics} and ][]{Padovani16}.

\subsection{Brightness temperature}
\label{sec:Tb}

The bases of jets are expected to have high brightness temperatures (i.e.\ high levels of radio surface brightness) due to non-thermal processes from relativistic electrons \citep[see e.g.][]{Neff83,Preuss83,Blundell98,Ulvestad05,Alexandroff12}. We calculate brightness temperature ($T_\mathrm{B}$) in K for our sources following the standard equation: 
\begin{equation}
    T_\textrm{B}=\frac{1.8\times 10^9(1+z)S_\nu}{\nu^2\Theta_\textrm{maj}\Theta_\textrm{min}} ,
\end{equation}
where $S_\nu$ is the peak flux density in mJy/beam, $\nu$ is the frequency in GHz, and $\Theta_\textrm{maj}$ and $\Theta_\textrm{min}$ are the source major and minor axis sizes respectively, deconvolved from the beam, in milliarcsec \citep[following][]{Ulvestad05,Doi13,Berton18}. 

We calculated ${T}_\textrm{B}$ from our HR images using {\sc casa} Gaussian fits (using the `imfit' routine) to the central component to get the flux density, size and associated errors with each. For those with multiple central components in our HR image, either the brightest or the one with $\alpha>-0.6$ was used (as in Section \ref{sec:core}). Four sources in our sample (J0749$+$4510, J1045$+$0843, J1100$+$0846 and J1300$+$0355) are reported as point sources using `imfit' on the candidate core components. For these we used 1/2 of the beam size to derive lower limits on our $T_\textrm{B}$ calculation \citep[following e.g.][]{Doi13}.\footnote{For J1108$+$0659 we use the core properties from \citet{Bondi16} at 8.5GHz.} Two sources in the sample (J1356$+$1026 and J1553$+$4407) have errors on their minor axis size from `imfit' which are larger than the value (giving sizes consistent with zero); for these sources we give a lower limit on the brightness temperature calculated using the `imfit' major and minor sizes plus their associated errors.

The measured brightness temperatures for our sources span a broad range across $T_\textrm{B}=10^{2.5}$--$10^{8.5}$~K (ignoring lower limits) and are listed in Table \ref{tab:radio_properties}. To classify sources as Radio-AGN based upon their brightness temperatures we used the \citet{Condon91} theoretical upper limit due to compact starbursts. At the central wavelength of our C-band data this corresponds to values of $T_\textrm{B}>10^{4.6}$~K indicating a Radio-AGN.

We note that due to the relatively large spatial resolution of our observations ($\sim$0.3~arcsec) compared to the very long baseline interferometry (VLBI) observations that are typically used for investigating brightness temperatures \citep[e.g.][]{Falcke00,Nagar05}, even our deconvolved sizes could overestimate the true size of the radio cores, artificially decreasing the observed brightness temperature. Nonetheless, even given these limitations, seven of our sample have brightness temperatures indicative of AGN activity (see Table~\ref{tab:radio_properties}).

\subsection{Radio excess}
\label{sec:qir}

 %-------------------------------------------------------------
 \begin{figure}
 \centering
 \includegraphics[width=\hsize]{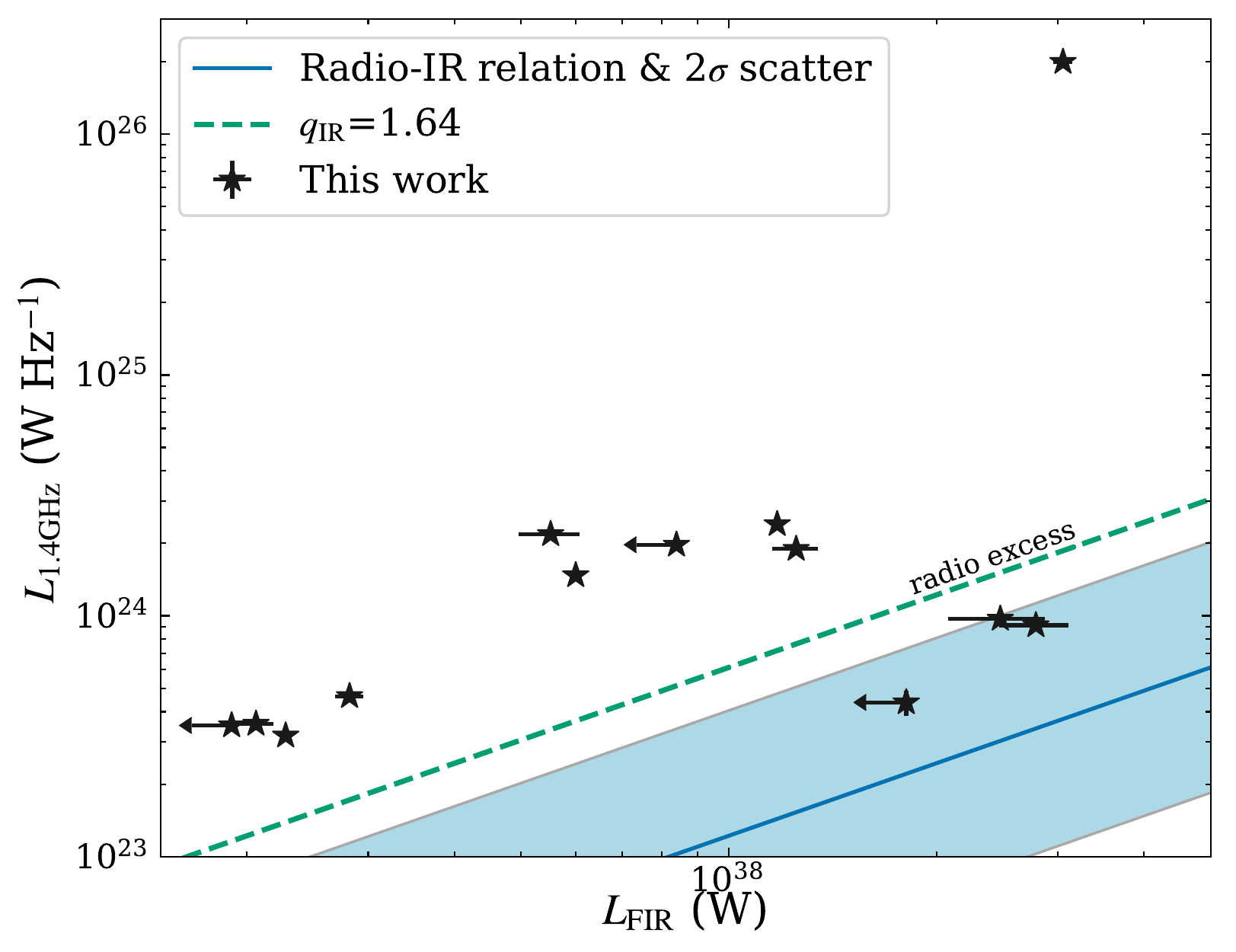}
 \caption{ The observed radio luminosity from NVSS compared to the far-infrared (42.5--122.5~$\mu$m) luminosity calculated from the observed IRAS / PACS photometry for the 13 sources with the required data (black stars; the errors on the 1.4~GHz luminosity are typically smaller than the point size). The green dashed line marks $q_\textrm{IR}=1.64$ and is the threshold where there is confirmed excess radio emission over that expected from star formation (see Section \ref{sec:qir}), with all but three of our sources falling in the radio excess region. The blue solid line and shaded region represent the radio--\emph{IR} correlation for normal star forming galaxies and the 2$\sigma$ scatter on the relation from \citet{Bell03}.
  }
  \label{fig:LIR_radio}
 \end{figure}

 %-------------------------------------------------------------

Both star formation and AGN related processes are capable of producing 1.4~GHz radio luminosities comparable to those observed in our sample \citep[see e.g.][]{Condon92}. However, sources with significant radio emission above the well-established radio--{\em IR} correlation of normal star-forming galaxies (\citealt[][]{Helou85,Bell03}) can be identified as Radio-AGN \citep[e.g.][]{DelMoro13,Padovani16,Smith20}.

We calculated the far infrared flux ($S_\textrm{FIR}$; 42.5--122.5~$\mu$m) for each of our sources following \citet{Helou85} \citep[see also][]{Marvil15}: 
\begin{equation}
\label{eq:FIR}
    S_\textrm{FIR} = 1.26\times10^{-14}(2.58 S_{60\mu \textrm{m}} + S_{100\mu\textrm{m}}) \textrm{W m}^{-2},
\end{equation}
where $S_{60\mu \textrm{m}}$ and $S_{100\mu \textrm{m}}$ are the rest frame 60 and 100~$\mu$m flux densities in Jy. We then calculated the offset of a source relative to the radio--\emph{IR} correlation, $q_\textrm{IR}$, following:
\begin{equation}
    q_\textrm{IR} = \log \left( \frac{S_\textrm{FIR}}{3.75\times10^{12} \textrm{W m}^{-2}} \right) - \log \left(\frac{ S_\textrm{1.4GHz}}{\textrm{W m}^{-2}\textrm{Hz}^{-1}}\right),
\end{equation}
where $S_\textrm{1.4GHz}$ is the K-corrected 1.4GHz radio flux density. We calculated the error on $q_\textrm{IR}$ from the reported catalogue errors on the flux density measurements.

We obtained the necessary rest-frame flux density measurements by linearly interpolating the Infrared Astronomical Satellite \citep[IRAS;][]{Neugebauer84} 60 and 100~$\mu$m flux densities or, if available, the ESA \emph{Herschel} Space Observatory PACS point source catalogue's 70 and 100~$\mu$m flux densities and associated errors \citep{Pilbratt10,Poglitsch10}. We used IRAS measurements from the faint source catalogue \citep[FSC;][]{Moshir92}. Following \citet{Wang14}, we followed a log likelihood method of matching IRAS to the closest Wide-field Infrared Survey Explorer (WISE) all-sky survey \citep{Wright10} source to the SDSS position ($<$ 2~arcsec) to account for the large and asymmetric IRAS beam. For sources detected in one band but not the other, we assume an upper limit on the \emph{FIR} flux. We also note that upper limits from IRAS for sources not detected at either 60 or 100~$\mu$m are not sufficiently constraining to reliably identify radio excess AGN (with the exception of the traditional `radio-loud' AGN); therefore, we only perform these analyses on the 13 sources with at least one detection across the relevant IRAS or PACS bands.

We show the position of our sources with measured \emph{IR} values compared to the relation for star-forming galaxies in Fig.~\ref{fig:LIR_radio}. It can be seen that the majority of our targets lie above the relationship. Following other works we use $q_\textrm{IR}<1.64$ to define a significant offset from the radio--\emph{IR} correlation and refer to these as `radio excess' (\citealt{Helou85}; see also \citealt{Marvil15}). Of the 13 sources, we find that ten are defined as radio excess, one has a non-constraining limit (J1355$+$2046 with $q_\textrm{IR}<2.04$) and two sources have $q_\textrm{IR}$ values that fall within the region occupied by star-forming galaxies. However, we cannot rule out that the two sources which appear to lie on the relation have an AGN contribution to their \emph{FIR} emission and hence should intrinsically lie above the relation \citep[see e.g.][]{Moric10,Bonzini13,Zakamska16,Wong16}. Indeed, one of these targets (J1108$+$0659) is identified as an AGN based on literature information (see Appendix~\ref*{sec:app:J1108+0659}), and another (J1509$+$1757) is borderline radio-loud by the criterion of \citet{Ivezic02} (with $R=1.0$; see Table \ref{tab:radio_properties} and Section~\ref{sec:radio_loud}).

Overall, we conclude that 10/13 (77~percent) of our sample for which we have the necessary constraints on the \emph{FIR} emission are classified as radio excess and, consequently, \emph{at least} 10/42 (24~percent) across the full sample have excess radio emission above what can be explained by star formation.

\subsection{Final identification of radio emission from AGN} 
\label{sec:final_classifications}

%-------------------------------------------------------------
 \begin{figure}
 \centering
 \includegraphics[width=\hsize]{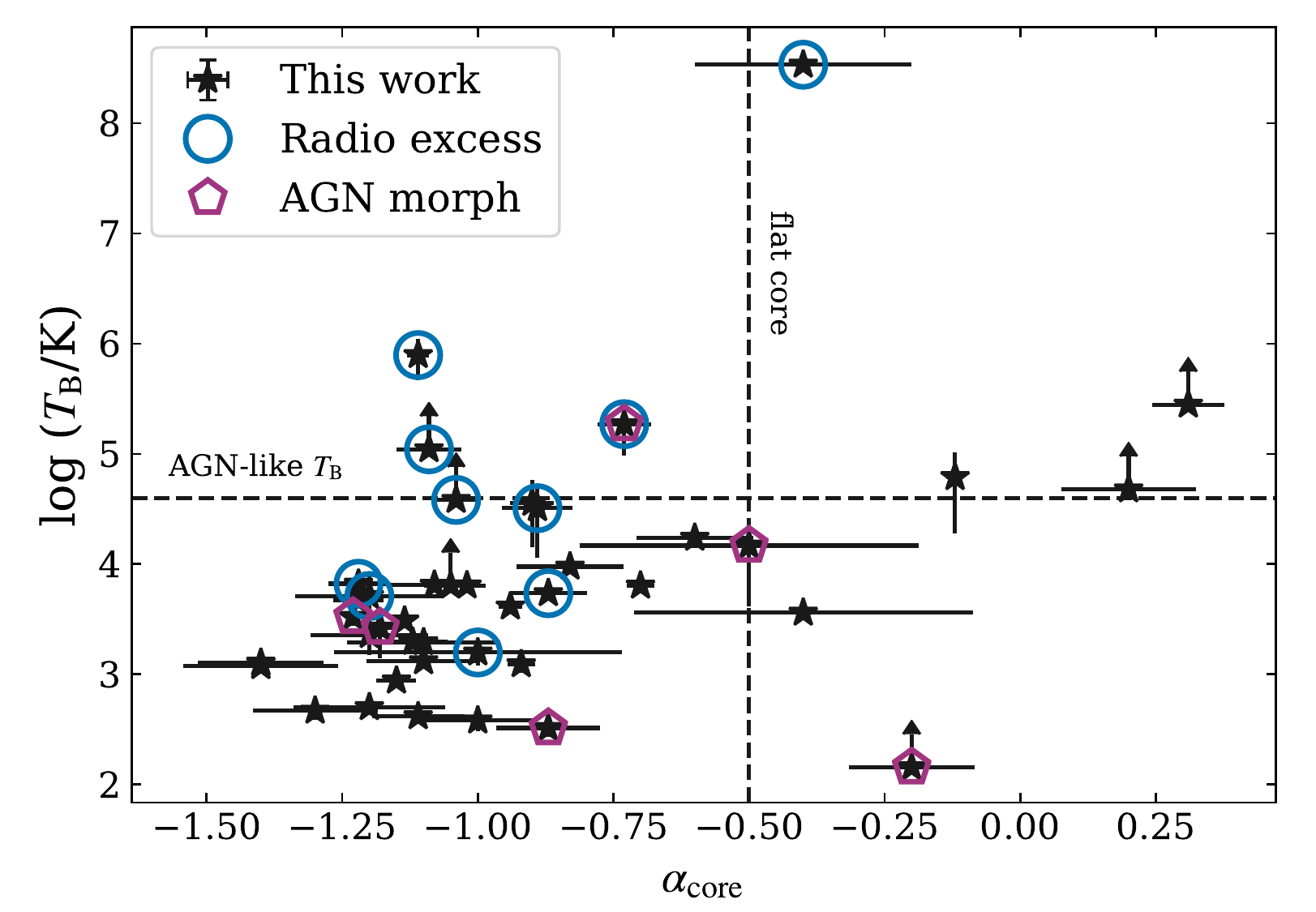}
 \caption{ This figure illustrates the four commonly used criteria used to separate targets with a Radio-AGN from those where the radio emission could be dominated by star formation which we combine in this work (see Section \ref{sec:results}). The logarithm of the brightness temperature is plotted against the spectral index of the radio component identified as the core (black stars). The vertical dashed line indicates a flat spectrum core (i.e.\ $\alpha_\textrm{core}\geq-0.5$) and the horizontal dashed line indicates high brightness temperatures (i.e.\ log $T_\textrm{B}$ $\geq$ 4.6), each of which imply an AGN origin of the radio emission (see Section \ref{sec:core} and \ref{sec:Tb}). Sources that were identified as radio excess (see Section \ref{sec:qir} and Fig.~\ref{fig:LIR_radio}) and those with radio morphology strongly suggestive of AGN radio emission (see Section \ref{sec:morph}) are additionally plotted with blue circles and/or magenta pentagons respectively, on top of the black stars used for the entire sample. Sources satisfying at least one of these criteria are considered Radio-AGN in this work with many satisfying multiple criteria. Due to a non detection we were unable to calculate a core spectral index or brightness temperature for J1016$+$0028; however, it is classified as a Radio-AGN based on its morphology. 
  }
  \label{fig:radio-AGN_selection}
 \end{figure}

 %-------------------------------------------------------------

By constraining the dominant physical mechanisms contributing to the radio emission in our sample we gain a deeper understanding of the energy balance in the galaxy and the possible drivers of outflows \citep[e.g.][]{Nims15,Harrison18}. Here we identify sources where we attribute the radio emission to an AGN, i.e.\ `Radio-AGN', by combining the four criteria described in the previous sub-sections. As previously discussed, these are all well established methods for identifying Radio-AGN, and we apply them here to our unique sample (see Section \ref{sec:target_context}). We put equal weight on all methods and consider any source which is deemed to be AGN-like in at least one to be a Radio-AGN. Our final classification process is presented in Fig.~\ref{fig:radio-AGN_selection}, and is summarised as follows:
\begin{enumerate}
    \item   Seven of our sources are Radio-AGN based on the identification of a flat spectrum core ($\alpha\geq-0.5$; see Section \ref{sec:core}).

    \item 10 of our 13 sources, for which we have meaningful constraints, have measured radio excess values indicative of AGN with $q_\textrm{IR}<1.64$ (see Section \ref{sec:qir}). Only one of these is classified as a Radio-AGN based on having a flat spectrum core, bringing our total number of Radio-AGN in this sample up to 16. 

    \item Seven sources have brightness temperatures indicating that their core radio emission is dominated by AGN related processes (Section~\ref{sec:Tb}). All seven of these have been identified as Radio-AGN based on either their spectral index and/or their $q_\textrm{IR}$ values.
   
    \item Seven sources are classified as Radio-AGN based upon their morphology (Section~\ref{sec:morph}), adding four sources which were not classified as Radio-AGN based on the criteria described above. This brings the total number of Radio-AGN up to 20 and these are labelled with a `yes' in the final column of Table~\ref{tab:radio_properties}. 
\end{enumerate}

The limitations of each of the methods we applied are discussed in the relevant sub-sections. In each case we conclude that our approach is conservative and the true number of Radio-AGN is likely to be higher than that finally quoted here. In particular, more extensive far-infrared data is likely to increase the number of sources identified as `radio excess' and higher spatial resolution and/or higher frequency radio data, would likely increase the number of  identified flat spectrum or high brightness temperature sources \citep[see e.g.][]{Smith20}. Indeed, four of the sources in our sample which are not classified as Radio-AGN by the method described above, have been previously identified as hosting compact jets in the literature by using higher spatial resolution radio imaging, classical radio-loudness and/or a more detailed treatment of the far-infrared data. These are J0907$+$4620, J1010$+$1413, J1108$+$0659 and J1715$+$6008 and are labelled as `yes$^*$' in the final column of Table~\ref{tab:radio_properties} (see details in Appendix~\ref*{sec:app:J0907+4620},~\ref*{sec:app:J1010+1413},~\ref*{sec:app:J1108+0659} and \ref*{sec:app:J1715+6008}). In line with our overall goal of identifying as many Radio-AGN in our sample as possible, we include these four sources as Radio-AGN.

Overall, we conservatively conclude that {\em at least} 24/42 (57~percent) of the sample can be classified as Radio-AGN (20/42 if only the data presented in this paper are considered). For the rest of the sample, the origin of the radio emission is unclear, with star formation or the AGN both as viable possibilities.

\section{Discussion}
\label{sec:discussion}

Using VLA observations of 42 luminous quasars (Fig.~\ref{fig:selection}) we have found that they have radio sizes ranging from LLS~$=$~0.08--66.5\,kpc (with three unresolved sources; see Section~\ref{sec:size} and Fig.~\ref{fig:size_vs_lum}). For 67\,percent of the sample we identified radio structures on 1--60\,kpc scales (Section~\ref{sec:morph}). Overall, by combining four different well-known diagnostics we have found that at least 57\,percent of our sample can be classified as `Radio-AGN' (Section~\ref{sec:final_classifications}). Here we discuss these results in the context of commonly used selection criteria of selecting `radio-loud' AGN (Section~\ref{sec:RL_diagnostics}) and compare the radio properties of this sample to typical `radio-loud' samples (Section~\ref{sec:radioSamples_comp}). Finally, we explore the relationship between the ionised gas and radio properties of our sources and discuss what this might mean for understanding the feedback processes in quasar host galaxies (Section \ref{sec:oiii-radio}).

\subsection{Comparison to diagnostics to identify radio-loud AGN}
\label{sec:RL_diagnostics}

%-------------------------------------------------------------
 \begin{figure*}
 \centering
 \includegraphics[width=18cm]{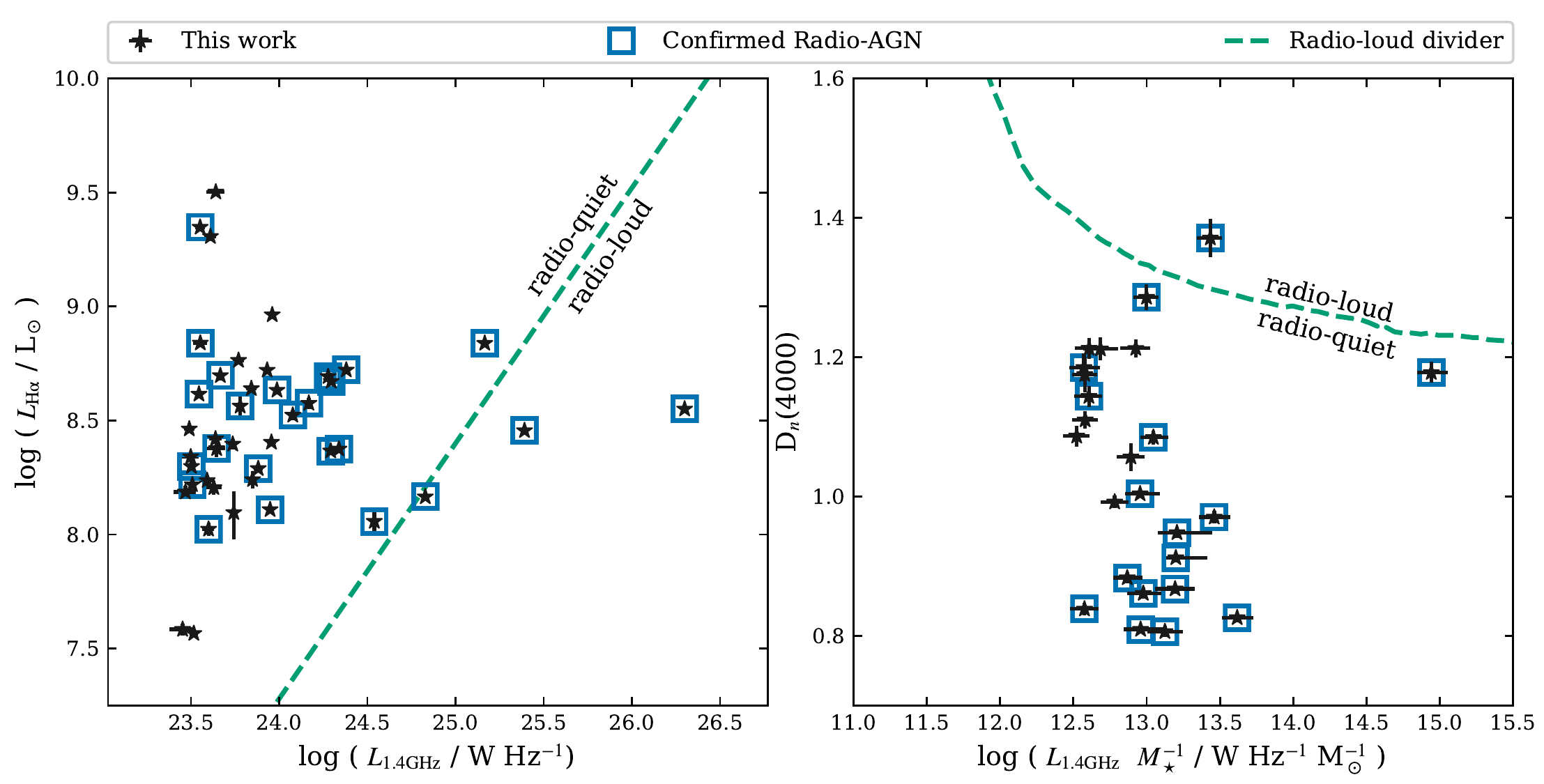}
 \caption{Our targets plotted in two diagnostic plots that are designed to separate radio-loud and radio-quiet AGN  (black stars; see Section \ref{sec:RL_diagnostics}). \emph{Left:} H$\alpha$ versus 1.4~GHz radio luminosity. \emph{Right:} D$_{n}$4000 versus the ratio of the 1.4~GHz radio luminosity to the stellar mass for the Type 2 AGN in our sample. We note that the error-bars in both panels are often smaller than the point sizes. Our targets that are confirmed as Radio-AGN from our analyses (Section \ref{sec:final_classifications}) are further marked with blue squares in addition to the black stars used for the overall sample. All but four of our targets fall into the radio-quiet, candidate star-formation dominated, area of the figures, confirming that these diagnostics are not effective for finding low power Radio-AGN associated with powerful quasars.
  }
  \label{fig:radio_loudness}
 \end{figure*}

As already discussed in Section~\ref{sec:radio_loud}, the `radio-loud' division of \citet{Xu99} only identifies five sources in our sample (see Fig.~\ref{fig:selection}) and a further four of the Type 1 targets are moderately `radio-loud' based on the $R$ value (\citealt{Ivezic02}). Another widely adopted method for identifying `radio-loud' AGN are the criteria of \citet{Best12} \citep[see also][]{Sabater19}. Specifically, they used three diagnostics: one compares the observed radio emission to H$\alpha$ luminosity, because both of these are star-formation tracers for star-forming galaxies. Another diagnostic compares the strength of spectral break at 4000\AA\ (D$_n$4000; a tracer of mean stellar age), which correlates with the ratio of radio luminosity to stellar mass for star forming galaxies. Finally, they used emission-line ratio diagnostics as part of their selection criteria; however, we do not discuss this here because our sample was selected as AGN based on their optical emission line properties \citep{Mullaney13}.

We compare our classification from Section \ref{sec:final_classifications} to what we would have found using the \citet{Best12} diagnostics in Fig.~\ref{fig:radio_loudness}. For our targets we use the H$\alpha$ emission-line fluxes from \citet{Mullaney13} and we take both D$_n$4000 and stellar mass ($M_\star$) from MPA-JHU SDSS DR8 measurements\footnote{\url{https://www.sdss.org/dr12/spectro/galaxy_mpajhu/}} converted from a Kroupa to Chabrier IMF \citep{Madau14}.  
 We note that these stellar masses are calculated using optical SED fitting without considering AGN emission, and so are particularly unreliable for the Type 1 AGN in this sample (where the optical emission is expected to be dominated by the AGN). As such, the Type 1 sources are excluded from the diagnostic using stellar masses. However, we note that if we use the AGN-corrected stellar masses for the nine Type 2 AGN presented in \citet{Jarvis19}, the classifications do not change, and we therefore conclude that these mass measurements are suitable for the Type 2 sources.

 In Fig.~\ref{fig:radio_loudness} it can be seen that only one source (J0907$+$4620) based on the D$_n$4000 diagnostic and only three sources (J1223$+$5409, J1347$+$1217 and J1701$+$2226) based on the H$\alpha$ diagnostic are classified as radio-loud. These four are a subset of the five classical `radio-loud' sources identified by the \citet{Xu99} division (see Table \ref{tab:radio_properties} and Fig.~\ref{fig:selection}). This result is not surprising since these criteria are not designed to be complete but, instead, to cleanly select sources where the radio emission is definitely dominated by the AGN \citep{Best12, Sabater19}. In particular, the H$\alpha$ diagnostic is based on the assumption that for the galaxy population of interest, H$\alpha$ is a reliable star-formation diagnostic. It is therefore expected that this criterion will not be effective if applied to the most radiatively luminous AGN that are selected with emission lines, such as our sample, since the H$\alpha$ emission will have a strong, potentially dominant, contribution from the AGN (e.g.\ \citealt{Kewley01,Scholtz20}). This current work confirms that applying a combination of Radio-AGN selection methods, including the use of high-resolution radio imaging (Section~\ref{sec:results}), is important to obtain a more complete identification of Radio-AGN for the type of quasars studied here (i.e. moderate radio luminosity, \oiii\ luminous AGN).

\subsection{Comparison to traditional radio-loud AGN samples}
\label{sec:radioSamples_comp}

Combining our measured sizes with the total radio luminosities from NVSS ($L_\textrm{1.4GHz}$; see Table \ref{tab:sample_properties}) we compare the radio properties of our sample to those of traditional radio-loud AGN from \citet{An12} in Fig.~\ref{fig:size_vs_lum}. As can be seen in this figure, the sizes we measure for our sample ($\sim$0.1--10~kpc) are broadly consistent with those found in samples of compact `radio-loud' AGN with jets (in this case represented primarily by GHz‐Peaked Spectrum -- GPS -- and Compact Steep Spectrum -- CSS -- sources). Most of our sources would be excluded from such samples due to their lower radio luminosities and not being classified as radio-loud AGN by traditional methods (Section~\ref{sec:RL_diagnostics}).

The largest sources in our sample (i.e.\ those with radio sizes $\gtrsim$10~kpc) overlap with the tail of the smallest, dimmest Fanaroff-Riley class I radio galaxies \citep[FRI;][see Fig.~\ref{fig:size_vs_lum}]{Fanaroff74}. The smallest FRI sources in the \citet{An12} sample are from \citet{Fanti87} which has spatial resolutions down to $\sim$1~arcsec (in comparison, our smallest resolution is $\sim$0.3~arcsec). This, combined with differences in sample selection and how the radio sizes are measured between this work and \citet{Fanti87}, implies that the smaller sizes in our sample compared to typical FRI sources could be mostly driven by a resolution / measurement effect.

Overall, our observations are consistent with other radio studies of quasars, using both low and high frequency data, which show a wide range of radio properties \citep[e.g.\ we see no clear bimodality in the size and luminosity distributions;][]{Mahony12,Gurkan19}. Our sources have an interesting moderate radio luminosity range, with median $L_{\rm 1.4GHz}\approx10^{23.8}$, where AGN may be expected to start to dominate the radio emission, but where they are excluded from traditional `radio-loud' samples \citep[][]{Kimball11b,White15,Kellermann16,Zakamska16,White17}. Indeed, the majority of our targets do not fit neatly within the traditional FRI and FRII radio classifications. Although deeper data may help determine the relative contributions of jets, lobes and hot spots, our results and recent results in the literature demonstrate that high spatial resolution images of low power Radio-AGN can reveal much more complexity in radio morphologies than that seen in the traditional `FRI' and `FRII' radio galaxies \citep[e.g.][]{Mingo19}. Many of our sources have featureless radio morphology (i.e.\ our `C' classification), particularly in the lower resolution images and as such could be classified as `FR0' galaxies according to some classification schemes \citep[e.g.][]{Baldi15}. However, higher spatial resolution data may reveal more complex morphologies on smaller scales \citep[see discussion in][]{Hardcastle20}.

We also note that several of our sources show radio morphologies similar to those typically seen in Seyfert galaxies. Specifically, Seyferts often show `C shaped', `S shaped' or `figure-8 shaped' radio structures on 1-10~kpc scales \citep[see e.g.][]{Duric88,Wehrle88,Kharb06,Kharb16}. Similar structures are seen in this sample in e.g.\ J0752$+$1935, J0907$+$4620, J1000$+$1242, 1302$+$1624 and J1430+1339 (see Figs \ref*{fig:app:J0752+1935}, \ref*{fig:app:J0907+4620}, \ref*{fig:app:J1000+1242}, \ref*{fig:app:J1302+1624} and \ref*{fig:app:J1430+1339}). This suggests that there is a continuity in the radio properties of quasars and Seyferts. 

 \subsection{The ionised gas -- radio connection }
 \label{sec:oiii-radio}

 Here, we compare the radio sizes of our sample to the \oiii\ luminosities and line widths. We use this to explore the connection between the ionised gas and the radio emission in our sample and compare this to other AGN samples in the literature.

\subsubsection[OIII luminosity]{Prevalence of radio structures with high \oiii\ luminosity} 
\label{sec:Loiii}

In Section \ref{sec:morph} we found that 67~percent of our sample show spatially extended features on kiloparsec scales. This is a marginally higher fraction than has been found in previous spatially resolved radio studies of `radio-quiet' quasars with similar radio luminosities (specifically \citealt{Kukula98} which found extended features in $\sim$50~percent of their sample and 44~percent in \citealt{Pierce20}). Compared to these works the main difference of this study is the selection of sources with \loiii$>10^{42.1}$~erg~s$^{-1}$. Specifically, the selection criteria in \citet{Pierce20} are almost identical to those in this work with the exception that their sample all have \loiii$<$10$^{42}$~erg~s$^{-1}$. Importantly, their radio observations are comparable to ours (i.e.\ VLA L- and C-band A-array configuration data with similar observation times and rms noise levels).

\begin{figure}
 \centering
 \includegraphics[width=\hsize]{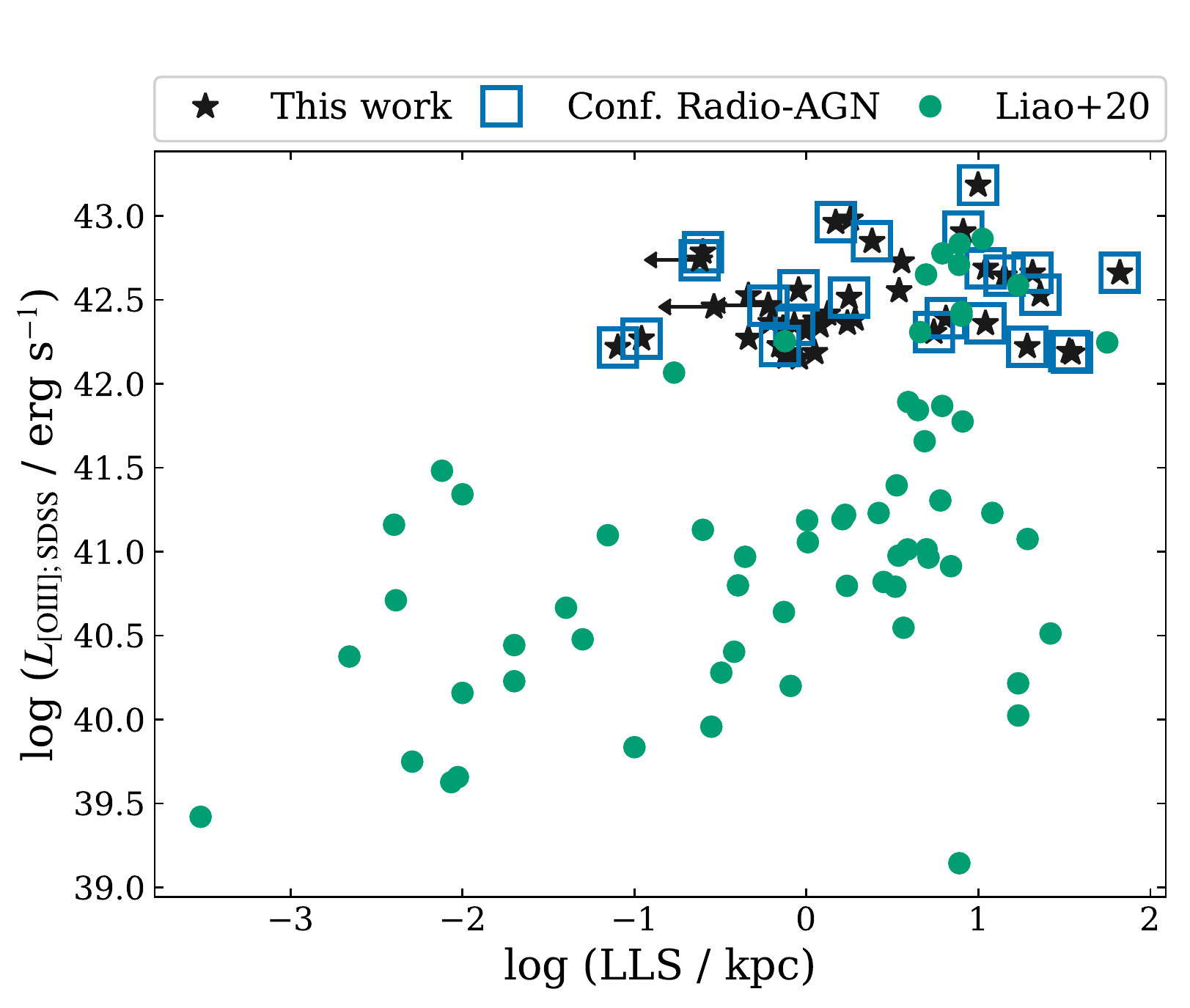}
 \caption{ \oiii\ luminosity (from SDSS single Gaussian fits) versus largest linear size (LLS) of the radio emission. We plot the sample presented here (black stars) and the SDSS matched radio-loud sample of \citet{Liao20} (green circles; see Section \ref{sec:Loiii}). The sources in our sample which we were able to confirm as Radio-AGN (see Section \ref{sec:final_classifications}) are additionally marked with blue squares on top of the black stars used for the entire sample. High \loiii\ AGN seem to be preferentially associated with larger radio sources, where less \oiii\ luminous AGN can be associated with a wide range of radio sizes (see Section \ref{sec:Loiii}). 
  }
  \label{fig:size_vs_OIII}
 \end{figure}

For compact powerful radio galaxies (e.g.\ CSS / GPS sources), there is a generally accepted link between \oiii\ luminosity and radio size (see review in \citealt{O'Dea20}). This has traditionally been reported as a trend for larger radio sizes to be associated with higher \oiii\ luminosities (see e.g.\ \citealt{ODea98, Labiano08}). Although this trend is weak \cite[see][]{Kunert-Bajraszewska10}, there is a distinct lack of the most compact radio sources (i.e.\ $\lesssim$0.1~kpc) at the most extreme \oiii\ luminosities (i.e.\ $L_\textrm{ \oiii}\gtrsim10^{42}$~erg\,s$^{-1}$). To investigate this further, in Fig.~\ref{fig:size_vs_OIII} we show the radio sizes and \oiii\ luminosities from our pre-dominantly radio-quiet sources and those of the radio-loud sources from \citet{Liao20}, who also used SDSS spectroscopy to make their \oiii\ emission-line measurements.

As can be seen in Fig.~\ref{fig:size_vs_OIII}, our data suggest that the trend between radio size and \oiii\ luminosity observed in radio-loud populations also holds for radio-quiet AGN (see also \citealt{Leipski06}). The \oiii\ bright sources in the \citet{Liao20} radio-loud sample (i.e.\ those with \loiii$>$10$^{42}$~erg~s$^{-1}$) all have sizes $\geq$0.75\,kpc, similar to the smallest sizes measured in this work (see Fig.~\ref{fig:size_vs_OIII}). This provides a possible explanation for why we find a larger fraction of spatially extended sources than the radio-quiet, lower \oiii\ luminosity sample of \citet{Pierce20}. This could also explain why we found a higher fraction of spatially extended sources in our pilot study of \citet{Jarvis19} ($\sim$80--90~percent), since the subset of targets in that work all fall at the higher \oiii\ luminosity end of the current sample (median \loiii~=~4.7$\times$10$^{42}$~erg s$^{-1}$, compared to 2.1$\times$10$^{42}$~erg s$^{-1}$ for the complete sample presented here; see Fig.~\ref{fig:selection}).

%-------------------------------------------------------------

 %-------------------------------------------------------------

  \subsubsection{Radio sizes related to ionised gas kinematics}
 \label{sec:FWHM}
 
 %-------------------------------------------------------------
 \begin{figure}
 \centering
 \includegraphics[width=\hsize]{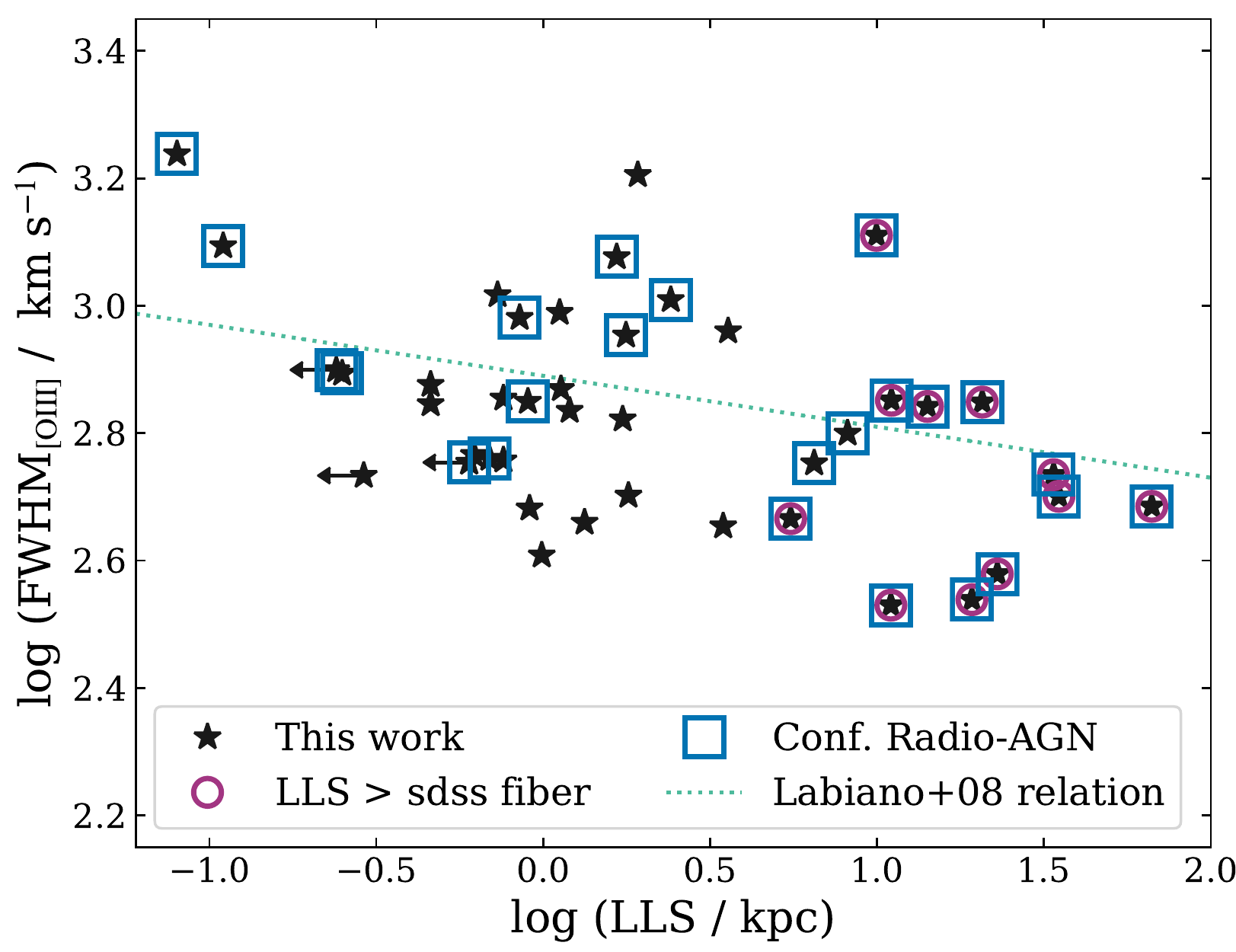}
 \caption{ Flux-weighted FWHM of the [O~{\sc iii}] line from the two Gaussian fits of \citet{Mullaney13} (see Table \ref{tab:sample_properties}) compared to the largest linear size (LLS) calculated in this work (see Table \ref{tab:radio_properties}; black stars). The sources for which we have a confirmed Radio-AGN classification (see Section \ref{sec:final_classifications}) are further marked with blue squares on top of the black stars used for the entire sample. Magenta circles on top of the black stars additionally mark those whose LLS is larger than the SDSS fibre. For comparison, we include the linear fit to archival compact radio-loud AGN from \citet{Labiano08} (green dotted line). Larger radio sources are preferentially associated with galaxies with less extreme \oiii\ kinematics than more compact sources. 
  }
  \label{fig:size_vs_fwhm}
 \end{figure}

 %-------------------------------------------------------------

An association between the most extreme ionised outflows and the most compact radio galaxies has previously been reported for `radio-loud' populations \citep[see e.g.][]{Gelderman94,Best00,Holt08,Kim13}. \citet{Molyneux19} provided indications of such a link in radio-quiet AGN. Here we use our measurements, that use $\sim$20 times better spatial resolution radio images than \citet{Molyneux19}, to test this result on our pre-dominantly radio-quiet quasar sample.
 
 In Fig.~\ref{fig:size_vs_fwhm} we compare the radio sizes of the sources from our sample (see Table \ref{tab:radio_properties}) to the flux-weighted FWHM from the \citet{Mullaney13} two Gaussian fits to the [O~{\sc iii}] line profile (Table \ref{tab:sample_properties}). We find evidence for a negative correlation between the FWHM and the radio size. Although there is a large scatter, the existence of this correlation is supported by a Spearman rank correlation coefficient of $-$0.42 and p-value for a null hypothesis of uncorrelated data of 0.008 (excluding upper limits). We note that some of the scatter we find in this relation could be the result of different levels of projection effects affecting the observed radio sizes. The results presented in Fig.~\ref{fig:size_vs_fwhm} are in qualitative agreement with the results of \citet{Molyneux19}.

A thorough, quantitative comparison between the trend demonstrated in this work and that found in the literature is difficult due to the relatively small samples and/or inhomogeneous datasets involved. Nonetheless, \citet{Holt08} find that extended radio-loud AGN (radio sizes greater than 30~kpc) typically have \oiii\ FWHM values around 200-300~\kms, while compact radio sources (radio sizes less than 30~kpc) have typical \oiii\ FWHM values of around 1000~\kms. These values are consistent with the results presented here. A more direct comparison can also be made with the work of \citet{Labiano08}, whose radio-loud CSS / GPS sample spans a similar range of \oiii\ luminosities and radio sizes to the sample presented here (approximately $10^{41}<$~$L_{\rm [O~III]}<10^{44}$~erg~s$^{-1}$ and 0.1~$<$~LLS~$<$~10~kpc ) but with significantly higher radio luminosities ($L_\textrm{1.4GHz}\gtrsim10^{25}$~W~Hz$^{-1}$). We show the relationship from \citet{Labiano08} in Fig.~\ref{fig:size_vs_fwhm}, which is qualitatively similar to the trend seen for our sample. Nonetheless, the relation between the width of the \oiii\ emission line and the radio size found in both samples is weak and larger, more homogeneous samples are required for a robust quantification of the slope and scatter of this relationship.

Overall, we conclude that the connection between ionised gas kinematics and radio emission, seen in powerful `radio-loud' AGN is also seen in the lower radio luminosity systems of our sample.

\subsubsection{Implications for how quasars interact with the ISM}

Across our sample the high fraction of extended radio sources (Fig.~\ref{fig:size_vs_lum}) and the observed connection between both the \oiii\ luminosity and ionised gas kinematics with the size of the radio emission (Fig.~\ref{fig:size_vs_OIII} and Fig.~\ref{fig:size_vs_fwhm}) could be explained by either jets or quasar-driven winds interacting with the ISM. Both mechanisms would be able to ionise larger volumes of gas as they expand / age \citep{Bicknell97,Moy02,Vink06,Jiang10,Nims15,Mukherjee18}, resulting in more extreme \oiii\ luminosities preferentially associated with extended radio emission. Additionally, as the jet / wind expands, interactions with the ISM could cause deceleration, which would result in the observed lower velocities (traced by the FWHM of the \oiii\ emission) for larger radio sources \citep[see e.g.][]{Wagner12,Labiano08,Dugan17,Bicknell18,Mukherjee18,Mukherjee20}. Furthermore, both winds and low power jets will preferentially take the path of least resistance through a clumpy medium resulting in less interaction, and hence a smaller velocity dispersion, with the gas at larger scales where the gas density drops \citep{Mukherjee18,Costa14,Costa18b}.

Therefore, the observations we have made could provide evidence for a strong interaction between either radio jets or quasar-driven winds and the interstellar medium. Indeed, direct evidence for this is shown by combining radio imaging with integral field spectroscopy for nine of the targets in \citet{Jarvis19}. Furthermore, it is interesting to note that over half of our sources (24) have steep core spectral indices with $\alpha\le-1$ (see Table~\ref{tab:radio_properties}), steeper than the typically observed value of around $-0.7$. Radio emission originating from shocked quasar winds has been predicted to have spectral indices closer to $-$1 \citep{Nims15}. Strong radiative losses due to interactions between jets and the ISM could also result in a steep spectral index \citep[see e.g.][]{Congiu17}. Spatially-resolved studies of both the radio properties and the multi-phase gas kinematics, will greatly increase our understanding of how typical, radio-quiet AGN and quasars interact with their host galaxies \citep[e.g.][]{Venturi21}.

\section{Conclusions}
\label{sec:conclusion}

We have presented the first results on our sample of 42 $z<0.2$, \oiii\ luminous AGN ($L_{\rm [O~III]}>10^{42.1}$~ergs~s$^{-1}$) from our Quasar Feedback Survey. The targets represent the 30~percent of the parent AGN population, in this redshift and [O~{\sc iii}] luminosity range, which have radio luminosities of $L_{\rm 1.4GHz}> 10^{23.4}$~W~Hz$^{-1}$ (see Section~\ref{sec:target_selection}). Overall they have moderate radio luminosities (median $L_{\rm 1.4GHz}=5.9\times10^{23}$~W~Hz$^{-1}$). Here we have presented spatially resolved radio images at 1.5~GHz and 6~GHz with around 1 and 0.3~arcsec resolution for the full sample (e.g.\ Fig.~\ref{fig:app:J1553+4407}). We have confirmed the importance of using high resolution radio images, i.e.\ beyond that typically available for all-sky radio surveys, to identify `Radio-AGN' (Section~\ref{sec:morph}). By combining our new radio images with archival information, we investigated the source of the radio emission and the relationship between the radio emission and the [O~{\sc iii}] emission-line properties. 

Our main conclusions are as follows:
\begin{enumerate}
    \item 28 out of the 42 targets (i.e.\ 67~percent) show morphologically distinct spatially extended radio features on scales of $\gtrsim$0.3~arcsec with a diverse range of morphologies (see Fig.~\ref{fig:morph}). We measured the radio sizes of our sources to be 0.08~$<$~LLS~$<$~66.5~kpc. These sizes are comparable to those observed for compact radio galaxy samples (e.g.\ CSS / GPS sources) that typically have radio luminosities that are at least 2 orders of magnitude higher than our sample (see Section \ref{sec:radioSamples_comp} and Fig.~\ref{fig:size_vs_lum}). 
    
    \item Only 4--9 of the sample are radio-loud by traditional criteria (see Section~\ref{sec:radio_loud}, Section~\ref{sec:RL_diagnostics}, Figs~\ref{fig:selection} and \ref{fig:radio_loudness}). However, {\em at least} 24/42 (57~percent) have a strong indication of radio emission associated with an AGN. This is determined by combining a series of well-known Radio-AGN selection techniques that use measurements of: the radio--infrared excess parameter; the core spectral index; the brightness temperatures and radio morphologies (see Section \ref{sec:final_classifications}, Fig.~\ref{fig:LIR_radio} and Fig.~\ref{fig:radio-AGN_selection}). We are unable to draw definitive conclusions on the origin of the radio emission in the remainder of the targets without further data. 
    
    \item We confirmed that there is a link between both the luminosity and the width of the \oiii\ emission line and the radio size in `radio-quiet' quasars that is similar to that which has previously been observed for `radio-loud' AGN. Specifically, \oiii\ luminous sources are more likely to have extended radio emission and larger radio sources tend to have narrower \oiii\ lines. This suggests that the radio and ionised gas properties in radio-quiet quasars are closely connected (see Section \ref{sec:oiii-radio}, Fig.~\ref{fig:size_vs_OIII} and Fig.~\ref{fig:size_vs_fwhm}).

\end{enumerate}

This work is the first to examine the properties of the full Quasar Feedback Survey sample (Fig.~\ref{fig:selection}). We find that the AGN contribute significantly to the radio emission even though the sample are bolometrically luminous quasars and the majority of the targets are classically `radio-quiet'. This result suggests that care should be taken when considering the division of feedback into `radio’ and `quasar’ modes.

Some of the ongoing work we are pursuing in the Quasar Feedback survey includes exploring the radio SEDs and obtaining higher spatial resolution radio observations to distinguish the forms of radio emission in this sample (e.g.\ jets verses shocks from quasar-driven winds). We will also present spatially-resolved ionised gas and molecular gas measurements for subsets of the sample, to explore in more detail the multi-phase outflows and ISM properties (e.g.\ Girdhar et al.\ in prep), building on the preliminary work presented in \citet{Jarvis19,Jarvis20} on the pilot sample. As such, this paper represents a first step towards a detailed understanding of the mechanisms and impact of AGN feedback in this unique sample of quasars. 

\section*{Acknowledgements}

Much of this work was presented in a preliminary form in MEJ's PhD thesis \citep{Jarvis_thesis}. The authors would like to thank the referee for their comments that improved the quality of this manuscript. The National Radio Astronomy Observatory is a facility of the National Science Foundation operated under cooperative agreement by Associated Universities, Inc.
This research made use of Photutils, an Astropy package for detection and photometry of astronomical sources \citep{Bradley19}. DMA and ACE thank the Science Technology and Facilities Council (STFC) for support through grant ST/T000244/1.

\section*{Data Availability}
 
The data underlying this article were accessed from the NRAO Science Data Archive (\url{https://archive.nrao.edu/archive/advquery.jsp}) using the proposal ids: 13B-127 and 18A-300. The radio images produced for this work are available from Newcastle University's data repository (\url{https://data.ncl.ac.uk/projects/Quasar_Feedback_Survey/92264}) and can also be accessed through our website: (\url{https://blogs.ncl.ac.uk/quasarfeedbacksurvey/data/}).

%%%%%%%%%%%%%%%%%%%%%%%%%%%%%%%%%%%%%%%%%%%%%%%%%%



\bibliographystyle{mnras} % 
\bibliography{VLA_paper.bib} 

\begin{thebibliography}{}
\makeatletter
\relax
\def\mn@urlcharsother{\let\do\@makeother \do\$\do\&\do\#\do\^\do\_\do\%\do\~}
\def\mn@doi{\begingroup\mn@urlcharsother \@ifnextchar [ {\mn@doi@}
  {\mn@doi@[]}}
\def\mn@doi@[#1]#2{\def\@tempa{#1}\ifx\@tempa\@empty \href
  {http://dx.doi.org/#2} {doi:#2}\else \href {http://dx.doi.org/#2} {#1}\fi
  \endgroup}
\def\mn@eprint#1#2{\mn@eprint@#1:#2::\@nil}
\def\mn@eprint@arXiv#1{\href {http://arxiv.org/abs/#1} {{\tt arXiv:#1}}}
\def\mn@eprint@dblp#1{\href {http://dblp.uni-trier.de/rec/bibtex/#1.xml}
  {dblp:#1}}
\def\mn@eprint@#1:#2:#3:#4\@nil{\def\@tempa {#1}\def\@tempb {#2}\def\@tempc
  {#3}\ifx \@tempc \@empty \let \@tempc \@tempb \let \@tempb \@tempa \fi \ifx
  \@tempb \@empty \def\@tempb {arXiv}\fi \@ifundefined
  {mn@eprint@\@tempb}{\@tempb:\@tempc}{\expandafter \expandafter \csname
  mn@eprint@\@tempb\endcsname \expandafter{\@tempc}}}

\bibitem[\protect\citeauthoryear{{Ahumada} et~al.,}{{Ahumada}
  et~al.}{2020}]{Ahumada20}
{Ahumada} R.,  et~al., 2020, \mn@doi [\apjs] {10.3847/1538-4365/ab929e}, \href
  {https://ui.adsabs.harvard.edu/abs/2020ApJS..249....3A} {249, 3}

\bibitem[\protect\citeauthoryear{{Alexander} \& {Hickox}}{{Alexander} \&
  {Hickox}}{2012}]{Alexander12}
{Alexander} D.~M.,  {Hickox} R.~C.,  2012, \mn@doi [\nar]
  {10.1016/j.newar.2011.11.003}, \href
  {http://adsabs.harvard.edu/abs/2012NewAR..56...93A} {56, 93}

\bibitem[\protect\citeauthoryear{{Alexandroff} et~al.,}{{Alexandroff}
  et~al.}{2012}]{Alexandroff12}
{Alexandroff} R.,  et~al., 2012, \mn@doi [\mnras]
  {10.1111/j.1365-2966.2012.20959.x}, \href
  {https://ui.adsabs.harvard.edu/abs/2012MNRAS.423.1325A} {423, 1325}

\bibitem[\protect\citeauthoryear{{Alexandroff}, {Zakamska}, {van Velzen},
  {Greene}  \& {Strauss}}{{Alexandroff} et~al.}{2016}]{Alexandroff+16}
{Alexandroff} R.~M.,  {Zakamska} N.~L.,  {van Velzen} S.,  {Greene} J.~E.,
  {Strauss} M.~A.,  2016, \mn@doi [\mnras] {10.1093/mnras/stw2124}, \href
  {http://adsabs.harvard.edu/abs/2016MNRAS.463.3056A} {463, 3056}

\bibitem[\protect\citeauthoryear{{An} \& {Baan}}{{An} \& {Baan}}{2012}]{An12}
{An} T.,  {Baan} W.~A.,  2012, \mn@doi [\apj] {10.1088/0004-637X/760/1/77},
  \href {https://ui.adsabs.harvard.edu/abs/2012ApJ...760...77A} {760, 77}

\bibitem[\protect\citeauthoryear{{Baldi}, {Capetti}  \& {Giovannini}}{{Baldi}
  et~al.}{2015}]{Baldi15}
{Baldi} R.~D.,  {Capetti} A.,   {Giovannini} G.,  2015, \mn@doi [\aap]
  {10.1051/0004-6361/201425426}, \href
  {https://ui.adsabs.harvard.edu/abs/2015A&A...576A..38B} {576, A38}

\bibitem[\protect\citeauthoryear{{Baldi} et~al.,}{{Baldi}
  et~al.}{2018}]{Baldi18}
{Baldi} R.~D.,  et~al., 2018, \mn@doi [\mnras] {10.1093/mnras/sty342}, \href
  {https://ui.adsabs.harvard.edu/abs/2018MNRAS.476.3478B} {476, 3478}

\bibitem[\protect\citeauthoryear{{Balmaverde} et~al.,}{{Balmaverde}
  et~al.}{2020}]{Balmaverde20}
{Balmaverde} B.,  et~al., 2020, arXiv e-prints, \href
  {https://ui.adsabs.harvard.edu/abs/2020arXiv201011195B} {p. arXiv:2010.11195}

\bibitem[\protect\citeauthoryear{{Baron} et~al.,}{{Baron}
  et~al.}{2018}]{Baron18}
{Baron} D.,  et~al., 2018, \mn@doi [\mnras] {10.1093/mnras/sty2113}, \href
  {https://ui.adsabs.harvard.edu/abs/2018MNRAS.480.3993B} {480, 3993}

\bibitem[\protect\citeauthoryear{{Becker}, {White}  \& {Helfand}}{{Becker}
  et~al.}{1995}]{Becker95}
{Becker} R.~H.,  {White} R.~L.,   {Helfand} D.~J.,  1995, \mn@doi [\apj]
  {10.1086/176166}, \href {http://adsabs.harvard.edu/abs/1995ApJ...450..559B}
  {450, 559}

\bibitem[\protect\citeauthoryear{{Begelman} \& {McKee}}{{Begelman} \&
  {McKee}}{1983}]{Begelman83}
{Begelman} M.~C.,  {McKee} C.~F.,  1983, \mn@doi [\apj] {10.1086/161179}, \href
  {https://ui.adsabs.harvard.edu/abs/1983ApJ...271...89B} {271, 89}

\bibitem[\protect\citeauthoryear{{Behar}, {Vogel}, {Baldi}, {Smith}  \&
  {Mushotzky}}{{Behar} et~al.}{2018}]{Behar18}
{Behar} E.,  {Vogel} S.,  {Baldi} R.~D.,  {Smith} K.~L.,   {Mushotzky} R.~F.,
  2018, \mn@doi [\mnras] {10.1093/mnras/sty850}, \href
  {https://ui.adsabs.harvard.edu/abs/2018MNRAS.478..399B} {478, 399}

\bibitem[\protect\citeauthoryear{{Bell}}{{Bell}}{2003}]{Bell03}
{Bell} E.~F.,  2003, \mn@doi [\apj] {10.1086/367829}, \href
  {http://adsabs.harvard.edu/abs/2003ApJ...586..794B} {586, 794}

\bibitem[\protect\citeauthoryear{{Berton} et~al.,}{{Berton}
  et~al.}{2018}]{Berton18}
{Berton} M.,  et~al., 2018, \mn@doi [\aap] {10.1051/0004-6361/201832612}, \href
  {https://ui.adsabs.harvard.edu/abs/2018A&A...614A..87B} {614, A87}

\bibitem[\protect\citeauthoryear{{Best} \& {Heckman}}{{Best} \&
  {Heckman}}{2012}]{Best12}
{Best} P.~N.,  {Heckman} T.~M.,  2012, \mn@doi [\mnras]
  {10.1111/j.1365-2966.2012.20414.x}, \href
  {https://ui.adsabs.harvard.edu/abs/2012MNRAS.421.1569B} {421, 1569}

\bibitem[\protect\citeauthoryear{{Best}, {R{\"o}ttgering}  \& {Longair}}{{Best}
  et~al.}{2000}]{Best00}
{Best} P.~N.,  {R{\"o}ttgering} H.~J.~A.,   {Longair} M.~S.,  2000, \mn@doi
  [\mnras] {10.1046/j.1365-8711.2000.03028.x}, \href
  {https://ui.adsabs.harvard.edu/abs/2000MNRAS.311...23B} {311, 23}

\bibitem[\protect\citeauthoryear{{Bicknell}, {Dopita}  \& {O'Dea}}{{Bicknell}
  et~al.}{1997}]{Bicknell97}
{Bicknell} G.~V.,  {Dopita} M.~A.,   {O'Dea} C. P.~O.,  1997, \mn@doi [\apj]
  {10.1086/304400}, \href
  {https://ui.adsabs.harvard.edu/abs/1997ApJ...485..112B} {485, 112}

\bibitem[\protect\citeauthoryear{{Bicknell}, {Mukherjee}, {Wagner},
  {Sutherland}  \& {Nesvadba}}{{Bicknell} et~al.}{2018}]{Bicknell18}
{Bicknell} G.~V.,  {Mukherjee} D.,  {Wagner} A. e.~Y.,  {Sutherland} R.~S.,
  {Nesvadba} N. P.~H.,  2018, \mn@doi [\mnras] {10.1093/mnras/sty070}, \href
  {https://ui.adsabs.harvard.edu/abs/2018MNRAS.475.3493B} {475, 3493}

\bibitem[\protect\citeauthoryear{{Bischetti} et~al.,}{{Bischetti}
  et~al.}{2020}]{Bischetti20}
{Bischetti} M.,  et~al., 2020, arXiv e-prints, \href
  {https://ui.adsabs.harvard.edu/abs/2020arXiv200901112B} {p. arXiv:2009.01112}

\bibitem[\protect\citeauthoryear{{Blandford} \& {Payne}}{{Blandford} \&
  {Payne}}{1982}]{Blandford82}
{Blandford} R.~D.,  {Payne} D.~G.,  1982, \mn@doi [\mnras]
  {10.1093/mnras/199.4.883}, \href
  {https://ui.adsabs.harvard.edu/abs/1982MNRAS.199..883B} {199, 883}

\bibitem[\protect\citeauthoryear{{Bluck} et~al.,}{{Bluck}
  et~al.}{2020}]{Bluck20}
{Bluck} A. F.~L.,  et~al., 2020, arXiv e-prints, \href
  {https://ui.adsabs.harvard.edu/abs/2020arXiv200905341B} {p. arXiv:2009.05341}

\bibitem[\protect\citeauthoryear{{Blundell} \& {Beasley}}{{Blundell} \&
  {Beasley}}{1998}]{Blundell98}
{Blundell} K.~M.,  {Beasley} A.~J.,  1998, \mn@doi [\mnras]
  {10.1046/j.1365-8711.1998.01752.x}, \href
  {https://ui.adsabs.harvard.edu/abs/1998MNRAS.299..165B} {299, 165}

\bibitem[\protect\citeauthoryear{{Bondi}, {P{\'e}rez-Torres}, {Piconcelli}  \&
  {Fu}}{{Bondi} et~al.}{2016}]{Bondi16}
{Bondi} M.,  {P{\'e}rez-Torres} M.~A.,  {Piconcelli} E.,   {Fu} H.,  2016,
  \mn@doi [\aap] {10.1051/0004-6361/201528021}, \href
  {https://ui.adsabs.harvard.edu/abs/2016A&A...588A.102B} {588, A102}

\bibitem[\protect\citeauthoryear{{Bonzini}, {Padovani}, {Mainieri},
  {Kellermann}, {Miller}, {Rosati}, {Tozzi}  \& {Vattakunnel}}{{Bonzini}
  et~al.}{2013}]{Bonzini13}
{Bonzini} M.,  {Padovani} P.,  {Mainieri} V.,  {Kellermann} K.~I.,  {Miller}
  N.,  {Rosati} P.,  {Tozzi} P.,   {Vattakunnel} S.,  2013, \mn@doi [\mnras]
  {10.1093/mnras/stt1879}, \href
  {https://ui.adsabs.harvard.edu/abs/2013MNRAS.436.3759B} {436, 3759}

\bibitem[\protect\citeauthoryear{Bradley et~al.,}{Bradley
  et~al.}{2019}]{Bradley19}
Bradley L.,  et~al., 2019, astropy/photutils: v0.6.
Zenodo, \mn@doi{10.5281/zenodo.2533376}

\bibitem[\protect\citeauthoryear{Briggs}{Briggs}{1995}]{Briggs95}
Briggs D.,  1995, PhD thesis, The New Mexico Institute of Mining and
  Technology, Socorro, New Mexico, -

\bibitem[\protect\citeauthoryear{{Brownson}, {Belfiore}, {Maiolino}, {Lin}  \&
  {Carniani}}{{Brownson} et~al.}{2020}]{Brownson20}
{Brownson} S.,  {Belfiore} F.,  {Maiolino} R.,  {Lin} L.,   {Carniani} S.,
  2020, \mn@doi [\mnras] {10.1093/mnrasl/slaa128}, \href
  {https://ui.adsabs.harvard.edu/abs/2020MNRAS.498L..66B} {498, L66}

\bibitem[\protect\citeauthoryear{{Brusa} et~al.,}{{Brusa}
  et~al.}{2015}]{Brusa15}
{Brusa} M.,  et~al., 2015, \mn@doi [\mnras] {10.1093/mnras/stu2117}, \href
  {https://ui.adsabs.harvard.edu/abs/2015MNRAS.446.2394B} {446, 2394}

\bibitem[\protect\citeauthoryear{{Carniani} et~al.,}{{Carniani}
  et~al.}{2015}]{Carniani15}
{Carniani} S.,  et~al., 2015, \mn@doi [\aap] {10.1051/0004-6361/201526557},
  \href {https://ui.adsabs.harvard.edu/abs/2015A&A...580A.102C} {580, A102}

\bibitem[\protect\citeauthoryear{{Chabrier}}{{Chabrier}}{2003}]{Chabrier03}
{Chabrier} G.,  2003, \mn@doi [\pasp] {10.1086/376392}, \href
  {http://adsabs.harvard.edu/abs/2003PASP..115..763C} {115, 763}

\bibitem[\protect\citeauthoryear{{Chartas} et~al.,}{{Chartas}
  et~al.}{2020}]{Chartas20}
{Chartas} G.,  et~al., 2020, \mn@doi [\mnras] {10.1093/mnras/staa1534}, \href
  {https://ui.adsabs.harvard.edu/abs/2020MNRAS.496..598C} {496, 598}

\bibitem[\protect\citeauthoryear{{Chen} et~al.,}{{Chen} et~al.}{2020}]{Chen20}
{Chen} S.,  et~al., 2020, arXiv e-prints, \href
  {https://ui.adsabs.harvard.edu/abs/2020arXiv200601700C} {p. arXiv:2006.01700}

\bibitem[\protect\citeauthoryear{{Circosta} et~al.,}{{Circosta}
  et~al.}{2018}]{Circosta18}
{Circosta} C.,  et~al., 2018, \mn@doi [\aap] {10.1051/0004-6361/201833520},
  \href {https://ui.adsabs.harvard.edu/abs/2018A&A...620A..82C} {620, A82}

\bibitem[\protect\citeauthoryear{{Condon}}{{Condon}}{1992}]{Condon92}
{Condon} J.~J.,  1992, \mn@doi [\araa] {10.1146/annurev.aa.30.090192.003043},
  \href {https://ui.adsabs.harvard.edu/abs/1992ARA&A..30..575C} {30, 575}

\bibitem[\protect\citeauthoryear{{Condon}, {Huang}, {Yin}  \& {Thuan}}{{Condon}
  et~al.}{1991}]{Condon91}
{Condon} J.~J.,  {Huang} Z.~P.,  {Yin} Q.~F.,   {Thuan} T.~X.,  1991, \mn@doi
  [\apj] {10.1086/170407}, \href
  {https://ui.adsabs.harvard.edu/abs/1991ApJ...378...65C} {378, 65}

\bibitem[\protect\citeauthoryear{{Condon}, {Cotton}, {Greisen}, {Yin},
  {Perley}, {Taylor}  \& {Broderick}}{{Condon} et~al.}{1998}]{Condon98}
{Condon} J.~J.,  {Cotton} W.~D.,  {Greisen} E.~W.,  {Yin} Q.~F.,  {Perley}
  R.~A.,  {Taylor} G.~B.,   {Broderick} J.~J.,  1998, \mn@doi [\aj]
  {10.1086/300337}, \href {http://adsabs.harvard.edu/abs/1998AJ....115.1693C}
  {115, 1693}

\bibitem[\protect\citeauthoryear{{Condon}, {Kellermann}, {Kimball},
  {Ivezi{\'c}}  \& {Perley}}{{Condon} et~al.}{2013}]{Condon13}
{Condon} J.~J.,  {Kellermann} K.~I.,  {Kimball} A.~E.,  {Ivezi{\'c}} {\v{Z}}.,
   {Perley} R.~A.,  2013, \mn@doi [\apj] {10.1088/0004-637X/768/1/37}, \href
  {https://ui.adsabs.harvard.edu/abs/2013ApJ...768...37C} {768, 37}

\bibitem[\protect\citeauthoryear{{Congiu} et~al.,}{{Congiu}
  et~al.}{2017}]{Congiu17}
{Congiu} E.,  et~al., 2017, \mn@doi [\aap] {10.1051/0004-6361/201730616}, \href
  {https://ui.adsabs.harvard.edu/abs/2017A&A...603A..32C} {603, A32}

\bibitem[\protect\citeauthoryear{{Costa}, {Sijacki}  \& {Haehnelt}}{{Costa}
  et~al.}{2014}]{Costa14}
{Costa} T.,  {Sijacki} D.,   {Haehnelt} M.~G.,  2014, \mn@doi [\mnras]
  {10.1093/mnras/stu1632}, \href
  {https://ui.adsabs.harvard.edu/abs/2014MNRAS.444.2355C} {444, 2355}

\bibitem[\protect\citeauthoryear{{Costa}, {Rosdahl}, {Sijacki}  \&
  {Haehnelt}}{{Costa} et~al.}{2018}]{Costa18b}
{Costa} T.,  {Rosdahl} J.,  {Sijacki} D.,   {Haehnelt} M.~G.,  2018, \mn@doi
  [\mnras] {10.1093/mnras/sty1514}, \href
  {https://ui.adsabs.harvard.edu/abs/2018MNRAS.479.2079C} {479, 2079}

\bibitem[\protect\citeauthoryear{{Cresci} \& {Maiolino}}{{Cresci} \&
  {Maiolino}}{2018}]{Cresci18}
{Cresci} G.,  {Maiolino} R.,  2018, \mn@doi [Nature Astronomy]
  {10.1038/s41550-018-0404-5}, \href
  {https://ui.adsabs.harvard.edu/abs/2018NatAs...2..179C} {2, 179}

\bibitem[\protect\citeauthoryear{{Davies} et~al.,}{{Davies}
  et~al.}{2020a}]{Davies20RIC}
{Davies} R.,  et~al., 2020a, \mn@doi [\mnras] {10.1093/mnras/staa2413}, \href
  {https://ui.adsabs.harvard.edu/abs/2020MNRAS.498.4150D} {498, 4150}

\bibitem[\protect\citeauthoryear{{Davies} et~al.,}{{Davies}
  et~al.}{2020b}]{Davies20}
{Davies} R.~L.,  et~al., 2020b, \mn@doi [\apj] {10.3847/1538-4357/ab86ad},
  \href {https://ui.adsabs.harvard.edu/abs/2020ApJ...894...28D} {894, 28}

\bibitem[\protect\citeauthoryear{{Del Moro} et~al.,}{{Del Moro}
  et~al.}{2013}]{DelMoro13}
{Del Moro} A.,  et~al., 2013, \mn@doi [\aap] {10.1051/0004-6361/201219880},
  \href {https://ui.adsabs.harvard.edu/abs/2013A&A...549A..59D} {549, A59}

\bibitem[\protect\citeauthoryear{{Dey} et~al.,}{{Dey} et~al.}{2019}]{Dey19}
{Dey} A.,  et~al., 2019, \mn@doi [\aj] {10.3847/1538-3881/ab089d}, \href
  {https://ui.adsabs.harvard.edu/abs/2019AJ....157..168D} {157, 168}

\bibitem[\protect\citeauthoryear{{Doi}, {Asada}, {Fujisawa}, {Nagai},
  {Hagiwara}, {Wajima}  \& {Inoue}}{{Doi} et~al.}{2013}]{Doi13}
{Doi} A.,  {Asada} K.,  {Fujisawa} K.,  {Nagai} H.,  {Hagiwara} Y.,  {Wajima}
  K.,   {Inoue} M.,  2013, \mn@doi [\apj] {10.1088/0004-637X/765/1/69}, \href
  {https://ui.adsabs.harvard.edu/abs/2013ApJ...765...69D} {765, 69}

\bibitem[\protect\citeauthoryear{{Dugan}, {Gaibler}  \& {Silk}}{{Dugan}
  et~al.}{2017}]{Dugan17}
{Dugan} Z.,  {Gaibler} V.,   {Silk} J.,  2017, \mn@doi [\apj]
  {10.3847/1538-4357/aa7566}, \href
  {https://ui.adsabs.harvard.edu/abs/2017ApJ...844...37D} {844, 37}

\bibitem[\protect\citeauthoryear{{Duric} \& {Seaquist}}{{Duric} \&
  {Seaquist}}{1988}]{Duric88}
{Duric} N.,  {Seaquist} E.~R.,  1988, \mn@doi [\apj] {10.1086/166118}, \href
  {https://ui.adsabs.harvard.edu/abs/1988ApJ...326..574D} {326, 574}

\bibitem[\protect\citeauthoryear{{Everett}}{{Everett}}{2005}]{Everett05}
{Everett} J.~E.,  2005, \mn@doi [\apj] {10.1086/432678}, \href
  {https://ui.adsabs.harvard.edu/abs/2005ApJ...631..689E} {631, 689}

\bibitem[\protect\citeauthoryear{{Fabian}}{{Fabian}}{2012}]{Fabian12}
{Fabian} A.~C.,  2012, \mn@doi [\araa] {10.1146/annurev-astro-081811-125521},
  \href {http://adsabs.harvard.edu/abs/2012ARA%26A..50..455F} {50, 455}

\bibitem[\protect\citeauthoryear{{Falcke}, {Nagar}, {Wilson}  \&
  {Ulvestad}}{{Falcke} et~al.}{2000}]{Falcke00}
{Falcke} H.,  {Nagar} N.~M.,  {Wilson} A.~S.,   {Ulvestad} J.~S.,  2000,
  \mn@doi [\apj] {10.1086/309543}, \href
  {https://ui.adsabs.harvard.edu/abs/2000ApJ...542..197F} {542, 197}

\bibitem[\protect\citeauthoryear{{Fanaroff} \& {Riley}}{{Fanaroff} \&
  {Riley}}{1974}]{Fanaroff74}
{Fanaroff} B.~L.,  {Riley} J.~M.,  1974, \mn@doi [\mnras]
  {10.1093/mnras/167.1.31P}, \href
  {http://adsabs.harvard.edu/abs/1974MNRAS.167P..31F} {167, 31P}

\bibitem[\protect\citeauthoryear{{Fanti}, {Fanti}, {de Ruiter}  \&
  {Parma}}{{Fanti} et~al.}{1987}]{Fanti87}
{Fanti} C.,  {Fanti} R.,  {de Ruiter} H.~R.,   {Parma} P.,  1987, \aaps, \href
  {https://ui.adsabs.harvard.edu/abs/1987A&AS...69...57F} {69, 57}

\bibitem[\protect\citeauthoryear{{Fischer} et~al.,}{{Fischer}
  et~al.}{2019}]{Fischer19}
{Fischer} T.~C.,  et~al., 2019, \mn@doi [\apj] {10.3847/1538-4357/ab11c3},
  \href {https://ui.adsabs.harvard.edu/abs/2019ApJ...875..102F} {875, 102}

\bibitem[\protect\citeauthoryear{{Fukumura}, {Kazanas}, {Contopoulos}  \&
  {Behar}}{{Fukumura} et~al.}{2010}]{Fukumura10}
{Fukumura} K.,  {Kazanas} D.,  {Contopoulos} I.,   {Behar} E.,  2010, \mn@doi
  [\apj] {10.1088/0004-637X/715/1/636}, \href
  {https://ui.adsabs.harvard.edu/abs/2010ApJ...715..636F} {715, 636}

\bibitem[\protect\citeauthoryear{{Gallimore}, {Axon}, {O'Dea}, {Baum}  \&
  {Pedlar}}{{Gallimore} et~al.}{2006}]{Gallimore06}
{Gallimore} J.~F.,  {Axon} D.~J.,  {O'Dea} C.~P.,  {Baum} S.~A.,   {Pedlar} A.,
   2006, \mn@doi [\aj] {10.1086/504593}, \href
  {http://adsabs.harvard.edu/abs/2006AJ....132..546G} {132, 546}

\bibitem[\protect\citeauthoryear{{Gelderman} \& {Whittle}}{{Gelderman} \&
  {Whittle}}{1994}]{Gelderman94}
{Gelderman} R.,  {Whittle} M.,  1994, \mn@doi [\apjs] {10.1086/191946}, \href
  {http://adsabs.harvard.edu/abs/1994ApJS...91..491G} {91, 491}

\bibitem[\protect\citeauthoryear{{Greene}, {Setton}, {Bezanson}, {Suess},
  {Kriek}, {Spilker}, {Goulding}  \& {Feldmann}}{{Greene}
  et~al.}{2020}]{Greene20}
{Greene} J.~E.,  {Setton} D.,  {Bezanson} R.,  {Suess} K.~A.,  {Kriek} M.,
  {Spilker} J.~S.,  {Goulding} A.~D.,   {Feldmann} R.,  2020, \mn@doi [\apjl]
  {10.3847/2041-8213/aba534}, \href
  {https://ui.adsabs.harvard.edu/abs/2020ApJ...899L...9G} {899, L9}

\bibitem[\protect\citeauthoryear{{G{\"u}rkan} et~al.,}{{G{\"u}rkan}
  et~al.}{2019}]{Gurkan19}
{G{\"u}rkan} G.,  et~al., 2019, \mn@doi [\aap] {10.1051/0004-6361/201833892},
  \href {https://ui.adsabs.harvard.edu/abs/2019A&A...622A..11G} {622, A11}

\bibitem[\protect\citeauthoryear{{Hainline}, {Hickox}, {Greene}, {Myers}  \&
  {Zakamska}}{{Hainline} et~al.}{2013}]{Hainline13}
{Hainline} K.~N.,  {Hickox} R.,  {Greene} J.~E.,  {Myers} A.~D.,   {Zakamska}
  N.~L.,  2013, \mn@doi [\apj] {10.1088/0004-637X/774/2/145}, \href
  {https://ui.adsabs.harvard.edu/abs/2013ApJ...774..145H} {774, 145}

\bibitem[\protect\citeauthoryear{{Hardcastle} \& {Croston}}{{Hardcastle} \&
  {Croston}}{2020}]{Hardcastle20}
{Hardcastle} M.~J.,  {Croston} J.~H.,  2020, \mn@doi [\nar]
  {10.1016/j.newar.2020.101539}, \href
  {https://ui.adsabs.harvard.edu/abs/2020NewAR..8801539H} {88, 101539}

\bibitem[\protect\citeauthoryear{{Harrison}}{{Harrison}}{2017}]{Harrison17}
{Harrison} C.~M.,  2017, \mn@doi [Nature Astronomy] {10.1038/s41550-017-0165},
  \href {http://adsabs.harvard.edu/abs/2017NatAs...1E.165H} {1, 0165}

\bibitem[\protect\citeauthoryear{{Harrison} et~al.,}{{Harrison}
  et~al.}{2012}]{Harrison12a}
{Harrison} C.~M.,  et~al., 2012, \mn@doi [\mnras]
  {10.1111/j.1365-2966.2012.21723.x}, \href
  {https://ui.adsabs.harvard.edu/abs/2012MNRAS.426.1073H} {426, 1073}

\bibitem[\protect\citeauthoryear{{Harrison}, {Alexander}, {Mullaney}  \&
  {Swinbank}}{{Harrison} et~al.}{2014}]{Harrison14}
{Harrison} C.~M.,  {Alexander} D.~M.,  {Mullaney} J.~R.,   {Swinbank} A.~M.,
  2014, \mn@doi [\mnras] {10.1093/mnras/stu515}, \href
  {http://adsabs.harvard.edu/abs/2014MNRAS.441.3306H} {441, 3306}

\bibitem[\protect\citeauthoryear{{Harrison}, {Thomson}, {Alexander}, {Bauer},
  {Edge}, {Hogan}, {Mullaney}  \& {Swinbank}}{{Harrison}
  et~al.}{2015}]{Harrison15}
{Harrison} C.~M.,  {Thomson} A.~P.,  {Alexander} D.~M.,  {Bauer} F.~E.,  {Edge}
  A.~C.,  {Hogan} M.~T.,  {Mullaney} J.~R.,   {Swinbank} A.~M.,  2015, \mn@doi
  [\apj] {10.1088/0004-637X/800/1/45}, \href
  {http://adsabs.harvard.edu/abs/2015ApJ...800...45H} {800, 45}

\bibitem[\protect\citeauthoryear{{Harrison} et~al.,}{{Harrison}
  et~al.}{2016}]{Harrison16}
{Harrison} C.~M.,  et~al., 2016, \mn@doi [\mnras] {10.1093/mnras/stv2727},
  \href {https://ui.adsabs.harvard.edu/abs/2016MNRAS.456.1195H} {456, 1195}

\bibitem[\protect\citeauthoryear{{Harrison}, {Costa}, {Tadhunter},
  {Fl{\"u}tsch}, {Kakkad}, {Perna}  \& {Vietri}}{{Harrison}
  et~al.}{2018}]{Harrison18}
{Harrison} C.~M.,  {Costa} T.,  {Tadhunter} C.~N.,  {Fl{\"u}tsch} A.,  {Kakkad}
  D.,  {Perna} M.,   {Vietri} G.,  2018, \mn@doi [Nature Astronomy]
  {10.1038/s41550-018-0403-6}, \href
  {https://ui.adsabs.harvard.edu/abs/2018NatAs...2..198H} {2, 198}

\bibitem[\protect\citeauthoryear{{Heckman}, {Miley}, {van Breugel}  \&
  {Butcher}}{{Heckman} et~al.}{1981}]{Heckman81}
{Heckman} T.~M.,  {Miley} G.~K.,  {van Breugel} W.~J.~M.,   {Butcher} H.~R.,
  1981, \mn@doi [\apj] {10.1086/159050}, \href
  {https://ui.adsabs.harvard.edu/abs/1981ApJ...247..403H} {247, 403}

\bibitem[\protect\citeauthoryear{{Heckman}, {Kauffmann}, {Brinchmann},
  {Charlot}, {Tremonti}  \& {White}}{{Heckman} et~al.}{2004}]{Heckman04}
{Heckman} T.~M.,  {Kauffmann} G.,  {Brinchmann} J.,  {Charlot} S.,  {Tremonti}
  C.,   {White} S. D.~M.,  2004, \mn@doi [\apj] {10.1086/422872}, \href
  {https://ui.adsabs.harvard.edu/abs/2004ApJ...613..109H} {613, 109}

\bibitem[\protect\citeauthoryear{{Helou}, {Soifer}  \&
  {Rowan-Robinson}}{{Helou} et~al.}{1985}]{Helou85}
{Helou} G.,  {Soifer} B.~T.,   {Rowan-Robinson} M.,  1985, \mn@doi [\apjl]
  {10.1086/184556}, \href
  {https://ui.adsabs.harvard.edu/abs/1985ApJ...298L...7H} {298, L7}

\bibitem[\protect\citeauthoryear{{Holt}, {Tadhunter}  \& {Morganti}}{{Holt}
  et~al.}{2008}]{Holt08}
{Holt} J.,  {Tadhunter} C.~N.,   {Morganti} R.,  2008, \mn@doi [\mnras]
  {10.1111/j.1365-2966.2008.13089.x}, \href
  {https://ui.adsabs.harvard.edu/abs/2008MNRAS.387..639H} {387, 639}

\bibitem[\protect\citeauthoryear{{Hopkins}, {Richards}  \&
  {Hernquist}}{{Hopkins} et~al.}{2007}]{Hopkins07}
{Hopkins} P.~F.,  {Richards} G.~T.,   {Hernquist} L.,  2007, \mn@doi [\apj]
  {10.1086/509629}, \href
  {https://ui.adsabs.harvard.edu/abs/2007ApJ...654..731H} {654, 731}

\bibitem[\protect\citeauthoryear{{Husemann}, {Wisotzki}, {S{\'a}nchez}  \&
  {Jahnke}}{{Husemann} et~al.}{2013}]{Husemann13}
{Husemann} B.,  {Wisotzki} L.,  {S{\'a}nchez} S.~F.,   {Jahnke} K.,  2013,
  \mn@doi [\aap] {10.1051/0004-6361/201220076}, \href
  {https://ui.adsabs.harvard.edu/abs/2013A&A...549A..43H} {549, A43}

\bibitem[\protect\citeauthoryear{{Husemann}, {Scharw{\"a}chter}, {Bennert},
  {Mainieri}, {Woo}  \& {Kakkad}}{{Husemann} et~al.}{2016}]{Husemann16}
{Husemann} B.,  {Scharw{\"a}chter} J.,  {Bennert} V.~N.,  {Mainieri} V.,  {Woo}
  J.~H.,   {Kakkad} D.,  2016, \mn@doi [\aap] {10.1051/0004-6361/201527992},
  \href {https://ui.adsabs.harvard.edu/abs/2016A&A...594A..44H} {594, A44}

\bibitem[\protect\citeauthoryear{{Husemann} et~al.,}{{Husemann}
  et~al.}{2017}]{HusemannCARS}
{Husemann} B.,  et~al., 2017, \mn@doi [The Messenger]
  {10.18727/0722-6691/5038}, \href
  {https://ui.adsabs.harvard.edu/abs/2017Msngr.169...42H} {169, 42}

\bibitem[\protect\citeauthoryear{{Ivezi{\'c}} et~al.,}{{Ivezi{\'c}}
  et~al.}{2002}]{Ivezic02}
{Ivezi{\'c}} {\v{Z}}.,  et~al., 2002, \mn@doi [\aj] {10.1086/344069}, \href
  {https://ui.adsabs.harvard.edu/abs/2002AJ....124.2364I} {124, 2364}

\bibitem[\protect\citeauthoryear{Jarvis}{Jarvis}{2020}]{Jarvis_thesis}
Jarvis M.~E.,  2020, Multi-wavelength analysis to constrain the role of AGN in
  galaxy evolution, \url {http://nbn-resolving.de/urn:nbn:de:bvb:19-267601}

\bibitem[\protect\citeauthoryear{{Jarvis} et~al.,}{{Jarvis}
  et~al.}{2019}]{Jarvis19}
{Jarvis} M.~E.,  et~al., 2019, \mn@doi [\mnras] {10.1093/mnras/stz556}, \href
  {https://ui.adsabs.harvard.edu/abs/2019MNRAS.485.2710J} {485, 2710}

\bibitem[\protect\citeauthoryear{{Jarvis} et~al.,}{{Jarvis}
  et~al.}{2020}]{Jarvis20}
{Jarvis} M.~E.,  et~al., 2020, \mn@doi [\mnras] {10.1093/mnras/staa2196}, \href
  {https://ui.adsabs.harvard.edu/abs/2020MNRAS.498.1560J} {498, 1560}

\bibitem[\protect\citeauthoryear{{Jiang}, {Ciotti}, {Ostriker}  \&
  {Spitkovsky}}{{Jiang} et~al.}{2010}]{Jiang10}
{Jiang} Y.-F.,  {Ciotti} L.,  {Ostriker} J.~P.,   {Spitkovsky} A.,  2010,
  \mn@doi [\apj] {10.1088/0004-637X/711/1/125}, \href
  {https://ui.adsabs.harvard.edu/abs/2010ApJ...711..125J} {711, 125}

\bibitem[\protect\citeauthoryear{{Kakkad} et~al.,}{{Kakkad}
  et~al.}{2016}]{Kakkad16}
{Kakkad} D.,  et~al., 2016, \mn@doi [\aap] {10.1051/0004-6361/201527968}, \href
  {https://ui.adsabs.harvard.edu/abs/2016A&A...592A.148K} {592, A148}

\bibitem[\protect\citeauthoryear{{Kakkad} et~al.,}{{Kakkad}
  et~al.}{2020}]{Kakkad20}
{Kakkad} D.,  et~al., 2020, arXiv e-prints, \href
  {https://ui.adsabs.harvard.edu/abs/2020arXiv200801728K} {p. arXiv:2008.01728}

\bibitem[\protect\citeauthoryear{{Karouzos}, {Woo}  \& {Bae}}{{Karouzos}
  et~al.}{2016}]{Karouzos16}
{Karouzos} M.,  {Woo} J.-H.,   {Bae} H.-J.,  2016, \mn@doi [\apj]
  {10.3847/1538-4357/833/2/171}, \href
  {https://ui.adsabs.harvard.edu/abs/2016ApJ...833..171K} {833, 171}

\bibitem[\protect\citeauthoryear{{Kauffmann} \& {Maraston}}{{Kauffmann} \&
  {Maraston}}{2019}]{Kauffmann19}
{Kauffmann} G.,  {Maraston} C.,  2019, \mn@doi [\mnras]
  {10.1093/mnras/stz2271}, \href
  {https://ui.adsabs.harvard.edu/abs/2019MNRAS.489.1973K} {489, 1973}

\bibitem[\protect\citeauthoryear{{Kellermann}, {Sramek}, {Schmidt}, {Shaffer}
  \& {Green}}{{Kellermann} et~al.}{1989}]{Kellermann89}
{Kellermann} K.~I.,  {Sramek} R.,  {Schmidt} M.,  {Shaffer} D.~B.,   {Green}
  R.,  1989, \mn@doi [\aj] {10.1086/115207}, \href
  {http://adsabs.harvard.edu/abs/1989AJ.....98.1195K} {98, 1195}

\bibitem[\protect\citeauthoryear{{Kellermann}, {Condon}, {Kimball}, {Perley}
  \& {Ivezi{\'c}}}{{Kellermann} et~al.}{2016}]{Kellermann16}
{Kellermann} K.~I.,  {Condon} J.~J.,  {Kimball} A.~E.,  {Perley} R.~A.,
  {Ivezi{\'c}} {\v{Z}}.,  2016, \mn@doi [\apj] {10.3847/0004-637X/831/2/168},
  \href {https://ui.adsabs.harvard.edu/abs/2016ApJ...831..168K} {831, 168}

\bibitem[\protect\citeauthoryear{{Kewley}, {Heisler}, {Dopita}  \&
  {Lumsden}}{{Kewley} et~al.}{2001}]{Kewley01}
{Kewley} L.~J.,  {Heisler} C.~A.,  {Dopita} M.~A.,   {Lumsden} S.,  2001,
  \mn@doi [\apjs] {10.1086/318944}, \href
  {https://ui.adsabs.harvard.edu/abs/2001ApJS..132...37K} {132, 37}

\bibitem[\protect\citeauthoryear{{Kharb}, {O'Dea}, {Baum}, {Colbert}  \&
  {Xu}}{{Kharb} et~al.}{2006}]{Kharb06}
{Kharb} P.,  {O'Dea} C.~P.,  {Baum} S.~A.,  {Colbert} E.~J.~M.,   {Xu} C.,
  2006, \mn@doi [\apj] {10.1086/507945}, \href
  {https://ui.adsabs.harvard.edu/abs/2006ApJ...652..177K} {652, 177}

\bibitem[\protect\citeauthoryear{{Kharb}, {O'Dea}, {Baum}, {Hardcastle},
  {Dicken}, {Croston}, {Mingo}  \& {Noel-Storr}}{{Kharb}
  et~al.}{2014}]{Kharb14}
{Kharb} P.,  {O'Dea} C.~P.,  {Baum} S.~A.,  {Hardcastle} M.~J.,  {Dicken} D.,
  {Croston} J.~H.,  {Mingo} B.,   {Noel-Storr} J.,  2014, \mn@doi [\mnras]
  {10.1093/mnras/stu421}, \href
  {https://ui.adsabs.harvard.edu/abs/2014MNRAS.440.2976K} {440, 2976}

\bibitem[\protect\citeauthoryear{{Kharb}, {Das}, {Paragi}, {Subramanian}  \&
  {Chitta}}{{Kharb} et~al.}{2015}]{Kharb15}
{Kharb} P.,  {Das} M.,  {Paragi} Z.,  {Subramanian} S.,   {Chitta} L.~P.,
  2015, \mn@doi [\apj] {10.1088/0004-637X/799/2/161}, \href
  {https://ui.adsabs.harvard.edu/abs/2015ApJ...799..161K} {799, 161}

\bibitem[\protect\citeauthoryear{{Kharb}, {Srivastava}, {Singh}, {Gallimore},
  {Ishwara-Chandra}  \& {Ananda}}{{Kharb} et~al.}{2016}]{Kharb16}
{Kharb} P.,  {Srivastava} S.,  {Singh} V.,  {Gallimore} J.~F.,
  {Ishwara-Chandra} C.~H.,   {Ananda} H.,  2016, \mn@doi [\mnras]
  {10.1093/mnras/stw699}, \href
  {https://ui.adsabs.harvard.edu/abs/2016MNRAS.459.1310K} {459, 1310}

\bibitem[\protect\citeauthoryear{{Kharb}, {Subramanian}, {Vaddi}, {Das}  \&
  {Paragi}}{{Kharb} et~al.}{2017}]{Kharb17}
{Kharb} P.,  {Subramanian} S.,  {Vaddi} S.,  {Das} M.,   {Paragi} Z.,  2017,
  \mn@doi [\apj] {10.3847/1538-4357/aa8321}, \href
  {https://ui.adsabs.harvard.edu/abs/2017ApJ...846...12K} {846, 12}

\bibitem[\protect\citeauthoryear{{Kim}, {Ho}, {Lonsdale}, {Lacy}, {Blain}  \&
  {Kimball}}{{Kim} et~al.}{2013}]{Kim13}
{Kim} M.,  {Ho} L.~C.,  {Lonsdale} C.~J.,  {Lacy} M.,  {Blain} A.~W.,
  {Kimball} A.~E.,  2013, \mn@doi [\apjl] {10.1088/2041-8205/768/1/L9}, \href
  {https://ui.adsabs.harvard.edu/abs/2013ApJ...768L...9K} {768, L9}

\bibitem[\protect\citeauthoryear{{Kimball}, {Ivezi{\'c}}, {Wiita}  \&
  {Schneider}}{{Kimball} et~al.}{2011a}]{Kimball11a}
{Kimball} A.~E.,  {Ivezi{\'c}} {\v{Z}}.,  {Wiita} P.~J.,   {Schneider} D.~P.,
  2011a, \mn@doi [\aj] {10.1088/0004-6256/141/6/182}, \href
  {https://ui.adsabs.harvard.edu/abs/2011AJ....141..182K} {141, 182}

\bibitem[\protect\citeauthoryear{{Kimball}, {Kellermann}, {Condon},
  {Ivezi{\'c}}  \& {Perley}}{{Kimball} et~al.}{2011b}]{Kimball11b}
{Kimball} A.~E.,  {Kellermann} K.~I.,  {Condon} J.~J.,  {Ivezi{\'c}} {\v{Z}}.,
   {Perley} R.~A.,  2011b, \mn@doi [\apjl] {10.1088/2041-8205/739/1/L29}, \href
  {https://ui.adsabs.harvard.edu/abs/2011ApJ...739L..29K} {739, L29}

\bibitem[\protect\citeauthoryear{{King} \& {Pounds}}{{King} \&
  {Pounds}}{2015}]{King15}
{King} A.,  {Pounds} K.,  2015, \mn@doi [\araa]
  {10.1146/annurev-astro-082214-122316}, \href
  {https://ui.adsabs.harvard.edu/abs/2015ARA&A..53..115K} {53, 115}

\bibitem[\protect\citeauthoryear{{Kukula}, {Dunlop}, {Hughes}  \&
  {Rawlings}}{{Kukula} et~al.}{1998}]{Kukula98}
{Kukula} M.~J.,  {Dunlop} J.~S.,  {Hughes} D.~H.,   {Rawlings} S.,  1998,
  \mn@doi [\mnras] {10.1046/j.1365-8711.1998.01481.x}, \href
  {http://adsabs.harvard.edu/abs/1998MNRAS.297..366K} {297, 366}

\bibitem[\protect\citeauthoryear{{Kunert-Bajraszewska} \&
  {Labiano}}{{Kunert-Bajraszewska} \& {Labiano}}{2010}]{Kunert-Bajraszewska10}
{Kunert-Bajraszewska} M.,  {Labiano} A.,  2010, \mn@doi [\mnras]
  {10.1111/j.1365-2966.2010.17300.x}, \href
  {https://ui.adsabs.harvard.edu/abs/2010MNRAS.408.2279K} {408, 2279}

\bibitem[\protect\citeauthoryear{{Labiano}}{{Labiano}}{2008}]{Labiano08}
{Labiano} A.,  2008, \mn@doi [\aap] {10.1051/0004-6361:200810399}, \href
  {https://ui.adsabs.harvard.edu/abs/2008A&A...488L..59L} {488, L59}

\bibitem[\protect\citeauthoryear{{Lansbury}, {Jarvis}, {Harrison}, {Alexander},
  {Del Moro}, {Edge}, {Mullaney}  \& {Thomson}}{{Lansbury}
  et~al.}{2018}]{Lansbury18}
{Lansbury} G.~B.,  {Jarvis} M.~E.,  {Harrison} C.~M.,  {Alexander} D.~M.,  {Del
  Moro} A.,  {Edge} A.~C.,  {Mullaney} J.~R.,   {Thomson} A.~P.,  2018, \mn@doi
  [\apjl] {10.3847/2041-8213/aab357}, \href
  {https://ui.adsabs.harvard.edu/abs/2018ApJ...856L...1L} {856, L1}

\bibitem[\protect\citeauthoryear{{Laor} \& {Behar}}{{Laor} \&
  {Behar}}{2008}]{Laor08}
{Laor} A.,  {Behar} E.,  2008, \mn@doi [\mnras]
  {10.1111/j.1365-2966.2008.13806.x}, \href
  {https://ui.adsabs.harvard.edu/abs/2008MNRAS.390..847L} {390, 847}

\bibitem[\protect\citeauthoryear{{Leipski}, {Falcke}, {Bennert}  \&
  {H{\"u}ttemeister}}{{Leipski} et~al.}{2006}]{Leipski06}
{Leipski} C.,  {Falcke} H.,  {Bennert} N.,   {H{\"u}ttemeister} S.,  2006,
  \mn@doi [\aap] {10.1051/0004-6361:20054311}, \href
  {https://ui.adsabs.harvard.edu/abs/2006A&A...455..161L} {455, 161}

\bibitem[\protect\citeauthoryear{{Liao} \& {Gu}}{{Liao} \& {Gu}}{2020}]{Liao20}
{Liao} M.,  {Gu} M.,  2020, \mn@doi [\mnras] {10.1093/mnras/stz2981}, \href
  {https://ui.adsabs.harvard.edu/abs/2020MNRAS.491...92L} {491, 92}

\bibitem[\protect\citeauthoryear{{Liu}, {Zakamska}, {Greene}, {Nesvadba}  \&
  {Liu}}{{Liu} et~al.}{2013}]{Liu13b}
{Liu} G.,  {Zakamska} N.~L.,  {Greene} J.~E.,  {Nesvadba} N. P.~H.,   {Liu} X.,
   2013, \mn@doi [\mnras] {10.1093/mnras/stt051}, \href
  {https://ui.adsabs.harvard.edu/abs/2013MNRAS.430.2327L} {430, 2327}

\bibitem[\protect\citeauthoryear{{Liu}, {Zakamska}  \& {Greene}}{{Liu}
  et~al.}{2014}]{Liu14}
{Liu} G.,  {Zakamska} N.~L.,   {Greene} J.~E.,  2014, \mn@doi [\mnras]
  {10.1093/mnras/stu974}, \href
  {https://ui.adsabs.harvard.edu/abs/2014MNRAS.442.1303L} {442, 1303}

\bibitem[\protect\citeauthoryear{{Madau} \& {Dickinson}}{{Madau} \&
  {Dickinson}}{2014}]{Madau14}
{Madau} P.,  {Dickinson} M.,  2014, \mn@doi [\araa]
  {10.1146/annurev-astro-081811-125615}, \href
  {https://ui.adsabs.harvard.edu/abs/2014ARA&A..52..415M} {52, 415}

\bibitem[\protect\citeauthoryear{{Mahony}, {Sadler}, {Croom}, {Ekers}, {Feain}
  \& {Murphy}}{{Mahony} et~al.}{2012}]{Mahony12}
{Mahony} E.~K.,  {Sadler} E.~M.,  {Croom} S.~M.,  {Ekers} R.~D.,  {Feain}
  I.~J.,   {Murphy} T.,  2012, \mn@doi [\apj] {10.1088/0004-637X/754/1/12},
  \href {https://ui.adsabs.harvard.edu/abs/2012ApJ...754...12M} {754, 12}

\bibitem[\protect\citeauthoryear{{Maiolino} et~al.,}{{Maiolino}
  et~al.}{2017}]{Maiolino17}
{Maiolino} R.,  et~al., 2017, \mn@doi [\nat] {10.1038/nature21677}, \href
  {https://ui.adsabs.harvard.edu/abs/2017Natur.544..202M} {544, 202}

\bibitem[\protect\citeauthoryear{{Marvil}, {Owen}  \& {Eilek}}{{Marvil}
  et~al.}{2015}]{Marvil15}
{Marvil} J.,  {Owen} F.,   {Eilek} J.,  2015, \mn@doi [\aj]
  {10.1088/0004-6256/149/1/32}, \href
  {https://ui.adsabs.harvard.edu/abs/2015AJ....149...32M} {149, 32}

\bibitem[\protect\citeauthoryear{{McCarthy}, {Schaye}, {Bower}, {Ponman},
  {Booth}, {Dalla Vecchia}  \& {Springel}}{{McCarthy}
  et~al.}{2011}]{McCarthy11}
{McCarthy} I.~G.,  {Schaye} J.,  {Bower} R.~G.,  {Ponman} T.~J.,  {Booth}
  C.~M.,  {Dalla Vecchia} C.,   {Springel} V.,  2011, \mn@doi [\mnras]
  {10.1111/j.1365-2966.2010.18033.x}, \href
  {https://ui.adsabs.harvard.edu/abs/2011MNRAS.412.1965M} {412, 1965}

\bibitem[\protect\citeauthoryear{{McNamara} \& {Nulsen}}{{McNamara} \&
  {Nulsen}}{2012}]{McNamara12}
{McNamara} B.~R.,  {Nulsen} P.~E.~J.,  2012, \mn@doi [New Journal of Physics]
  {10.1088/1367-2630/14/5/055023}, \href
  {https://ui.adsabs.harvard.edu/abs/2012NJPh...14e5023M} {14, 055023}

\bibitem[\protect\citeauthoryear{{Mingo} et~al.,}{{Mingo}
  et~al.}{2019}]{Mingo19}
{Mingo} B.,  et~al., 2019, \mn@doi [\mnras] {10.1093/mnras/stz1901}, \href
  {https://ui.adsabs.harvard.edu/abs/2019MNRAS.488.2701M} {488, 2701}

\bibitem[\protect\citeauthoryear{{Mingozzi} et~al.,}{{Mingozzi}
  et~al.}{2019}]{Mingozzi19}
{Mingozzi} M.,  et~al., 2019, \mn@doi [\aap] {10.1051/0004-6361/201834372},
  \href {https://ui.adsabs.harvard.edu/abs/2019A&A...622A.146M} {622, A146}

\bibitem[\protect\citeauthoryear{{Mizumoto}, {Done}, {Tomaru}  \&
  {Edwards}}{{Mizumoto} et~al.}{2019}]{Mizumoto19}
{Mizumoto} M.,  {Done} C.,  {Tomaru} R.,   {Edwards} I.,  2019, \mn@doi
  [\mnras] {10.1093/mnras/stz2225}, \href
  {https://ui.adsabs.harvard.edu/abs/2019MNRAS.489.1152M} {489, 1152}

\bibitem[\protect\citeauthoryear{{Molyneux}, {Harrison}  \&
  {Jarvis}}{{Molyneux} et~al.}{2019}]{Molyneux19}
{Molyneux} S.~J.,  {Harrison} C.~M.,   {Jarvis} M.~E.,  2019, \mn@doi [\aap]
  {10.1051/0004-6361/201936408}, \href
  {https://ui.adsabs.harvard.edu/abs/2019A&A...631A.132M} {631, A132}

\bibitem[\protect\citeauthoryear{{Mori{\'c}}, {Smol{\v c}i{\'c}}, {Kimball},
  {Riechers}, {Ivezi{\'c}}  \& {Scoville}}{{Mori{\'c}} et~al.}{2010}]{Moric10}
{Mori{\'c}} I.,  {Smol{\v c}i{\'c}} V.,  {Kimball} A.,  {Riechers} D.~A.,
  {Ivezi{\'c}} {\v Z}.,   {Scoville} N.,  2010, \mn@doi [\apj]
  {10.1088/0004-637X/724/1/779}, \href
  {http://adsabs.harvard.edu/abs/2010ApJ...724..779M} {724, 779}

\bibitem[\protect\citeauthoryear{{Moshir}, {Kopman}  \& {Conrow}}{{Moshir}
  et~al.}{1992}]{Moshir92}
{Moshir} M.,  {Kopman} G.,   {Conrow} T.~A.~O.,  1992, {IRAS Faint Source
  Survey, Explanatory supplement version 2}.
Infrared Processing and Analysis Center

\bibitem[\protect\citeauthoryear{{Moy} \& {Rocca-Volmerange}}{{Moy} \&
  {Rocca-Volmerange}}{2002}]{Moy02}
{Moy} E.,  {Rocca-Volmerange} B.,  2002, \mn@doi [\aap]
  {10.1051/0004-6361:20011727}, \href
  {https://ui.adsabs.harvard.edu/abs/2002A&A...383...46M} {383, 46}

\bibitem[\protect\citeauthoryear{{Mukherjee}, {Bicknell}, {Wagner},
  {Sutherland}  \& {Silk}}{{Mukherjee} et~al.}{2018}]{Mukherjee18}
{Mukherjee} D.,  {Bicknell} G.~V.,  {Wagner} A.~Y.,  {Sutherland} R.~S.,
  {Silk} J.,  2018, \mn@doi [\mnras] {10.1093/mnras/sty1776}, \href
  {http://adsabs.harvard.edu/abs/2018MNRAS.479.5544M} {479, 5544}

\bibitem[\protect\citeauthoryear{Mukherjee, Bodo, Mignone, Rossi  \&
  Vaidya}{Mukherjee et~al.}{2020}]{Mukherjee20}
Mukherjee D.,  Bodo G.,  Mignone A.,  Rossi P.,   Vaidya B.,  2020, \mn@doi
  [Monthly Notices of the Royal Astronomical Society] {10.1093/mnras/staa2934},
  499, 681–701

\bibitem[\protect\citeauthoryear{{Mullaney}, {Alexander}, {Fine}, {Goulding},
  {Harrison}  \& {Hickox}}{{Mullaney} et~al.}{2013}]{Mullaney13}
{Mullaney} J.~R.,  {Alexander} D.~M.,  {Fine} S.,  {Goulding} A.~D.,
  {Harrison} C.~M.,   {Hickox} R.~C.,  2013, \mn@doi [\mnras]
  {10.1093/mnras/stt751}, \href
  {http://adsabs.harvard.edu/abs/2013MNRAS.433..622M} {433, 622}

\bibitem[\protect\citeauthoryear{{Nagar}, {Falcke}  \& {Wilson}}{{Nagar}
  et~al.}{2005}]{Nagar05}
{Nagar} N.~M.,  {Falcke} H.,   {Wilson} A.~S.,  2005, \mn@doi [\aap]
  {10.1051/0004-6361:20042277}, \href
  {https://ui.adsabs.harvard.edu/abs/2005A&A...435..521N} {435, 521}

\bibitem[\protect\citeauthoryear{{Neff} \& {de Bruyn}}{{Neff} \& {de
  Bruyn}}{1983}]{Neff83}
{Neff} S.~G.,  {de Bruyn} A.~G.,  1983, \aap, \href
  {https://ui.adsabs.harvard.edu/abs/1983A&A...128..318N} {128, 318}

\bibitem[\protect\citeauthoryear{{Nelson} \& {Whittle}}{{Nelson} \&
  {Whittle}}{1996}]{Nelson96}
{Nelson} C.~H.,  {Whittle} M.,  1996, \mn@doi [\apj] {10.1086/177405}, \href
  {https://ui.adsabs.harvard.edu/abs/1996ApJ...465...96N} {465, 96}

\bibitem[\protect\citeauthoryear{{Nesvadba}, {De Breuck}, {Lehnert}, {Best}  \&
  {Collet}}{{Nesvadba} et~al.}{2017}]{Nesvadba17a}
{Nesvadba} N.~P.~H.,  {De Breuck} C.,  {Lehnert} M.~D.,  {Best} P.~N.,
  {Collet} C.,  2017, \mn@doi [\aap] {10.1051/0004-6361/201528040}, \href
  {https://ui.adsabs.harvard.edu/abs/2017A&A...599A.123N} {599, A123}

\bibitem[\protect\citeauthoryear{{Neugebauer} et~al.,}{{Neugebauer}
  et~al.}{1984}]{Neugebauer84}
{Neugebauer} G.,  et~al., 1984, \mn@doi [\apjl] {10.1086/184209}, \href
  {http://adsabs.harvard.edu/abs/1984ApJ...278L...1N} {278, L1}

\bibitem[\protect\citeauthoryear{{Nims}, {Quataert}  \&
  {Faucher-Gigu{\`e}re}}{{Nims} et~al.}{2015}]{Nims15}
{Nims} J.,  {Quataert} E.,   {Faucher-Gigu{\`e}re} C.-A.,  2015, \mn@doi
  [\mnras] {10.1093/mnras/stu2648}, \href
  {http://adsabs.harvard.edu/abs/2015MNRAS.447.3612N} {447, 3612}

\bibitem[\protect\citeauthoryear{{O'Dea}}{{O'Dea}}{1998}]{ODea98}
{O'Dea} C.~P.,  1998, \mn@doi [\pasp] {10.1086/316162}, \href
  {https://ui.adsabs.harvard.edu/abs/1998PASP..110..493O} {110, 493}

\bibitem[\protect\citeauthoryear{{O'Dea} \& {Saikia}}{{O'Dea} \&
  {Saikia}}{2020}]{O'Dea20}
{O'Dea} C.~P.,  {Saikia} D.~J.,  2020, arXiv e-prints, \href
  {https://ui.adsabs.harvard.edu/abs/2020arXiv200902750O} {p. arXiv:2009.02750}

\bibitem[\protect\citeauthoryear{{Orienti} \& {Dallacasa}}{{Orienti} \&
  {Dallacasa}}{2014}]{Orienti14}
{Orienti} M.,  {Dallacasa} D.,  2014, \mn@doi [\mnras] {10.1093/mnras/stt2217},
  \href {https://ui.adsabs.harvard.edu/abs/2014MNRAS.438..463O} {438, 463}

\bibitem[\protect\citeauthoryear{{Padovani}}{{Padovani}}{2016}]{Padovani16}
{Padovani} P.,  2016, \mn@doi [\aapr] {10.1007/s00159-016-0098-6}, \href
  {http://adsabs.harvard.edu/abs/2016A%26ARv..24...13P} {24, 13}

\bibitem[\protect\citeauthoryear{{Padovani}}{{Padovani}}{2017}]{Padovani17b}
{Padovani} P.,  2017, \mn@doi [Nature Astronomy] {10.1038/s41550-017-0194},
  \href {http://adsabs.harvard.edu/abs/2017NatAs...1E.194P} {1, 0194}

\bibitem[\protect\citeauthoryear{{Padovani}, {Bonzini}, {Kellermann}, {Miller},
  {Mainieri}  \& {Tozzi}}{{Padovani} et~al.}{2015}]{Padovani15}
{Padovani} P.,  {Bonzini} M.,  {Kellermann} K.~I.,  {Miller} N.,  {Mainieri}
  V.,   {Tozzi} P.,  2015, \mn@doi [\mnras] {10.1093/mnras/stv1375}, \href
  {https://ui.adsabs.harvard.edu/abs/2015MNRAS.452.1263P} {452, 1263}

\bibitem[\protect\citeauthoryear{{Panessa}, {Baldi}, {Laor}, {Padovani},
  {Behar}  \& {McHardy}}{{Panessa} et~al.}{2019}]{Panessa19}
{Panessa} F.,  {Baldi} R.~D.,  {Laor} A.,  {Padovani} P.,  {Behar} E.,
  {McHardy} I.,  2019, \mn@doi [Nature Astronomy] {10.1038/s41550-019-0765-4},
  \href {https://ui.adsabs.harvard.edu/abs/2019NatAs...3..387P} {3, 387}

\bibitem[\protect\citeauthoryear{{Perna} et~al.,}{{Perna}
  et~al.}{2015}]{Perna15}
{Perna} M.,  et~al., 2015, \mn@doi [\aap] {10.1051/0004-6361/201425035}, \href
  {https://ui.adsabs.harvard.edu/abs/2015A&A...574A..82P} {574, A82}

\bibitem[\protect\citeauthoryear{{Perna} et~al.,}{{Perna}
  et~al.}{2018}]{Perna18}
{Perna} M.,  et~al., 2018, \mn@doi [\aap] {10.1051/0004-6361/201833040}, \href
  {https://ui.adsabs.harvard.edu/abs/2018A&A...619A..90P} {619, A90}

\bibitem[\protect\citeauthoryear{{Pierce}, {Tadhunter}  \& {Morganti}}{{Pierce}
  et~al.}{2020}]{Pierce20}
{Pierce} J.~C.~S.,  {Tadhunter} C.~N.,   {Morganti} R.,  2020, \mn@doi [\mnras]
  {10.1093/mnras/staa531}, \href
  {https://ui.adsabs.harvard.edu/abs/2020MNRAS.494.2053P} {494, 2053}

\bibitem[\protect\citeauthoryear{{Pilbratt} et~al.,}{{Pilbratt}
  et~al.}{2010}]{Pilbratt10}
{Pilbratt} G.~L.,  et~al., 2010, \mn@doi [\aap] {10.1051/0004-6361/201014759},
  \href {http://adsabs.harvard.edu/abs/2010A%26A...518L...1P} {518, L1}

\bibitem[\protect\citeauthoryear{{Poglitsch} et~al.,}{{Poglitsch}
  et~al.}{2010}]{Poglitsch10}
{Poglitsch} A.,  et~al., 2010, \mn@doi [\aap] {10.1051/0004-6361/201014535},
  \href {http://adsabs.harvard.edu/abs/2010A%26A...518L...2P} {518, L2}

\bibitem[\protect\citeauthoryear{{Preuss} \& {Fosbury}}{{Preuss} \&
  {Fosbury}}{1983}]{Preuss83}
{Preuss} E.,  {Fosbury} R.~A.~E.,  1983, \mn@doi [\mnras]
  {10.1093/mnras/204.3.783}, \href
  {https://ui.adsabs.harvard.edu/abs/1983MNRAS.204..783P} {204, 783}

\bibitem[\protect\citeauthoryear{{Rakshit} \& {Woo}}{{Rakshit} \&
  {Woo}}{2018a}]{Rakshit18}
{Rakshit} S.,  {Woo} J.-H.,  2018a, \mn@doi [\apj] {10.3847/1538-4357/aad9f8},
  \href {https://ui.adsabs.harvard.edu/abs/2018ApJ...865....5R} {865, 5}

\bibitem[\protect\citeauthoryear{{Rakshit} \& {Woo}}{{Rakshit} \&
  {Woo}}{2018b}]{Rackshit18}
{Rakshit} S.,  {Woo} J.-H.,  2018b, \mn@doi [\apj] {10.3847/1538-4357/aad9f8},
  \href {https://ui.adsabs.harvard.edu/abs/2018ApJ...865....5R} {865, 5}

\bibitem[\protect\citeauthoryear{{Ramakrishnan} et~al.,}{{Ramakrishnan}
  et~al.}{2019}]{Ramakrishnan19}
{Ramakrishnan} V.,  et~al., 2019, \mn@doi [\mnras] {10.1093/mnras/stz1244},
  \href {https://ui.adsabs.harvard.edu/abs/2019MNRAS.487..444R} {487, 444}

\bibitem[\protect\citeauthoryear{{Reyes} et~al.,}{{Reyes}
  et~al.}{2008}]{Reyes08}
{Reyes} R.,  et~al., 2008, \mn@doi [\aj] {10.1088/0004-6256/136/6/2373}, \href
  {https://ui.adsabs.harvard.edu/abs/2008AJ....136.2373R} {136, 2373}

\bibitem[\protect\citeauthoryear{{Rosario} et~al.,}{{Rosario}
  et~al.}{2018}]{Rosario18}
{Rosario} D.~J.,  et~al., 2018, \mn@doi [\mnras] {10.1093/mnras/stx2670}, \href
  {https://ui.adsabs.harvard.edu/abs/2018MNRAS.473.5658R} {473, 5658}

\bibitem[\protect\citeauthoryear{{Rose}, {Tadhunter}, {Ramos Almeida},
  {Rodr{\'\i}guez Zaur{\'\i}n}, {Santoro}  \& {Spence}}{{Rose}
  et~al.}{2018}]{Rose18}
{Rose} M.,  {Tadhunter} C.,  {Ramos Almeida} C.,  {Rodr{\'\i}guez Zaur{\'\i}n}
  J.,  {Santoro} F.,   {Spence} R.,  2018, \mn@doi [\mnras]
  {10.1093/mnras/stx2590}, \href
  {https://ui.adsabs.harvard.edu/abs/2018MNRAS.474..128R} {474, 128}

\bibitem[\protect\citeauthoryear{{Rupke}, {G{\"u}ltekin}  \&
  {Veilleux}}{{Rupke} et~al.}{2017}]{Rupke17}
{Rupke} D. S.~N.,  {G{\"u}ltekin} K.,   {Veilleux} S.,  2017, \mn@doi [\apj]
  {10.3847/1538-4357/aa94d1}, \href
  {https://ui.adsabs.harvard.edu/abs/2017ApJ...850...40R} {850, 40}

\bibitem[\protect\citeauthoryear{{Sabater} et~al.,}{{Sabater}
  et~al.}{2019}]{Sabater19}
{Sabater} J.,  et~al., 2019, \mn@doi [\aap] {10.1051/0004-6361/201833883},
  \href {https://ui.adsabs.harvard.edu/abs/2019A&A...622A..17S} {622, A17}

\bibitem[\protect\citeauthoryear{{Santoro}, {Tadhunter}, {Baron}, {Morganti}
  \& {Holt}}{{Santoro} et~al.}{2020}]{Santoro20}
{Santoro} F.,  {Tadhunter} C.,  {Baron} D.,  {Morganti} R.,   {Holt} J.,  2020,
  arXiv e-prints, \href {https://ui.adsabs.harvard.edu/abs/2020arXiv200911175S}
  {p. arXiv:2009.11175}

\bibitem[\protect\citeauthoryear{{Schirmer}, {Diaz}, {Holhjem}, {Levenson}  \&
  {Winge}}{{Schirmer} et~al.}{2013}]{Schirmer13}
{Schirmer} M.,  {Diaz} R.,  {Holhjem} K.,  {Levenson} N.~A.,   {Winge} C.,
  2013, \mn@doi [\apj] {10.1088/0004-637X/763/1/60}, \href
  {https://ui.adsabs.harvard.edu/abs/2013ApJ...763...60S} {763, 60}

\bibitem[\protect\citeauthoryear{{Scholtz} et~al.,}{{Scholtz}
  et~al.}{2018}]{Scholtz18}
{Scholtz} J.,  et~al., 2018, \mn@doi [\mnras] {10.1093/mnras/stx3177}, \href
  {https://ui.adsabs.harvard.edu/abs/2018MNRAS.475.1288S} {475, 1288}

\bibitem[\protect\citeauthoryear{{Scholtz} et~al.,}{{Scholtz}
  et~al.}{2020}]{Scholtz20}
{Scholtz} J.,  et~al., 2020, \mn@doi [\mnras] {10.1093/mnras/staa030}, \href
  {https://ui.adsabs.harvard.edu/abs/2020MNRAS.492.3194S} {492, 3194}

\bibitem[\protect\citeauthoryear{{Sch{\"o}nell}, {Storchi-Bergmann}, {Riffel},
  {Riffel}, {Bianchin}, {Dahmer-Hahn}, {Diniz}  \& {Dametto}}{{Sch{\"o}nell}
  et~al.}{2019}]{Schonell19}
{Sch{\"o}nell} A.~J.,  {Storchi-Bergmann} T.,  {Riffel} R.~A.,  {Riffel} R.,
  {Bianchin} M.,  {Dahmer-Hahn} L.~G.,  {Diniz} M.~R.,   {Dametto} N.~Z.,
  2019, \mn@doi [\mnras] {10.1093/mnras/stz523}, \href
  {https://ui.adsabs.harvard.edu/abs/2019MNRAS.485.2054S} {485, 2054}

\bibitem[\protect\citeauthoryear{{Shen}, {Hopkins}, {Faucher-Gigu{\`e}re},
  {Alexander}, {Richards}, {Ross}  \& {Hickox}}{{Shen} et~al.}{2020}]{Shen20}
{Shen} X.,  {Hopkins} P.~F.,  {Faucher-Gigu{\`e}re} C.-A.,  {Alexander} D.~M.,
  {Richards} G.~T.,  {Ross} N.~P.,   {Hickox} R.~C.,  2020, \mn@doi [\mnras]
  {10.1093/mnras/staa1381}, \href
  {https://ui.adsabs.harvard.edu/abs/2020MNRAS.495.3252S} {495, 3252}

\bibitem[\protect\citeauthoryear{{Smith} et~al.,}{{Smith}
  et~al.}{2020}]{Smith20}
{Smith} K.~L.,  et~al., 2020, \mn@doi [\mnras] {10.1093/mnras/stz3608}, \href
  {https://ui.adsabs.harvard.edu/abs/2020MNRAS.492.4216S} {492, 4216}

\bibitem[\protect\citeauthoryear{{Ulvestad}, {Antonucci}  \&
  {Barvainis}}{{Ulvestad} et~al.}{2005}]{Ulvestad05}
{Ulvestad} J.~S.,  {Antonucci} R. R.~J.,   {Barvainis} R.,  2005, \mn@doi
  [\apj] {10.1086/427426}, \href
  {https://ui.adsabs.harvard.edu/abs/2005ApJ...621..123U} {621, 123}

\bibitem[\protect\citeauthoryear{{Veilleux}}{{Veilleux}}{1991}]{Veilleux91}
{Veilleux} S.,  1991, \mn@doi [\apjs] {10.1086/191535}, \href
  {https://ui.adsabs.harvard.edu/abs/1991ApJS...75..383V} {75, 383}

\bibitem[\protect\citeauthoryear{{Venturi} et~al.,}{{Venturi}
  et~al.}{2018}]{Venturi18}
{Venturi} G.,  et~al., 2018, \mn@doi [\aap] {10.1051/0004-6361/201833668},
  \href {https://ui.adsabs.harvard.edu/abs/2018A&A...619A..74V} {619, A74}

\bibitem[\protect\citeauthoryear{{Venturi} et~al.,}{{Venturi}
  et~al.}{2020}]{Venturi21}
{Venturi} G.,  et~al., 2020, arXiv e-prints, \href
  {https://ui.adsabs.harvard.edu/abs/2020arXiv201104677V} {p. arXiv:2011.04677}

\bibitem[\protect\citeauthoryear{{Vietri} et~al.,}{{Vietri}
  et~al.}{2018}]{Vietri18}
{Vietri} G.,  et~al., 2018, \mn@doi [\aap] {10.1051/0004-6361/201732335}, \href
  {https://ui.adsabs.harvard.edu/abs/2018A&A...617A..81V} {617, A81}

\bibitem[\protect\citeauthoryear{{Villar Mart{\'{\i}}n}, {Emonts}, {Humphrey},
  {Cabrera Lavers}  \& {Binette}}{{Villar Mart{\'{\i}}n}
  et~al.}{2014}]{VillarMartin14}
{Villar Mart{\'{\i}}n} M.,  {Emonts} B.,  {Humphrey} A.,  {Cabrera Lavers} A.,
   {Binette} L.,  2014, \mn@doi [\mnras] {10.1093/mnras/stu448}, \href
  {http://adsabs.harvard.edu/abs/2014MNRAS.440.3202V} {440, 3202}

\bibitem[\protect\citeauthoryear{{Villar-Mart{\'\i}n}, {Arribas}, {Emonts},
  {Humphrey}, {Tadhunter}, {Bessiere}, {Cabrera Lavers}  \& {Ramos
  Almeida}}{{Villar-Mart{\'\i}n} et~al.}{2016}]{VillarMartin16}
{Villar-Mart{\'\i}n} M.,  {Arribas} S.,  {Emonts} B.,  {Humphrey} A.,
  {Tadhunter} C.,  {Bessiere} P.,  {Cabrera Lavers} A.,   {Ramos Almeida} C.,
  2016, \mn@doi [\mnras] {10.1093/mnras/stw901}, \href
  {https://ui.adsabs.harvard.edu/abs/2016MNRAS.460..130V} {460, 130}

\bibitem[\protect\citeauthoryear{{Vink}, {Snellen}, {Mack}  \&
  {Schilizzi}}{{Vink} et~al.}{2006}]{Vink06}
{Vink} J.,  {Snellen} I.,  {Mack} K.-H.,   {Schilizzi} R.,  2006, \mn@doi
  [\mnras] {10.1111/j.1365-2966.2006.10036.x}, \href
  {https://ui.adsabs.harvard.edu/abs/2006MNRAS.367..928V} {367, 928}

\bibitem[\protect\citeauthoryear{{Wagner}, {Bicknell}  \& {Umemura}}{{Wagner}
  et~al.}{2012}]{Wagner12}
{Wagner} A.~Y.,  {Bicknell} G.~V.,   {Umemura} M.,  2012, \mn@doi [\apj]
  {10.1088/0004-637X/757/2/136}, \href
  {https://ui.adsabs.harvard.edu/abs/2012ApJ...757..136W} {757, 136}

\bibitem[\protect\citeauthoryear{{Wang}, {Rowan-Robinson}, {Norberg}, {Heinis}
  \& {Han}}{{Wang} et~al.}{2014}]{Wang14}
{Wang} L.,  {Rowan-Robinson} M.,  {Norberg} P.,  {Heinis} S.,   {Han} J.,
  2014, \mn@doi [\mnras] {10.1093/mnras/stu915}, \href
  {http://adsabs.harvard.edu/abs/2014MNRAS.442.2739W} {442, 2739}

\bibitem[\protect\citeauthoryear{{Wang}, {Xu}  \& {Wei}}{{Wang}
  et~al.}{2018}]{Wang18}
{Wang} J.,  {Xu} D.~W.,   {Wei} J.~Y.,  2018, \mn@doi [\apj]
  {10.3847/1538-4357/aa9d1b}, \href
  {https://ui.adsabs.harvard.edu/abs/2018ApJ...852...26W} {852, 26}

\bibitem[\protect\citeauthoryear{{Wehrle} \& {Morris}}{{Wehrle} \&
  {Morris}}{1988}]{Wehrle88}
{Wehrle} A.~E.,  {Morris} M.,  1988, \mn@doi [\aj] {10.1086/114765}, \href
  {https://ui.adsabs.harvard.edu/abs/1988AJ.....95.1689W} {95, 1689}

\bibitem[\protect\citeauthoryear{{White}, {Jarvis}, {H{\"a}u{\ss}ler}  \&
  {Maddox}}{{White} et~al.}{2015}]{White15}
{White} S.~V.,  {Jarvis} M.~J.,  {H{\"a}u{\ss}ler} B.,   {Maddox} N.,  2015,
  \mn@doi [\mnras] {10.1093/mnras/stv134}, \href
  {https://ui.adsabs.harvard.edu/abs/2015MNRAS.448.2665W} {448, 2665}

\bibitem[\protect\citeauthoryear{{White}, {Jarvis}, {Kalfountzou},
  {Hardcastle}, {Verma}, {Cao Orjales}  \& {Stevens}}{{White}
  et~al.}{2017}]{White17}
{White} S.~V.,  {Jarvis} M.~J.,  {Kalfountzou} E.,  {Hardcastle} M.~J.,
  {Verma} A.,  {Cao Orjales} J.~M.,   {Stevens} J.,  2017, \mn@doi [\mnras]
  {10.1093/mnras/stx284}, \href
  {https://ui.adsabs.harvard.edu/abs/2017MNRAS.468..217W} {468, 217}

\bibitem[\protect\citeauthoryear{{Whittle}}{{Whittle}}{1992}]{Whittle92}
{Whittle} M.,  1992, \mn@doi [\apj] {10.1086/171064}, \href
  {https://ui.adsabs.harvard.edu/abs/1992ApJ...387..109W} {387, 109}

\bibitem[\protect\citeauthoryear{{Wilson} \& {Heckman}}{{Wilson} \&
  {Heckman}}{1985}]{Wilson85}
{Wilson} A.~S.,  {Heckman} T.~M.,  1985, in {Miller} J.~S.,  ed., Astrophysics
  of Active Galaxies and Quasi-Stellar Objects. pp 39--109

\bibitem[\protect\citeauthoryear{{Wong} et~al.,}{{Wong} et~al.}{2016}]{Wong16}
{Wong} O.~I.,  et~al., 2016, \mn@doi [\mnras] {10.1093/mnras/stw957}, \href
  {https://ui.adsabs.harvard.edu/abs/2016MNRAS.460.1588W} {460, 1588}

\bibitem[\protect\citeauthoryear{{Woods}, {Klein}, {Castor}, {McKee}  \&
  {Bell}}{{Woods} et~al.}{1996}]{Woods96}
{Woods} D.~T.,  {Klein} R.~I.,  {Castor} J.~I.,  {McKee} C.~F.,   {Bell} J.~B.,
   1996, \mn@doi [\apj] {10.1086/177101}, \href
  {https://ui.adsabs.harvard.edu/abs/1996ApJ...461..767W} {461, 767}

\bibitem[\protect\citeauthoryear{{Wright} et~al.,}{{Wright}
  et~al.}{2010}]{Wright10}
{Wright} E.~L.,  et~al., 2010, \mn@doi [\aj] {10.1088/0004-6256/140/6/1868},
  \href {http://adsabs.harvard.edu/abs/2010AJ....140.1868W} {140, 1868}

\bibitem[\protect\citeauthoryear{{Wylezalek} \& {Morganti}}{{Wylezalek} \&
  {Morganti}}{2018}]{Wylezalek18}
{Wylezalek} D.,  {Morganti} R.,  2018, \mn@doi [Nature Astronomy]
  {10.1038/s41550-018-0409-0}, \href
  {https://ui.adsabs.harvard.edu/abs/2018NatAs...2..181W} {2, 181}

\bibitem[\protect\citeauthoryear{{Wylezalek}, {Flores}, {Zakamska}, {Greene}
  \& {Riffel}}{{Wylezalek} et~al.}{2020}]{Wylezalek20}
{Wylezalek} D.,  {Flores} A.~M.,  {Zakamska} N.~L.,  {Greene} J.~E.,   {Riffel}
  R.~A.,  2020, \mn@doi [\mnras] {10.1093/mnras/staa062}, \href
  {https://ui.adsabs.harvard.edu/abs/2020MNRAS.492.4680W} {492, 4680}

\bibitem[\protect\citeauthoryear{{Xu}, {Livio}  \& {Baum}}{{Xu}
  et~al.}{1999}]{Xu99}
{Xu} C.,  {Livio} M.,   {Baum} S.,  1999, \mn@doi [\aj] {10.1086/301007}, \href
  {http://adsabs.harvard.edu/abs/1999AJ....118.1169X} {118, 1169}

\bibitem[\protect\citeauthoryear{{Yesuf} \& {Ho}}{{Yesuf} \&
  {Ho}}{2020}]{Yesuf20}
{Yesuf} H.~M.,  {Ho} L.~C.,  2020, arXiv e-prints, \href
  {https://ui.adsabs.harvard.edu/abs/2020arXiv200712026Y} {p. arXiv:2007.12026}

\bibitem[\protect\citeauthoryear{{Zakamska} \& {Greene}}{{Zakamska} \&
  {Greene}}{2014}]{Zakamska14}
{Zakamska} N.~L.,  {Greene} J.~E.,  2014, \mn@doi [\mnras]
  {10.1093/mnras/stu842}, \href
  {http://adsabs.harvard.edu/abs/2014MNRAS.442..784Z} {442, 784}

\bibitem[\protect\citeauthoryear{{Zakamska}, {Strauss}, {Heckman}, {Ivezi{\'c}}
   \& {Krolik}}{{Zakamska} et~al.}{2004}]{Zakamska04}
{Zakamska} N.~L.,  {Strauss} M.~A.,  {Heckman} T.~M.,  {Ivezi{\'c}} {\v{Z}}.,
  {Krolik} J.~H.,  2004, \mn@doi [\aj] {10.1086/423220}, \href
  {https://ui.adsabs.harvard.edu/abs/2004AJ....128.1002Z} {128, 1002}

\bibitem[\protect\citeauthoryear{{Zakamska} et~al.,}{{Zakamska}
  et~al.}{2016a}]{Zakamska16}
{Zakamska} N.~L.,  et~al., 2016a, \mn@doi [\mnras] {10.1093/mnras/stv2571},
  \href {http://adsabs.harvard.edu/abs/2016MNRAS.455.4191Z} {455, 4191}

\bibitem[\protect\citeauthoryear{{Zakamska} et~al.,}{{Zakamska}
  et~al.}{2016b}]{Zakamska16b}
{Zakamska} N.~L.,  et~al., 2016b, \mn@doi [\mnras] {10.1093/mnras/stw718},
  \href {https://ui.adsabs.harvard.edu/abs/2016MNRAS.459.3144Z} {459, 3144}

\bibitem[\protect\citeauthoryear{{do Nascimento} et~al.,}{{do Nascimento}
  et~al.}{2019}]{doNascimento19}
{do Nascimento} J.~C.,  et~al., 2019, \mn@doi [\mnras] {10.1093/mnras/stz1083},
  \href {https://ui.adsabs.harvard.edu/abs/2019MNRAS.486.5075D} {486, 5075}

\makeatother
\end{thebibliography}

%%%%%%%%%%%%%%%%%%%%%%%%%%%%%%%%%%%%%%%%%%%%%%%%%%

%%%%%%%%%%%%%%%%% APPENDICES %%%%%%%%%%%%%%%%%%%%%

\appendix

\section*{Supporting Information}

Supplementary data are presented in a companion PDF.

\smallskip

\noindent
\textbf{Appendix A.} Observations and Images.

\noindent
\textbf{Table A1.} The key properties of the VLA observations and images used in this work.

\noindent
\textbf{Appendix B.} Images and notes on individual objects.

\noindent
\textbf{Figures B1--B41.} As Fig.~\ref{fig:app:J1553+4407} but for remainder of sample.

\refstepcounter{section}
\label{app:obs}

%-------------------------------------------------------------

\refstepcounter{table}
    \label{tab:obs_tab} 

 %-------------------------------------------------------------

\refstepcounter{section}
\label{app:objects}

\refstepcounter{subsection}
   \label{sec:app:J0749+4510}

\refstepcounter{subsection}
   \label{sec:app:J0752+1935}

\refstepcounter{subsection}
   \label{sec:app:J0759+5050}

\refstepcounter{subsection}
   \label{sec:app:J0802+4643}

\refstepcounter{subsection}
   \label{sec:app:J0842+0759}

\refstepcounter{subsection}
   \label{sec:app:J0842+2048}

\refstepcounter{subsection}
   \label{sec:app:J0907+4620}

\refstepcounter{subsection}
   \label{sec:app:J0909+1052}

\refstepcounter{subsection}
   \label{sec:app:J0945+1737}

\refstepcounter{subsection}
   \label{sec:app:J0946+1319}

\refstepcounter{subsection}
   \label{sec:app:J0958+1439}

\refstepcounter{subsection}
   \label{sec:app:J1000+1242}

\refstepcounter{subsection}
   \label{sec:app:J1010+0612}

\refstepcounter{subsection}
   \label{sec:app:J1010+1413}

\refstepcounter{subsection}
   \label{sec:app:J1016+0028}

\refstepcounter{subsection}
   \label{sec:app:J1016+5358}

\refstepcounter{subsection}
   \label{sec:app:J1045+0843}

\refstepcounter{subsection}
   \label{sec:app:J1055+1102}

\refstepcounter{subsection}
   \label{sec:app:J1100+0846}

\refstepcounter{subsection}
   \label{sec:app:J1108+0659}

\refstepcounter{subsection}
   \label{sec:app:J1114+1939}

\refstepcounter{subsection}
   \label{sec:app:J1116+2200}

\refstepcounter{subsection}
   \label{sec:app:J1222-0007}

\refstepcounter{subsection}
   \label{sec:app:J1223+5409}

\refstepcounter{subsection}
   \label{sec:app:J1227+0419}

\refstepcounter{subsection}
   \label{sec:app:J1300+0355}

\refstepcounter{subsection}
   \label{sec:app:J1302+1624}

\refstepcounter{subsection}
   \label{sec:app:J1316+1753}

\refstepcounter{subsection}
   \label{sec:app:J1324+5849}

\refstepcounter{subsection}
   \label{sec:app:J1347+1217}

\refstepcounter{subsection}
   \label{sec:app:J1355+2046}

\refstepcounter{subsection}
   \label{sec:app:J1356+1026}

\refstepcounter{subsection}
   \label{sec:app:J1430+1339}

\refstepcounter{subsection}
   \label{sec:app:J1436+4928}

\refstepcounter{subsection}
   \label{sec:app:J1454+0803}

\refstepcounter{subsection}
   \label{sec:app:J1509+1757}

\refstepcounter{subsection}
   \label{sec:app:J1518+1403}

\refstepcounter{subsection}
   \label{sec:app:J1553+4407}

\refstepcounter{subsection}
   \label{sec:app:J1555+5403}

\refstepcounter{subsection}
   \label{sec:app:J1655+2146}

\refstepcounter{subsection}
   \label{sec:app:J1701+2226}

\refstepcounter{subsection}
   \label{sec:app:J1715+6008}

\refstepcounter{figure}
    \label{fig:app:J0749+4510}

\refstepcounter{figure}
    \label{fig:app:J0752+1935}

\refstepcounter{figure}
    \label{fig:app:J0759+5050}

\refstepcounter{figure}
    \label{fig:app:J0802+4643}

\refstepcounter{figure}
    \label{fig:app:J0842+0759}

\refstepcounter{figure}
    \label{fig:app:J0842+2048}

\refstepcounter{figure}
    \label{fig:app:J0907+4620}

\refstepcounter{figure}
    \label{fig:app:J0909+1052}

\refstepcounter{figure}
    \label{fig:app:J0945+1737}

\refstepcounter{figure}
    \label{fig:app:J0946+1319}

\refstepcounter{figure}
    \label{fig:app:J0958+1439}

\refstepcounter{figure}
    \label{fig:app:J1000+1242}

\refstepcounter{figure}
    \label{fig:app:J1010+0612}

\refstepcounter{figure}
    \label{fig:app:J1010+1413}

\refstepcounter{figure}
    \label{fig:app:J1016+0028}

\refstepcounter{figure}
    \label{fig:app:J1016+5358}

\refstepcounter{figure}
    \label{fig:app:J1045+0843}

\refstepcounter{figure}
    \label{fig:app:J1055+1102}

\refstepcounter{figure}
    \label{fig:app:J1100+0846}

\refstepcounter{figure}
    \label{fig:app:J1108+0659}

\refstepcounter{figure}
    \label{fig:app:J1114+1939}

\refstepcounter{figure}
    \label{fig:app:J1116+2200}

\refstepcounter{figure}
    \label{fig:app:J1222-0007}

\refstepcounter{figure}
    \label{fig:app:J1223+5409}

\refstepcounter{figure}
    \label{fig:app:J1227+0419}

\refstepcounter{figure}
    \label{fig:app:J1300+0355}

\refstepcounter{figure}
    \label{fig:app:J1302+1624}

\refstepcounter{figure}
    \label{fig:app:J1316+1753}

\refstepcounter{figure}
    \label{fig:app:J1324+5849}

\refstepcounter{figure}
    \label{fig:app:J1347+1217}

\refstepcounter{figure}
    \label{fig:app:J1355+2046}

\refstepcounter{figure}
    \label{fig:app:J1356+1026}

\refstepcounter{figure}
    \label{fig:app:J1430+1339}

\refstepcounter{figure}
    \label{fig:app:J1436+4928}

\refstepcounter{figure}
    \label{fig:app:J1454+0803}

\refstepcounter{figure}
    \label{fig:app:J1509+1757}

\refstepcounter{figure}
    \label{fig:app:J1518+1403}

\refstepcounter{figure}
    \label{fig:app:J1555+5403}

\refstepcounter{figure}
    \label{fig:app:J1655+2146}

\refstepcounter{figure}
    \label{fig:app:J1701+2226}

\refstepcounter{figure}
    \label{fig:app:J1715+6008}
    
%%%%%%%%%%%%%%%%%%%%%%%%%%%%%%%%%%%%%%%%%%%%%%%%%%

% Don't change these lines
\bsp	% typesetting comment
\label{lastpage}
\end{document}